\documentclass[prd,aps,preprintnumbers,11pt,nofootinbib,twoside,raggedbottom]{revtex4-1}

\usepackage[utf8]{inputenc}

\usepackage{xpatch}
\makeatletter
\patchcmd{\@ssect@ltx}
    {\addcontentsline{toc}{#1}{\protect\numberline{}#8}}
    {}
    {}
    {}
\makeatother

\usepackage[colorlinks,linktocpage=true]{hyperref}
\hypersetup{
	colorlinks   = true,
	citecolor    = blue,
	linkcolor    = blue
}

\usepackage{bbm}
\usepackage{amsfonts}
\usepackage{mathrsfs}
\usepackage{slashed}
\usepackage{bbold}
\usepackage{upgreek}
\usepackage{amsmath,amssymb}
\usepackage{graphicx}               
\usepackage{url}
\usepackage{multirow}
\usepackage[dvipsnames,table,xcdraw]{xcolor}
\usepackage[version=4]{mhchem}
\usepackage{booktabs}
\usepackage{fontawesome}

\makeatletter
\def\simgt{\mathrel{\lower2.5pt\vbox{\lineskip=0pt\baselineskip=0pt
           \hbox{$>$}\hbox{$\sim$}}}}
\def\simlt{\mathrel{\lower2.5pt\vbox{\lineskip=0pt\baselineskip=0pt
           \hbox{$<$}\hbox{$\sim$}}}}
\makeatother

\newcommand{\LL}{\mathcal{L}}
\newcommand{\OO}{\mathcal{O}}

\def\hl#1{{\color{BrickRed} #1}}

\def\vect#1{\boldsymbol{#1}}
\def\tens#1{{\bf #1}}
\def\epstens{\boldsymbol{\upvarepsilon}}

\def\edm{\text{edm}}
\def\mdm{\text{mdm}}
\def\ana{\text{ana}}

\def\ve{v_\text{e}}
\def\vve{\vect{v}_\text{e}}
\def\vesc{v_\text{esc}}

\newcommand\Tstrut{\rule{0pt}{3.5ex}}         
\newcommand\Bstrut{\rule[-2ex]{0pt}{0pt}}   
\renewcommand\tilde{\widetilde}

\allowdisplaybreaks

\begin{document}
	

\title{Effective Field Theory of Dark Matter Direct Detection\\ With Collective Excitations
}

\preprint{CALT-TH-2020-037}

\author{Tanner~Trickle}
\author{Zhengkang~Zhang}
\author{Kathryn~M.~Zurek}

\affiliation{Walter Burke Institute for Theoretical Physics, California Institute of Technology, Pasadena, CA 91125, USA}

\begin{abstract}
\vspace{10pt}
We develop a framework for computing light dark matter direct detection rates through single phonon and magnon excitations via general effective operators. Our work generalizes previous calculations focused on spin-independent interactions involving the total nucleon and electron numbers $N$ (the usual route to excite phonons) and spin-dependent interactions involving the total electron spin $\vect{S}$ (the usual route to excite magnons), leading us to identify new responses involving the orbital angular momenta $\vect{L}$, as well as spin-orbit couplings $\vect{L}\otimes\vect{S}$ in the target. All four types of responses can excite phonons, while couplings to electron's $\vect{S}$ and $\vect{L}$ can also excite magnons. We apply the effective field theory approach to a set of well-motivated relativistic benchmark models, including (pseudo-)scalar mediated interactions, and models where dark matter interacts via a multipole moment, such as a dark electric dipole, magnetic dipole or anapole moment. We find that couplings to point-like degrees of freedom $N$ and $\vect{S}$ often dominate dark matter detection rates, implying that exotic materials with orbital $\vect{L}$ order or large spin-orbit couplings $\vect{L}\otimes\vect{S}$ are not necessary to have strong reach to a broad class of DM models. We highlight that phonon based crystal experiments in active R\&D (such as SPICE) will probe light dark matter models well beyond those having a simple spin-independent interaction, including {\it e.g.}\ models with dipole and anapole interactions.  Lastly, we make publicly available a code, \href{https://phonodark.caltech.edu}{\textsf{PhonoDark}}, which computes single phonon production rates in a wide variety of materials with the effective field theory framework.
\end{abstract}

\maketitle

\tableofcontents
\clearpage

\section{Introduction}

Light dark matter (DM) with sub-GeV mass is theoretically well-motivated~\cite{Boehm:2003hm,Pospelov:2007mp,Hooper:2008im,Kumar:2009bw,ArkaniHamed:2008qp,Cheung:2009qd,Morrissey:2009ur,Kaplan:2009ag,Cohen:2010kn,Hall:2009bx,Hochberg:2014dra} but difficult to detect in traditional WIMP-focused experiments \cite{Aprile:2019xxb, Agnese:2018gze, Agnes:2018ves, Cui:2017nnn, Akerib:2018hck}. This can be understood from simple scattering kinematics: if the DM mass $m_\chi\lesssim$ GeV, the amount of energy deposited in a nuclear recoil process, $\omega = \frac{q^2}{2 m_N}$, is suppressed by the heavy target nucleus mass $m_N$ and limited by the possible momentum transfer $q \lesssim 2 m_\chi v$. This, along with a steady improvement to the energy sensitivity of detectors \cite{Pyle:2015pya,Maris:2017xvi, Rothe:2018bnc, Colantoni:2020cet,Fink:2020noh}, has motivated the study of excitation channels far outside the scope of standard nuclear recoil. Perhaps the most studied alternative is electronic excitations, in a variety of different targets, \textit{e.g.}\ individual atoms~\cite{Essig:2011nj,Graham:2012su,Lee:2015qva,Essig:2015cda,Essig:2017kqs,Catena:2019gfa}, semiconductors and scintillators~\cite{Essig:2011nj,Graham:2012su,Lee:2015qva,Essig:2015cda,Derenzo:2016fse,Agnese:2018col,Kurinsky:2019pgb,Abramoff:2019dfb,Aguilar-Arevalo:2019wdi,Trickle:2019nya,Griffin:2019mvc,Andersson:2020uwc,Barak:2020fql,Catena:2021qsr}, superconductors~\cite{Hochberg:2015pha, Hochberg:2015fth}, aromatic organic targets~\cite{Blanco:2019lrf}, graphene~\cite{Hochberg:2016ntt} and Dirac materials~\cite{Hochberg:2017wce,Coskuner:2019odd, Geilhufe:2019ndy}. The smallest DM mass that can be probed is limited by the band gap in these materials, typically $\mathcal{O}$(eV) corresponding to DM masses $\gtrsim$ MeV (the exceptions being superconductors and Dirac materials which typically have ${\cal O}$(meV) gaps and sensitivity to keV scale DM). 

For sensitivity to smaller energy deposits, and optimal reach to light DM and mediating particles, we look toward excitations at sub-eV energies.  Such excitations exist and are derived from collective behaviors of atoms, ions or electrons in condensed matter systems.  Phonons were proposed in Ref.~\cite{Schutz:2016tid} and further studied in Refs.~\cite{Knapen:2016cue,Acanfora:2019con,Caputo:2019cyg,Caputo:2019xum,Baym:2020uos} for direct detection in superfluid helium (where maxon and roton excitations also contribute), and were also discussed in the context of bosonic DM absorption in superconductors~\cite{Hochberg:2016ajh} and semiconductors~\cite{Hochberg:2016sqx}, though ultimately, acoustic and optical phonons in (polar) crystals were advanced~\cite{Knapen:2017ekk} and shown to have the best experimental prospects and sensitivity to light dark matter~\cite{Griffin:2018bjn,Trickle:2019nya,Griffin:2019mvc,Campbell-Deem:2019hdx,Griffin:2020lgd}.  Magnons -- quanta of collective spin excitations -- were also proposed in Ref.~\cite{Trickle:2019ovy}. Both phonons and magnons in crystal targets have typical energies up to $\mathcal{O}(100\,\text{meV})$. To date, the work in the literature has focused on demonstrating the sensitivity of phonons and magnons to simple DM models. Only spin-independent (SI) interactions, via couplings to linear combinations of the proton, neutron and electron numbers, have been considered for phonon excitations, while a few benchmark models have been studied for magnon excitations.

The goal of this paper is to extend these results to general types of DM interactions. Effective field theory (EFT) is well-suited for this purpose: we can match a relativistic theory of DM onto a nonrelativistic (NR) EFT, and then compute the target response to the EFT operators. Within this framework, starting from a UV model consisting of relativistic operators coupling the DM to the proton, neutron, and/or electron, we can systematically compute direct detection rates via single phonon and magnon excitations in various target materials. The idea is along the lines of previous works on EFT calculations of nuclear recoils~\cite{Fitzpatrick:2012ix,DelNobile:2013sia,Anand:2013yka,Gresham:2014vja,Anand:2014kea,DelNobile:2018dfg} (which extend earlier studies focused on standard SI and spin-dependent (SD) DM-nucleon interactions), and, more recently, of electron excitations in atoms~\cite{Catena:2019gfa} and crystals~\cite{Catena:2021qsr} (which extends earlier studies focused on SI DM-electron interactions), but technically there are important differences. Specifically, our EFT approach to DM-induced single phonon and magnon excitations consists of the following steps:
\begin{enumerate}
	\setlength{\itemsep}{3pt}
	\item Matching of a relativistic theory of DM interactions onto the NR EFT (DM model-specific). 
	\item Matching of NR operators onto DM couplings to lattice degrees of freedom (universal).
	\item Calculation of phonon or magnon excitation matrix elements (target- and excitation-specific).
\end{enumerate}

We explain each of these steps in the three subsections of Sec.~\ref{sec:EFT}. The first step -- matching relativistic DM theories to the NR EFT -- follows a similar procedure as previous works~\cite{Fitzpatrick:2012ix,DelNobile:2013sia,Anand:2013yka,Gresham:2014vja,Anand:2014kea,DelNobile:2018dfg,Catena:2019gfa,Catena:2021qsr}, but involves a larger set of independent operators due to the absence of Galilean invariance in a medium. For nuclear recoils, one then derives the nuclear responses to the EFT operators. Analogously, the key quantities in the present case are {\it crystal responses} which determines how DM couples to the collective excitations. (We emphasize, however, that despite the similar choice of terminology, collective excitations are associated with a different kinematic regime and degrees of freedom than nuclear recoils and therefore require a distinct EFT calculation.) Technically, for both phonon and magnon excitations in crystal targets, the second step listed above involves matching the NR EFT of DM-nucleon and DM-electron interactions onto an effective scattering potential that involves ionic degrees of freedom in the crystal lattice --- in the long wavelength (low momentum transfer) limit relevant for light DM scattering, these (as we will \hl{highlight} throughout) are quantities that characterize an ion as a whole, including the total particle numbers \hl{$\langle N_\psi\rangle$} for the proton, neutron and electron ($\psi=p,n,e$), total spins \hl{$\langle \vect{S}_\psi \rangle$}, orbital angular momenta \hl{$\langle \vect{L}_\psi \rangle$}, as well as spin-orbit couplings \hl{$\langle \vect{L}_\psi \otimes \vect{S}_\psi \rangle$} (a tensor with components $\langle L_\psi^i S_\psi^k \rangle$ summed over the constituent nucleons/electrons). Finally, in the third step, we quantize the scattering potential to obtain the phonon and magnon modes in a specific target material and compute the matrix elements for exciting them. All four types of crystal responses highlighted above can lead to phonon excitation in appropriately chosen targets, while \hl{$\langle \vect{S}_e \rangle$} and \hl{$\langle \vect{L}_e \rangle$} can also lead to magnon excitation. 

Our new results significantly extend the searchable DM model space via phonon and magnon excitations, which we showcase in Sec.~\ref{sec:benchmarks} with a variety of well-motivated benchmark models. We present full numerical calculations for several representative target materials, and apply simple analytic estimates to understand the results.  We compare for which operators and interactions one expects phonon versus magnon excitations to dominate the rate, quantifying and generalizing the discussion in Ref.~\cite{Trickle:2019ovy}. These calculations highlight the complementarity between phonon and magnon excitations, and between different targets, in probing the light DM theory space. Our code for computing single phonon excitation rates, \textsf{PhonoDark}, is publicly available~\cite{dm-phonon-scatter} and will be explained in detail in a forthcoming publication; it integrates the open-source phonon eigensystem solver \textsf{phonopy}~\cite{phonopy}, and takes general NR EFT operator coefficients, together with density functional theory (DFT) calculations of material properties, as input. Our magnon code, based on the Toth-Lake algorithm~\cite{Toth-Lake} for solving the magnon eigensystem, is also available upon request.

\section{Effective Field Theory Calculation of Dark Matter Induced Collective Excitations}
\label{sec:EFT}

Our goal is to present a framework for computing direct detection rates for general DM models, for the process where a DM particle scatters off a crystal target and induces a quasiparticle excitation in the crystal. This quantum mechanical process follows Fermi's golden rule which, when the incoming and outgoing DM particles are momentum eigenstates in free space, takes the form 
\begin{equation}
\Gamma (\vect{v}) = \frac{1}{V} \int \frac{d^3q}{(2\pi)^3} \sum_f \bigl|\langle f |\, \widetilde{\cal V} (-\vect{q},\vect{v})\, |i\rangle \bigr|^2 \,2\pi\,\delta\bigl(E_f-E_i-\omega_{\vect{q}}\bigr)\,,
\label{eq:Fermi_Gamma}
\end{equation}
where $\vect{v}$ is the incoming DM's velocity, $V$ is the total target volume, $|i\rangle$ and $|f\rangle$ are the initial and final states of the target system (defined with NR normalization: $\langle i | i \rangle$ = $\langle f | f \rangle = 1$), and $\widetilde{\cal V}$ is the Fourier transform of the scattering potential. The momentum transfer from the DM to the target, $\vect{q}$, is integrated over, while the energy deposition onto the target is constrained to be
\begin{equation}
\omega_{\vect{q}} = \frac{1}{2} m_\chi v^2 - \frac{(m_\chi \vect{v}-\vect{q})^2}{2m_\chi} = \vect{q}\cdot\vect{v}-\frac{q^2}{2m_\chi}\,.
\end{equation}
See Ref.~\cite{Trickle:2019nya} for a review of the general formalism.

We now need to specify the type of transitions $|i\rangle\to|f\rangle$ in the target system to calculate the matrix element $\langle f |\, \widetilde{\cal V} (-\vect{q},\vect{v})\, |i\rangle$. Here we focus on excitation of single phonon or magnon in a crystal target at zero temperature. We therefore take $|i\rangle$ to be the ground state $|0\rangle$, and $|f\rangle$ to be the one-phonon or one-magnon states $|\nu,\vect{k}\rangle$, labeled by branch $\nu$ and momentum $\vect{k}$ within the first Brillouin zone (1BZ). For a crystal target, we write the scattering potential as a sum of contributions from individual ions:\footnote{For simplicity, we will refer to either atoms or ions on lattice sites as ions.}
\begin{equation}
{\cal V} (\vect{x}, \vect{v}) = \sum_{lj} {\cal V}_{lj} (\vect{x}-\vect{x}_{lj} ,\vect{v})\,,
\end{equation}
where $l=1,\dots, N$ labels the primitive cells, $j=1,\dots, n$ labels the ions within each primitive cell, and $\vect{x}_{lj}$ is the position of the ion labeled by $l$, $j$. Therefore,
\begin{equation}
\widetilde{\cal V} (-\vect{q},\vect{v}) = \int d^3x \,e^{i\vect{q}\cdot\vect{x}}\, {\cal V} (\vect{x}, \vect{v})
= \sum_{l,j} e^{i\vect{q}\cdot\vect{x}_{lj}} \,\widetilde{\cal V}_{lj} (-\vect{q},\vect{v})\,,
\end{equation}
and we obtain
\begin{equation}
\Gamma(\vect{v}) = \frac{1}{V} \int \frac{d^3q}{(2\pi)^3} \sum_{\nu,\vect{k}} \biggl|\sum_{l,j} \langle \nu,\vect{k} |\,e^{i\vect{q}\cdot\vect{x}_{lj}} \,\widetilde{\cal V}_{lj} (-\vect{q},\vect{v}) |0\rangle \biggr|^2 \,2\pi\,\delta\bigl(\omega_{\nu,\vect{k}}-\omega_{\vect{q}}\bigr) \,.
\label{eq:Gamma_phononmagnon}
\end{equation}

The central quantity for the rate calculation is then the lattice potential $\widetilde{\cal V}_{lj}$ which the DM senses. This will depend on both the specific DM model and on the lattice degrees of freedom ({\em e.g.} the nucleon/electron number or total electronic spin of the ions) available to scatter from. We will determine the lattice potential $\widetilde{\cal V}_{lj}$ in two steps previously mentioned in the introduction: first, in Sec.~\ref{subsec:NR_matching}, we review the procedure of matching relativistic DM models onto NR effective operators; next, in Sec.~\ref{subsec:lattice_matching}, we further reduce these effective operators to DM couplings to the lattice degrees of freedom. In the simplest case of SI interactions, there is only one effective operator, $\OO_1=\mathbb{1}$, and $\widetilde{\cal V}_{lj}$ is a linear combination of $\hl{\langle N_p \rangle}_{lj}$, $\hl{\langle N_n \rangle}_{lj}$ and $\hl{\langle N_e \rangle}_{lj}$ (proton, neutron and electron numbers of the ions, respectively)~\cite{Trickle:2019nya}. More generally, a DM model can generate a larger set of effective operators that involve the spins, momentum transfer, and velocities. The resulting lattice potential $\widetilde{\cal V}_{lj}$ depends on  lattice degrees of freedom that include, in addition to the particle numbers $\hl{\langle N_\psi \rangle}_{lj}$ ($\psi = p,n,e$), also their spins $\hl{\langle \vect{S}_\psi \rangle}_{lj}$, orbital angular momenta $\hl{\langle \vect{L}_\psi \rangle}_{lj}$, as well as spin-orbit couplings $\hl{\langle \vect{L}_\psi \otimes \vect{S}_\psi \rangle}_{lj}$ (a tensor with components $\langle L^i_\psi S^k_\psi \rangle_{lj}$, see Eq.~\eqref{eq:LS_def} below). The last step in computing the scattering rate is to quantize $\widetilde{\cal V}_{lj}$ in terms of phonon/magnon creation and annihilation operators; we carry out this exercise in Sec.~\ref{subsec:quantization}. The framework in this section will provide the basis for concrete calculations of direct detection rates via single phonon and magnon excitations, and will be applied to a set of benchmark models in Sec.~\ref{sec:benchmarks}.

\subsection{From Dark Matter Models to Nonrelativistic Effective Operators}
\label{subsec:NR_matching}

In this subsection, we take a top-down approach in deriving the EFT, focusing on how the effective operators arise from NR matching of well-motivated relativistic models. While one can also take a bottom-up approach as in {\it e.g.}\ Ref.~\cite{Fitzpatrick:2012ix}, and construct the EFT by enumerating operators consistent with rotation and translation invariance, we find it useful to have a set of benchmark UV models to develop intuition on how realistic theories of DM, which often predict correlations between EFT operators \cite{Fitzpatrick:2010br,Gresham:2014vja}, can be probed by experiment.

Let us start from a relativistic model of a DM particle $\chi$ interacting with the proton ($p$), neutron ($n$) and electron ($e$);\footnote{The DM-proton and DM-neutron couplings follow from the DM-quark and DM-gluon couplings in the fundamental Lagrangian by standard methods, see {\it e.g.}\ Ref.~\cite{DelNobile:2013sia}.} we denote these Standard Model (SM) particles collectively by $\psi$ in the following. To compute the NR EFT, we take the NR limit of the relativistic theory and map it on to the appropriate NR degrees of freedom. The EFT consists of the NR fields $\chi^\pm$, $\psi^\pm$, generally defined by (using the SM fermion $\psi$ for example):
\begin{equation}
\psi^+(\vect{x}, t) \equiv \sum_I e^{-i\varepsilon_I t}\, \Psi_I(\vect{x})\,\hat b_I\,,\qquad
\psi^- \equiv (\psi^+)^\dagger\,.
\end{equation}
Here the sum is over energy eigenstates, $\varepsilon_I=E_I-m_\psi$ are the energy eigenvalues minus the rest mass, $\Psi_I(\vect{x})$ are the wavefunctions (which are two-component for spin-$\frac{1}{2}$ fermions) and $\hat b_I$ are the annihilation operators. In the familiar case of a fermion in free space, the energy eigenstates are labeled by momentum $\vect{k}$ and spin $s=\pm$, with eigenvalues $\varepsilon_{\vect{k},s}=\varepsilon_{\vect{k}}=\sqrt{\vect{k}^2 + m_\psi^2} - m_\psi \simeq \frac{\vect{k}^2}{2m_\psi}$, and therefore\footnote{In this and the next subsection, we shall use $\vect{k}$ to denote a SM fermion's momentum while deriving the lattice potential, which should not be confused with the phonon momentum in Eq.~\eqref{eq:Gamma_phononmagnon}. Afterward, starting from Sec.~\ref{subsec:quantization}, we will no longer need to deal with fermion momenta, and the notation $\vect{k}$ will be recycled for phonon momentum.}
\begin{equation}
\psi^+_\text{free}(\vect{x}, t) = \int\frac{d^3k}{(2\pi)^3}\, e^{-i\varepsilon_{\vect{k}} t}\, e^{i\vect{k}\cdot\vect{x}}\, \xi_s \,\hat b_{\vect{k},s}\,,
\label{eq:psi+free}
\end{equation}
where $\xi_+ = \left(\begin{smallmatrix}1 \\ 0\end{smallmatrix}\right)$, $\xi_- = \left(\begin{smallmatrix}0 \\ 1\end{smallmatrix}\right)$.

For a spin-$\frac{1}{2}$ fermion, the relation between the relativistic field $\psi$ and NR field $\psi^+$ is (see Appendix~\ref{app:fermion})
\begin{equation}
\psi(\vect{x},t) = e^{-im_\psi t} \,\frac{1}{\sqrt{2}}
\left(\begin{matrix}
        \Bigl(1-\frac{\vect{\sigma} \cdot\vect{k}}{2m_\psi+\varepsilon}\Bigr)\, \psi^+(\vect{x},t) \\
\Bigl(1+\frac{\vect{\sigma} \cdot\vect{k}}{2m_\psi+\varepsilon}\Bigr)\, \psi^+(\vect{x},t)
\end{matrix}\right),
\label{eq:NR_matching_fermion}
\end{equation}
at leading order in $m_\psi^{-1}$, where $\vect{k}$, $\varepsilon$ are operators acting on $\psi^+$. For a fermion in free space, we have $\vect{k}=-i\nabla$, $\varepsilon = i\partial_t$, which become simply numbers in momentum space. In the presence of an external potential $(\Phi,\vect{A})$ ({\it e.g.}\ electromagnetic fields from the ions), $\vect{k}=-i\nabla-\vect{A}$ is the kinematical momentum, while $\varepsilon = i\partial_t-\Phi$. Eq.~\eqref{eq:NR_matching_fermion} applies for the SM fermions $\psi=p,n,e$. If the DM $\chi$ is a spin-$\frac{1}{2}$ fermion, it also applies for the DM, with $\psi$ replaced by $\chi$. For a spin-0 DM, on the other hand, $\chi = e^{-im_\chi t}\chi^+$, with $\chi^+$ given by Eq.~\eqref{eq:psi+free} without the $\xi_s$ factor.

To demonstrate the procedure of matching a relativistic model onto the NR EFT, we focus on tree level DM scattering mediated by a spin-0 or abelian spin-1 particle, denoted by $\phi$ and $V_\mu$ respectively. While it should be kept in mind that the EFT is capable of describing a broader class of models, including {\em e.g.} loop-mediated scattering, we find it useful to organize our thinking by categorizing mediator couplings to fermion bilinears. In Table~\ref{tab:currents}, we list the commonly considered types of couplings at the level of the relativistic Lagrangian, and their NR limits. We explain the table in detail in the following two paragraphs.

\begin{table}[t]
\begin{tabular}{cccl}
\rowcolor[HTML]{EFEFEF} 
\hline
\hline
{\bf \, Lagrangian Term\,} & {\bf  \quad Coupling Type\quad} & \multicolumn{2}{c}{{\bf (Effective) Current $\to$ NR Limit}} \Tstrut\Bstrut\\
\hline
\hline
$g_S \phi \bar \psi \psi $ 
& Scalar 
& $J_S$ 
& $=\, \bar \psi \psi \;\to\; \mathbb{1}$ \Tstrut\Bstrut\\
\hline
$g_P \phi \bar \psi i\gamma^5 \psi $ 
& Pseudoscalar 
& $J_P$ 
& $=\, \bar \psi i\gamma^5 \psi \;\to\; -\frac{i\vect{q}}{m_\psi}\cdot\vect{S}_\psi$ \Tstrut\Bstrut\\
\hline
\hline
$g_V V_\mu \bar \psi \gamma^\mu \psi$ 
& Vector 
& $J_V^\mu$ 
& $=\, \bar \psi \gamma^\mu \psi$ \Tstrut\\
& & & $\to\; \Bigl( \,\mathbb{1} \,,\; \frac{\vect{K}}{2m_\psi}-\frac{i\vect{q}}{m_\psi}\times\vect{S}_\psi\Bigr)$ \Tstrut\\
\hline
$g_A V_\mu \bar \psi \gamma^\mu\gamma^5 \psi$ 
& Axial vector 
& $J_A^\mu$ 
& $=\, \bar \psi \gamma^\mu\gamma^5 \psi$ \Tstrut\\
& & & $\to\; \Bigl( \,\frac{\vect{K}}{m_\psi}\cdot\vect{S}_\psi \,,\; 2\vect{S}_\psi\Bigr)$ \Tstrut\Bstrut\\
\hline
\hline
$\frac{g_\edm}{4m_\psi} V_{\mu\nu} \bar \psi \sigma^{\mu\nu} i\gamma^5 \psi$ 
& Electric dipole 
& $J_\edm^\mu$ 
& $=\, \frac{1}{2m_\psi} \partial_\nu \bigl(\bar \psi \sigma^{\mu\nu} i\gamma^5 \psi\bigr)$ \Tstrut\\
& & & $\to\; \Bigl(\, -\frac{i\vect{q}}{m_\psi}\cdot\vect{S}_\psi \,,\; \frac{i\omega}{m_\psi}\vect{S}_\psi +\frac{i\vect{q}}{m_\psi}\times\bigl(\frac{\vect{K}}{2m_\psi}\times\vect{S}_\psi\bigr)  \Bigr)$ \Tstrut\Bstrut\\
\hline
$\frac{g_\mdm}{4m_\psi} V_{\mu\nu} \bar \psi \sigma^{\mu\nu} \psi$ 
& Magnetic dipole 
& $J_\mdm^\mu$ 
& $=\, \frac{1}{2m_\psi} \partial_\nu \bigl(\bar \psi \sigma^{\mu\nu} \psi\bigr)$ \Tstrut\\
& & & $\to\; \Bigl(\,\frac{i\vect{q}}{m_\psi}\cdot\bigl(\frac{\vect{K}}{2m_\psi}\times\vect{S}_\psi\bigr) -\frac{\vect{q}^2}{4m_\psi^2} \,,\; -\frac{i\vect{q}}{m_\psi}\times\vect{S}_\psi \Bigr)$ \Tstrut\Bstrut\\
\hline
\hline
$\frac{g_\ana}{4m_\psi^2} (\partial^\nu V_{\mu\nu}) \bigl(\bar \psi \gamma^\mu\gamma^5 \psi\bigr)$ \qquad
& Anapole 
& $J_\ana^\mu$ 
& $=\, -\frac{1}{4m_\psi^2} (g^{\mu\nu}\partial^2-\partial^\mu\partial^\nu) \bigl(\bar \psi \gamma_\nu\gamma^5 \psi\bigr)$ \Tstrut\\
& & & $\to\; -\frac{\vect{q}^2}{4m_\psi^2}J_A^\mu +\bigl(\frac{\vect{q}}{m_\psi}\cdot\vect{S}_\psi\bigr)\frac{q^\mu}{2m_\psi}$ \Tstrut\Bstrut\\
\hline
$\frac{g_{V2}}{4m_\psi^2} (\partial^\nu V_{\mu\nu}) \bigl(\bar \psi \gamma^\mu \psi\bigr)$ 
& Vector ($\OO(q^2)$)
& $J_{V2}^\mu$ 
& $=\, -\frac{1}{4m_\psi^2} \partial^2 \bigl(\bar \psi \gamma^\mu \psi\bigr) \;\to\; -\frac{\vect{q}^2}{4m_\psi^2}J_V^\mu$ \Tstrut\Bstrut\\
\hline
\hline
\end{tabular}
\caption{\label{tab:currents}
Types of couplings between a spin-$\frac{1}{2}$ fermion $\psi$ and a scalar (vector) mediator $\phi$ ($V_\mu$). The (effective) currents are defined by $\LL\supset g_X \phi J_X$ ($X=S,P$) or $g_X V_\mu J_X^\mu$ ($X=V,A,\edm,\mdm,\ana,V2$), upon integration by parts in the last four cases. The expressions following the arrows are the leading operators in the NR reduction of the currents (assuming scattering kinematics), which appear between the nonrelativistic fields $\psi^-$ and $\psi^+$ --- see {\it e.g.}\ Eq.~\eqref{eq:NR_matching_current_V}. These will be used to derive the NR operators generated by specific DM models involving tree-level exchange of a scalar or vector mediator in Table~\ref{tab:benchmarks}.
}
\end{table}

For a spin-0 mediator $\phi$, we consider its couplings to the scalar and pseudoscalar currents $J_S$, $J_P$. For a spin-1 mediator $V_\mu$, we consider both minimal coupling to the vector and axial-vector currents $J^\mu_V$, $J^\mu_A$, and non-minimal couplings to the field strength $V_{\mu\nu}$.\footnote{Other operators, such as those with derivatives acting on $\psi$ and those involving the dual field strength $\tilde V_{\mu\nu}$, are not independent --- see {\it e.g.}\ Ref.~\cite{Grzadkowski:2010es}.} The latter include a series of higher dimensional operators. At dimension five, we have the electric dipole moment ($\edm$) and magnetic dipole moment ($\mdm$) couplings. Upon integration by parts, they can be cast in the same form, $V_\mu J^\mu$, as the minimal coupling case, with effective currents $J_\edm^\mu$, $J_\mdm^\mu$ listed in the last column of Table~\ref{tab:currents}. Next, at dimension six, we consider $\partial^\nu V_{\mu\nu}$ coupling to the axial-vector and vector currents. The former represents a new type of coupling known as the anapole~\cite{Pospelov:2000bq,Fitzpatrick:2010br,Ho:2012bg}, and the corresponding effective current is denoted by $J_\ana^\mu$. On the other hand, $\partial^\nu V_{\mu\nu}$ coupling to the vector current gives an $\OO(q^2)$ contribution to the same form factors that $J^\mu_V$ induces ({\it i.e.}\ the familiar charge and magnetic dipole in quantum electrodynamics), so we denote the effective current by $J_{V2}^\mu$. It is useful to note that all the (effective) currents that couple to a spin-1 mediator, except $J_A^\mu$, are conserved: $q_\mu J_X^\mu=0$ ($X=V,\edm,\mdm,\ana,V2$).

In the NR limit, we can substitute Eq.~\eqref{eq:NR_matching_fermion} for the relativistic fermion field $\psi$ into the expressions for the (effective) currents in Table~\ref{tab:currents}, and expand in powers of $\frac{\vect{k}}{m_\psi}$ and $\frac{\varepsilon}{m_\psi}$. For example, for $J_V^\mu=(J_V^0, \vect{J}_V)$, we find, at leading order,
\begin{equation}
J_V^0 = \bar\psi\gamma^0\psi \to \psi^- \psi^+\,,\qquad
\vect{J}_V = \bar\psi\vect{\gamma}\psi \to \psi^- \biggl( \frac{\vect{K}}{2m_\psi}-\frac{i\vect{q}}{m_\psi}\times\vect{S}_\psi \biggr) \psi^+\,.
\label{eq:NR_matching_current_V}
\end{equation}
where $\vect{S}_\psi = \frac{\vect{\sigma}}{2}$ is the spin operator, and
\begin{equation}
\vect{K} \equiv \vect{k}'+\vect{k}\,,\qquad
\vect{q} \equiv \vect{k}'-\vect{k}\,,
\end{equation}
with $\vect{k}'$ defined as acting on the $\psi^-$ field on the left, $\vect{k}'=i\overleftarrow{\nabla} -\vect{A}$, giving the kinematical momentum of the final state $\psi$. We can carry out the same exercise for the other (effective) currents. The results, up to the first nonvanishing order, are listed after the arrows in the last column of Table~\ref{tab:currents}, with $\psi^-$ on the left and $\psi^+$ on the right implicit. We see that all currents reduce to operators involving $\vect{S}_\psi$, $\vect{K}$ and $i\vect{q}$; in the case of the electric dipole coupling, $\omega\equiv\varepsilon'-\varepsilon$ also appears, with $\varepsilon'$ defined as acting on $\psi^-$ on the left.

With Table~\ref{tab:currents}, it is straightforward to derive the NR effective operators generated by tree-level exchange of a spin-0 or spin-1 mediator between a DM current and a SM current. Concretely, let us consider a set of benchmark models of spin-$\frac{1}{2}$ DM~\cite{Gresham:2014vja}, listed in Table~\ref{tab:benchmarks}. In each model, the DM $\chi$ and a SM fermion $\psi$ each couple to the mediator via a linear combination of the currents in the last column of Table~\ref{tab:currents}, whose NR limits can be directly read off. Integrating out the mediator, we then arrive at a NR EFT for DM scattering of the form
\begin{equation}
\LL_\text{eff} = \chi^- \biggl[\varepsilon -\frac{\vect{p}^2}{2m_\chi} + \OO\bigl(m_\chi^{-2}\bigr)\biggr] \chi^+ + \psi^- \biggl[\varepsilon -\frac{\vect{k}^2}{2m_\psi} + \OO\bigl(m_\psi^{-2}\bigr)\biggr] \psi^+ +\sum_{i} \sum_{\psi=p,n,e} c_i^{(\psi)} \OO_i^{(\psi)} \chi^- \chi^+ \psi^- \psi^+\,.
\label{eq:EFT_Lag}
\end{equation}
For convenience, we reserve $\vect{k}$ and $\vect{k}'$ for the momentum operators acting on $\psi^\pm$, and write the same operators as $\vect{p}$ and $\vect{p}'$ when they act on $\chi^\pm$.  
We normalize the operators by powers of $m_\psi$ so that $\OO_i^{(\psi)}$ are dimensionless and their coefficients $c_i^{(\psi)}$ have dimension $-2$. 
For each UV model, the coefficients $c_i^{(\psi)}$ of the NR operators generated at leading order are given in Table~\ref{tab:benchmarks} (to be discussed in detail shortly). These coefficients contain all the information for constructing the lattice potential $\widetilde{\cal V}_{lj}$ for a given DM model, and will be exploited below for computing the DM detection rate.

\begin{table}[t]
	\centering
	\begin{footnotesize}
		\begin{tabular}{p{0.6in}|cccc}
			\rowcolor[HTML]{EFEFEF} 
			\hline\hline
			\multicolumn{2}{c}{{\bf Model}} & {\bf UV Lagrangian} & {\bf NR EFT} & {\bf Responses} \Tstrut\Bstrut\\
			\hline\hline
			\multicolumn{2}{c}{\multirow{2}{*}{Standard SI}} 
			& $\phi\, \bigl(g_\chi J_{S,\chi} + g_\psi J_{S,\psi}\bigr)$ or
			& \multirow{2}{*}{$c_1^{(\psi)} = \frac{g_\chi g_\psi^\text{eff}}{\vect{q}^2+m_{\phi,V}^2}$} 
			& \multirow{2}{*}{$N$} \Tstrut\\
			\multicolumn{2}{c}{} & $V_\mu \bigl( g_\chi J_{V,\chi}^\mu - g_\psi J_{V,\psi}^\mu \bigr)$ & & \Bstrut\\
			\hline\hline
			\multicolumn{2}{c}{\;\;\;Standard SD\,\footnote{Heavy mediator only.}} 
			& $V_\mu \bigl( g_\chi J_{A,\chi}^\mu + g_\psi J_{A,\psi}^\mu \bigr)$
			& $c_4^{(\psi)} = \frac{4g_\chi g_\psi}{\vect{q}^2+m_V^2}$ 
			& $S$ \Tstrut\Bstrut\\
			\hline\hline
			\multirow{3}{0.6in}{\centering \rule{0pt}{4ex} Other \rule{0pt}{4ex}\\scalar\\mediators\Bstrut} 
			& P\,$\times$\,S
			& $\phi\, \bigl(g_\chi J_{P,\chi} + g_\psi J_{S,\psi}\bigr)$ 
			& $c_{11}^{(\psi)} = \frac{m_\psi}{m_\chi} \frac{g_\chi g_\psi^\text{eff}}{\vect{q}^2+m_\phi^2}$ 
			& $N$ \Tstrut\Bstrut\\
			\cline{2-5}
			& S\,$\times$\,P 
			& $\phi\, \bigl(g_\chi J_{S,\chi} + g_\psi J_{P,\psi}\bigr)$ 
			& $c_{10}^{(\psi)} = -\frac{g_\chi g_\psi}{\vect{q}^2+m_\phi^2}$ 
			& $S$ \Tstrut\Bstrut\\
			\cline{2-5}
			& P\,$\times$\,P 
			& $\phi\, \bigl(g_\chi J_{P,\chi} + g_\psi J_{P,\psi}\bigr)$ 
			& $c_6^{(\psi)} = \frac{m_\psi}{m_\chi} \frac{g_\chi g_\psi}{\vect{q}^2+m_\phi^2}$ 
			& $S$ \Tstrut\Bstrut\\
			\hline\hline
			\multirow{7}{0.6in}{\centering \rule{0pt}{3.5ex} \;Multipole\;\rule{0pt}{3.5ex}\\DM\\models} 
			& Electric dipole
			& {\scriptsize $V_\mu \Bigl( g_\chi J_{\edm,\chi}^\mu + g_\psi \bigl(J_{V,\psi}^\mu +\delta\tilde\mu_\psi J_{\mdm,\psi}^\mu \bigr) \Bigr)$}
			& $c_{11}^{(\psi)} = -\frac{m_\psi}{m_\chi}\frac{g_\chi g_\psi^\text{eff}}{\vect{q}^2+m_V^2}$ 
			& $N$ \Tstrut\Bstrut\\
			\cline{2-5}
			& \multirow{5}{*}{\rule{0pt}{6ex} \;Magnetic dipole\; \rule{0pt}{6ex}}
			& \multirow{5}{*}{\centering \rule{0pt}{6ex}\scriptsize \;$V_\mu \Bigl( g_\chi J_{\mdm,\chi}^\mu + g_\psi \bigl(J_{V,\psi}^\mu +\delta\tilde\mu_\psi J_{\mdm,\psi}^\mu \bigr) \Bigr)$\;} 
			& $c_1^{(\psi)} = \frac{\vect{q}^2}{4m_\chi^2}\frac{g_\chi g_\psi^\text{eff}}{\vect{q}^2+m_V^2}$ 
			& \multirow{5}{*}{\rule{0pt}{6ex} $N,\,S,\,L$ \rule{0pt}{6ex}} \Tstrut\\
			& & & $c_4^{(\psi)} =  \text{{\scriptsize$\tilde\mu_\psi$}}\frac{\vect{q}^2}{m_\chi m_\psi}\frac{g_\chi g_\psi}{\vect{q}^2+m_V^2}$ & \Tstrut\\
			& & & $c_{5a}^{(\psi)} = \frac{m_\psi}{m_\chi}\frac{g_\chi g_\psi^\text{eff}}{\vect{q}^2+m_V^2}$ & \Tstrut\\
			& & & $c_{5b}^{(\psi)} = \frac{m_\psi}{m_\chi}\frac{g_\chi g_\psi}{\vect{q}^2+m_V^2}$ & \Tstrut\\
			& & & $c_6^{(\psi)} = -\text{{\scriptsize$\tilde\mu_\psi$}}\frac{m_\psi}{m_\chi}\frac{g_\chi g_\psi}{\vect{q}^2+m_V^2}$ & \Tstrut\Bstrut\\
			\cline{2-5}
			& \multirow{3}{*}{Anapole\Tstrut}
			& \multirow{3}{*}{{\scriptsize $V_\mu \Bigl( g_\chi J_{\ana,\chi}^\mu + g_\psi \bigl(J_{V,\psi}^\mu +\delta\tilde\mu_\psi J_{\mdm,\psi}^\mu \bigr) \Bigr)$}\Tstrut}
			& $c_{8a}^{(\psi)} = \frac{\vect{q}^2}{2m_\chi^2}\frac{g_\chi g_\psi^\text{eff}}{\vect{q}^2+m_V^2}$ 
			& \multirow{3}{*}{$N,\,S,\,L$\Tstrut} \Tstrut\\
			& & & $c_{8b}^{(\psi)} = \frac{\vect{q}^2}{2m_\chi^2}\frac{g_\chi g_\psi}{\vect{q}^2+m_V^2}$ & \Tstrut\\
			& & & $c_9^{(\psi)} = -\text{{\scriptsize$\tilde\mu_\psi$}}\frac{\vect{q}^2}{2m_\chi^2}\frac{g_\chi g_\psi}{\vect{q}^2+m_V^2}$ & \Tstrut\Bstrut\\
			\hline\hline
			\multicolumn{2}{c}{\multirow{4}{*}{\rule{0pt}{6ex} $(\vect{L}\cdot\vect{S})$-interacting \rule{0pt}{6ex}}}
			& \multirow{4}{*}{\rule{0pt}{6ex}\;\;$V_\mu \bigl( g_\chi J_{V,\chi}^\mu + g_\psi (J_{\mdm,\psi}^\mu +\kappa J_{V2,\psi}^\mu) \bigr)$\;\;\rule{0pt}{6ex}} 
			& \;\;$c_1^{(\psi)} = \text{{\scriptsize$(1+\kappa)$}} \frac{\vect{q}^2}{4m_\psi^2}\frac{g_\chi g_\psi}{\vect{q}^2+m_V^2}$\;\; 
			& \multirow{4}{*}{\rule{0pt}{6ex} $N, S,\, L\otimes S$ \rule{0pt}{6ex}} \Tstrut\\
			\multicolumn{2}{c}{} & & $c_{3a}^{(\psi)} = c_{3b}^{(\psi)} = \frac{g_\chi g_\psi}{\vect{q}^2+m_V^2}$ &\Tstrut\\
			\multicolumn{2}{c}{} & & $c_4^{(\psi)} = \frac{\vect{q}^2}{m_\chi m_\psi}\frac{g_\chi g_\psi}{\vect{q}^2+m_V^2}$ &\Tstrut\\
			\multicolumn{2}{c}{} & & $c_6^{(\psi)} = -\frac{m_\psi}{m_\chi}\frac{g_\chi g_\psi}{\vect{q}^2+m_V^2}$ &\Tstrut\Bstrut\\
			\hline\hline
		\end{tabular}
	\end{footnotesize}
	\caption{\label{tab:benchmarks}
		Benchmark models of spin-$\frac{1}{2}$ DM $\chi$ coupling to SM fermions $\psi=p,n,e$. For each model, the leading order nonvanishing coefficients $c_i^{(\psi)}$ for the NR EFT operators $\OO_i^{(\psi)}$ (defined in Table~\ref{tab:operators}) are listed in the second to last column. $g_\psi^\text{eff}$ are the screened couplings defined in Eq.~\eqref{eq:in-medium}, and $\tilde\mu_\psi = 1+\delta\tilde\mu_\psi$ is half the Land\'e $g$-factor of $\psi$ ($\tilde\mu_p\simeq 2.8$, $\tilde\mu_n\simeq -1.9$, $\tilde\mu_e\simeq 1$). The last column lists the lattice degrees of freedom which enter the scattering potential, Eq.~\eqref{eq:Vlj}. All models can excite phonons, and models with $S$ or $L$ response generated by DM-electron coupling can also excite magnons.
	} 
\end{table}

For kinematic conventions, we take
\begin{equation}
\vect{q} \equiv \vect{k}'-\vect{k} = \vect{p}-\vect{p}'
\end{equation}
to denote the momentum transfer from the DM to the target, which agrees with Refs.~\cite{DelNobile:2013sia,DelNobile:2018dfg} but has an opposite sign compared to the definitions in Refs.~\cite{Fitzpatrick:2012ix,Anand:2013yka,Gresham:2014vja,Anand:2014kea}. 
There are two other independent combinations of momenta:
\begin{equation}
\vect{v}_\chi \equiv \frac{\vect{P}}{2m_\chi} \,,\qquad\quad
\vect{v}_\psi \equiv \frac{\vect{K}}{2m_\psi} \,,
\label{eq:v_def}
\end{equation}
where $\vect{P}=\vect{p}'+\vect{p}$, $\vect{K}=\vect{k}'+\vect{k}$. Note that $\vect{v}_\chi$ should not be confused with the incoming DM's velocity, which we denote by $\vect{v}=\frac{\vect{p}}{m_\chi}$; the two are related by $\vect{v}_\chi = \vect{v}-\frac{\vect{q}}{2m_\chi}$.

The list of NR operators $\OO_i^{(\psi)}$ up to linear order in $\vect{v}_\chi$, $\vect{v}_\psi$ is presented in Table~\ref{tab:operators} (grouped into four categories to be explained below). 
These encompass all the operators generated at leading order in the benchmark models in Table~\ref{tab:benchmarks}. 
Our operator basis here is an extension of the familiar one from previous works on the EFT for direct detection via nuclear recoils~\cite{Fitzpatrick:2012ix,DelNobile:2013sia,Anand:2013yka,Gresham:2014vja,Anand:2014kea,DelNobile:2018dfg}. 
In the latter case, due to Galilean invariance, NR effective operators involve only the linear combination $\vect{v}^\perp\equiv \vect{v}_\chi -\vect{v}_\psi$. 
In contrast, for collective excitations considered in this work, in-medium effects, which break Galilean invariance, can be important, so $\vect{v}_\chi$ and $\vect{v}_\psi$ must be treated separately. 
We adopt the operator numbering convention of Ref.~\cite{DelNobile:2018dfg}, and split each $\vect{v}^\perp$-dependent operator into two; one term dependent on $\vect{v}_\chi$ and the other $\vect{v}_\psi$. For example, $\OO_7^{(\psi)} = \vect{S}_\psi \cdot \vect{v}^\perp = \vect{S}_\psi \cdot \vect{v}_\chi - \vect{S}_\psi \cdot \vect{v}_\psi \equiv \OO_{7a}^{(\psi)} -\OO_{7b}^{(\psi)}$, and we treat $\OO_{7a}^{(\psi)}$ and $\OO_{7b}^{(\psi)}$ as independent operators.

Among the benchmark models in Table~\ref{tab:benchmarks}, the standard SI and SD interactions correspond to $\OO_1$ and $\OO_4$, respectively.\footnote{Note that the standard SD interaction cannot be realized with a light mediator. In that case the leading interaction is induced by longitudinal vector exchange, and is proportional to $J_{P,\chi} J_{P,\psi}$ rather than $J_{A,\chi}^\mu {J_{A,\psi}}_\mu$.} Other types of scalar mediators generate $\OO_6$, $\OO_{10}$ and $\OO_{11}$. A well-motivated class of (hidden sector) models contain DM particles coupling to a vector mediator via a multipole moment, which in turn kinetically mixes with the photon (see {\it e.g.}\ Refs.~\cite{Pospelov:2000bq, Sigurdson:2004zp,Masso:2009mu,Kribs:2009fy,Fitzpatrick:2010br, Banks:2010eh,Ho:2012bg,Gresham:2014vja}). We consider the electric dipole, magnetic dipole and anapole DM models, which generate $\OO_{11}$, $\OO_{1,4,5a,5b,6}$ and $\OO_{8a,8b,9}$, respectively. Finally, Table~\ref{tab:benchmarks} includes a model where a vector mediator couples to the SM fermion's magnetic dipole moment $J_\mdm^\mu$, and as a result generates $\OO_{3b}$. Among other things, this leads to a coupling to the SM fermion's spin-orbit coupling, which can be the leading interaction if one simultaneously introduces a coupling to the ``$\OO(q^2)$ vector current'' $J^\mu_{V2}$ (see Table~\ref{tab:currents}), with a coefficient (relative to $J_\mdm^\mu$) tuned to $\kappa=-1$ to cancel the standard SI interaction $\OO_1$.

\begin{table}[t]
	\begin{tabular}{ccc}
		\rowcolor[HTML]{EFEFEF} 
		\hline
		\hline
		\textbf{Interaction Type} & \textbf{NR Operators} & \quad\textbf{Crystal Response}\quad \Tstrut\Bstrut\\
		\hline
		\hline
		\multirow{4}{*}{Coupling to {\it charge}, $\vect{v}_\psi$-{\it independent}} & $\OO_1^{(\psi)} = \mathbb{1}$ & \multirow{4}{*}{$N$}  \Tstrut  \\
		& $\OO_{5a}^{(\psi)} = \vect{S}_\chi \cdot \bigl(\frac{i\vect{q}}{m_\psi}\times\vect{v}_\chi\bigr)$ &  \\
		& $\OO_{8a}^{(\psi)} = \vect{S}_\chi\cdot\vect{v}_\chi$ &  \\
		& $\OO_{11}^{(\psi)} =\vect{S}_\chi\cdot \frac{i\vect{q}}{m_\psi}$ & \Bstrut \\
		\hline
		\multirow{10}{*}{Coupling to {\it spin}, $\vect{v}_\psi$-{\it independent}} & $\OO_{3a}^{(\psi)} = \vect{S}_\psi \cdot \bigl(\frac{i\vect{q}}{m_\psi}\times\vect{v}_\chi\bigr)$ & \multirow{10}{*}{$S$}  \Tstrut \\
		& $\OO_4^{(\psi)} = \vect{S}_\chi\cdot\vect{S}_\psi$ & \\
		& $\OO_6^{(\psi)} = \bigl(\vect{S}_\chi\cdot\frac{\vect{q}}{m_\psi}\bigr)\bigl(\vect{S}_\psi\cdot\frac{\vect{q}}{m_\psi}\bigr)$ & \\
		& $\OO_{7a}^{(\psi)} = \vect{S}_\psi\cdot\vect{v}_\chi$ & \\
		& $\OO_9^{(\psi)} = \vect{S}_\chi \cdot \bigl(\vect{S}_\psi\times \frac{i\vect{q}}{m_\psi}\bigr)$ & \\
		& $\OO_{10}^{(\psi)} = \vect{S}_\psi\cdot \frac{i\vect{q}}{m_\psi}$ & \\
		& $\OO_{12a}^{(\psi)} = \vect{S}_\chi\cdot \bigl(\vect{S}_\psi\times\vect{v}_\chi \bigr) $ & \\
		& $\OO_{13a}^{(\psi)} = \bigl(\vect{S}_\chi\cdot\vect{v}_\chi \bigr)\bigl(\vect{S}_\psi\cdot\frac{i\vect{q}}{m_\psi}\bigr)$ & \\
		& $\OO_{14a}^{(\psi)} = \bigl(\vect{S}_\psi\cdot\vect{v}_\chi \bigr)\bigl(\vect{S}_\chi\cdot\frac{i\vect{q}}{m_\psi}\bigr)$ & \\
		& $\OO_{15a}^{(\psi)} = \bigl(\vect{S}_\chi\cdot\bigl(\frac{i\vect{q}}{m_\psi}\times\vect{v}_\chi \bigr)\bigr)\bigl(\vect{S}_\psi\cdot\frac{i\vect{q}}{m_\psi}\bigr)$ & \Bstrut \\
		\hline
		\multirow{2}{*}{Coupling to {\it charge}, $\vect{v}_\psi$-{\it dependent}} & $\OO_{5b}^{(\psi)} = \vect{S}_\chi \cdot \bigl(\frac{i\vect{q}}{m_\psi}\times\vect{v}_\psi\bigr)$ &  \multirow{2}{*}{$L$} \Tstrut\\
		& $\OO_{8b}^{(\psi)} = \vect{S}_\chi\cdot\vect{v}_\psi$ & \Bstrut \\
		\hline
		\multirow{6}{*}{Coupling to {\it spin}, $\vect{v}_\psi$-{\it dependent}} & $\OO_{3b}^{(\psi)} = \vect{S}_\psi \cdot \bigl(\frac{i\vect{q}}{m_\psi}\times\vect{v}_\psi\bigr)$ &  \multirow{6}{*}{$L\otimes S$}  \Tstrut\\
		& $\OO_{7b}^{(\psi)} = \vect{S}_\psi\cdot\vect{v}_\psi$ & \\
		& $\OO_{12b}^{(\psi)} = \vect{S}_\chi\cdot \bigl(\vect{S}_\psi\times\vect{v}_\psi \bigr) $ & \\
		& $\OO_{13b}^{(\psi)} = \bigl(\vect{S}_\chi\cdot\vect{v}_\psi \bigr)\bigl(\vect{S}_\psi\cdot\frac{i\vect{q}}{m_\psi}\bigr)$ & \\
		& $\OO_{14b}^{(\psi)} = \bigl(\vect{S}_\psi\cdot\vect{v}_\psi \bigr)\bigl(\vect{S}_\chi\cdot\frac{i\vect{q}}{m_\psi}\bigr)$ & \\
		& $\OO_{15b}^{(\psi)} = \bigl(\vect{S}_\chi\cdot\bigl(\frac{i\vect{q}}{m_\psi}\times\vect{v}_\psi \bigr)\bigr)\bigl(\vect{S}_\psi\cdot\frac{i\vect{q}}{m_\psi}\bigr)$ & \Bstrut\\
		\hline
	\end{tabular}
	\caption{\label{tab:operators}
		NR effective operators relevant for DM scattering defined in Eq.~\eqref{eq:EFT_Lag}, organized into four categories, and the crystal responses generated. Here $\chi$ is the DM and $\psi$ is a SM particle that can be the proton, neutron or electron. $\vect{q}$ is the momentum transfer from the DM to the SM target, and $\vect{v}_\chi$, $\vect{v}_\psi$ are defined in Eq.~\eqref{eq:v_def}. Previous calculations~\cite{Knapen:2017ekk,Griffin:2018bjn,Trickle:2019nya, Trickle:2019ovy, Griffin:2019mvc} focused on phonon and magnon excitations via $\vect{v}_\psi$-independent couplings to charge and spin, corresponding to the first two categories listed here. In this work we extend the calculations to all operators.
	}
\end{table}

We also note that, in the case of a vector mediator coupling to the electron's vector current $J^\mu_{V,e}$, in-medium screening effects modify the effective couplings to the proton and electron~\cite{Hochberg:2015fth,Knapen:2017xzo,Hochberg:2017wce,Trickle:2019nya,Coskuner:2019odd}. For NR scattering, screening is negligible for transverse photon exchange, but can be significant for longitudinal photon exchange, which generates the first category of operators ($\OO_{1,5a,8a,11}$) in Table~\ref{tab:operators}. 
As shown in Refs.~\cite{Knapen:2017xzo,Trickle:2019nya}, this amounts to replacing
\begin{equation}
g_p \to g_p^\text{eff} = g_p +\biggl( 1-\frac{\vect{q}^2}{\vect{q}\cdot\epstens\cdot\vect{q}} \biggr) g_e \,,\qquad
g_e \to g_e^\text{eff} = \frac{\vect{q}^2}{\vect{q}\cdot\epstens\cdot\vect{q}}\, g_e \,,
\label{eq:in-medium}
\end{equation}
where $\epstens$ is the dielectric tensor, and $g_{p,e}$ are the tree-level (unscreened) couplings. The same is true for a scalar mediator coupling to the electron's scalar current $J_{S,e}$~\cite{Gelmini:2020xir}. For single phonon and magnon excitations below the electronic band gap that we focus on in this work, one can use the high-frequency dielectric $\epstens_\infty$, which captures the screening due to fast electron responses~\cite{Gonze1997a,Griffin:2018bjn,Trickle:2019nya}.

We will study the reach phonon and magnon detectors have to these benchmark models in Sec.~\ref{sec:benchmarks}, after developing the formalism of rate calculations within the EFT in the rest of this section.

\subsection{Matching Effective Operators Onto Lattice Degrees of Freedom}
\label{subsec:lattice_matching}

We now match the effective operators $\OO_i^{(\psi)}$ onto lattice degrees of freedom (\hl{highlighted} for clarity) that appear in the DM-ion scattering potentials $\tilde{\cal V}_{lj}$. In Table~\ref{tab:operators}, we have organized the operators into four categories, according to whether $\OO_i^{(\psi)}\propto \mathbb{1}_\psi$ (``coupling to charge'') or $\OO_i^{(\psi)}\propto \vect{S}_\psi$ (``coupling to spin''), and whether the operator involves $\vect{v}_\psi$. Since our focus is light DM that evades conventional searches via nuclear recoils and electronic excitations, we will work in the long wavelength limit, where the momentum transfer is small compared to the inverse ionic radius (corresponding to $m_\chi\lesssim 10$\,MeV), so at leading order, the only relevant degrees of freedom are those that characterize the ion as a whole. Intuitively, we expect couplings to charge and spin of a constituent particle $\psi=p,n,e$ to match onto couplings to the total number $\hl{\langle N_\psi\rangle}$ and spin $\hl{\langle\vect{S}_\psi\rangle}$ of that particle, respectively. These are point-like degrees of freedom that do not involve the internal motions of the ion constituents; they are the only degrees of freedom to which DM couples if the operator is velocity-independent. On the other hand, $\vect{v}_\psi$-dependent operators are expected to couple DM to the motion of $\psi$ particles inside an ion, manifest as the total orbital angular momenta $\hl{\langle\vect{L}_\psi\rangle}$ and spin-orbit couplings $\hl{\langle\vect{L}_\psi\otimes\vect{S}_\psi\rangle}$, which are ``composite'' degrees of freedom. In the rest of this subsection, we will see concretely how these intuitive expectations are borne out. The final result of this calculation is the lattice potential in terms of the NR EFT operator coefficients $c_i^{(\psi)}$, given below in Eq.~\eqref{eq:Vlj}.

Since the calculation proceeds in much the same way for all operators in the same category, to avoid tedious repetition we pick one operator from each category to explain the procedure: $\OO_1^{(\psi)}$, $\OO_4^{(\psi)}$, $\OO_{8b}^{(\psi)}$ and $\OO_{3b}^{(\psi)}$, with $\psi$ taken to be one of $p,n,e$. To obtain the DM-ion scattering potentials $\tilde{\cal V}_{lj}$, we need to compute the matrix elements of these operators between the incoming and outgoing states of the DM-ion system. Since the initial and final DM states are plane waves, the DM part of the matrix element simply yields a phase factor, so\\
\begin{eqnarray}
\widetilde{\cal V}_{lj}(-\vect{q},\vect{v}) &\supset& \sum_\alpha\biggl[
c_1^{(\psi)}\, \bigl\langle e^{i\vect{q}\cdot\vect{x}_\alpha} \bigr\rangle_{lj}
+\,c_4^{(\psi)}\, \vect{S}_\chi \cdot \bigl\langle e^{i\vect{q}\cdot\vect{x}_\alpha} \vect{S}_{\psi,\alpha} \bigr\rangle_{lj} \nonumber\\
&&\qquad\; +\,c_{8b}^{(\psi)}\, \vect{S}_\chi \cdot \bigl\langle e^{i\vect{q}\cdot\vect{x}_\alpha} \vect{v}_{\psi,\alpha} \bigr\rangle_{lj}
+\,c_{3b}^{(\psi)}\,\frac{i\vect{q}}{m_\psi} \cdot\bigl\langle e^{i\vect{q}\cdot\vect{x}_\alpha}\,\vect{v}_{\psi,\alpha}\times \vect{S}_{\psi,\alpha}\bigr\rangle_{lj}
\biggr]\,,
\label{eq:sampleVlj}
\end{eqnarray}
where $\alpha$ runs over all the $\psi$ fermions associated with the ion labeled by $l,j$, and $\langle\cdot \rangle$ represents the ionic expectation value (assuming the ionic state is unchanged for the low energy depositions of interest). Computing these expectation values in full generality is a tedious task that involves numerical integration over nuclear and electronic wavefunctions. However, the calculation is dramatically simplified in the long wavelength limit of interest here, where we can expand $e^{i\vect{q}\cdot\vect{x}_\alpha} = 1+i\vect{q}\cdot\vect{x}_\alpha+\dots$ and keep just the leading nonvanishing terms. In the following two paragraphs, we discuss in turn the $\vect{v}_\psi$-independent operators $\OO_1^{(\psi)}$, $\OO_4^{(\psi)}$ (first line of Eq.~\eqref{eq:sampleVlj}) and the $\vect{v}_\psi$-dependent operators $\OO_{8b}^{(\psi)}$, $\OO_{3b}^{(\psi)}$ (second line of Eq.~\eqref{eq:sampleVlj}).

\paragraph*{a) $\vect{v}_\psi$-independent operators: $\OO_1^{(\psi)}$, $\OO_4^{(\psi)}$.}
For these, it is sufficient to set $e^{i\vect{q}\cdot\vect{x}_\alpha} \to 1$:
\begin{eqnarray}
c_1^{(\psi)}\sum_\alpha\bigl\langle e^{i\vect{q}\cdot\vect{x}_\alpha} \bigr\rangle_{lj} &\simeq& c_1^{(\psi)}\sum_\alpha \langle 1\rangle_{lj} = c_1^{(\psi)}\hl{\langle N_\psi\rangle}_{lj}\,,\\
c_4^{(\psi)}\,\vect{S}_\chi \cdot\sum_\alpha\bigl\langle e^{i\vect{q}\cdot\vect{x}_\alpha} \vect{S}_{\psi,\alpha} \bigr\rangle_{lj} &\simeq& c_4^{(\psi)}\,\vect{S}_\chi \cdot\sum_\alpha \langle \vect{S}_{\psi,\alpha} \rangle_{lj} = c_4^{(\psi)}\,\vect{S}_\chi \cdot\hl{\langle \vect{S}_{\psi} \rangle}_{lj}\,.
\end{eqnarray}
So we obtain, respectively, the expectation values of the number and total spin of $\psi$ particles for ion $l,j$, as one would expect for the lowest order ``coupling to charge'' ($\mathcal{O}_1^{(\psi)}$) and ``coupling to spin'' ($\mathcal{O}_4^{(\psi)}$) operators. 
Note that $\hl{\langle \vect{S}_{\psi} \rangle}_{lj}$ should not be confused with the total nuclear or ionic spin, which may also contain orbital angular momentum components. 
We will see in the next subsection that the total ionic spin (from electrons) is relevant for magnon excitations, and we will need to work out its decomposition into spin and orbital components (see Eq.~\eqref{eq:proj} below); the total nuclear spin, on the other hand, does not enter the calculation of phonon or magnon excitations.

\paragraph*{b) $\vect{v}_\psi$-dependent operators: $\OO_{8b}^{(\psi)}$, $\OO_{3b}^{(\psi)}$.}
The operator $\vect{v}_{\psi,\alpha} = \frac{(\vect{k}+\vect{k}')_\alpha}{2m_\psi} = -\frac{i}{2m_\psi} \overleftrightarrow{\nabla}_\alpha$ is in fact the probability current, and its treatment is analogous to the nuclear recoil calculation~\cite{Fitzpatrick:2012ix}. 
Assuming the ionic states are energy eigenstates implies that the probability density is constant in time, and therefore by the continuity equation, $\partial_i \langle v^i_{\psi,\alpha} \rangle_{lj} = 0$. 
This means that $v^i_{\psi,\alpha}$ can be written as a total derivative, $v^i_{\psi,\alpha} = \partial_k \left( x^i_\alpha v^k_{\psi,\alpha} \right)$, and therefore has vanishing expectation value. The leading contribution then comes from expanding the $e^{i\vect{q}\cdot\vect{x}_\alpha}$ to the next order in $\vect{q}$:
\begin{eqnarray}
\sum_\alpha\bigl\langle e^{i\vect{q}\cdot\vect{x}_\alpha} \vect{v}_{\psi,\alpha} \bigr\rangle_{lj} &\simeq& i\sum_\alpha \bigl\langle ( \vect{q} \cdot \vect{x}_\alpha )\, \vect{v}_{\psi,\alpha} \bigr\rangle_{lj} \,,\\
\sum_\alpha\bigl\langle e^{i\vect{q}\cdot\vect{x}_\alpha}\, \vect{v}_{\psi,\alpha}\times \vect{S}_{\psi,\alpha} \bigr\rangle_{lj} &\simeq& i \sum_\alpha \bigl\langle ( \vect{q} \cdot \vect{x}_\alpha )\, \vect{v}_{\psi,\alpha} \times \vect{S}_{\psi,\alpha}\bigr\rangle_{lj}\,.
\end{eqnarray}
To go further, we note that $\bigl\langle x^i_\alpha v^k_{\psi,\alpha} \bigr\rangle_{lj}$ is anti-symmetric in $i \leftrightarrow  k$ since the symmetric part can be written as a total derivative, $x^i_\alpha v^k_{\psi,\alpha} + x^k_\alpha v^i_{\psi,\alpha} = \partial_{i'} \bigl( x^i_\alpha x^k_\alpha v^{i'}_{\psi,\alpha} \bigr)$ and therefore has vanishing expectation value. Expanding the anti-symmetric part gives
\begin{equation}
    \bigl\langle x^i_\alpha v^k_{\psi,\alpha} \bigr\rangle_{lj} = \frac{1}{2}\bigl\langle x^i_\alpha v^k_{\psi,\alpha} - x^k_\alpha v^i_{\psi,\alpha} \bigr\rangle_{lj} = -\frac{i}{4m_\psi} \Bigl( \bigl\langle x^i_\alpha \overrightarrow{\nabla}^k_\alpha \bigr\rangle_{lj} - \bigl\langle x^i_\alpha \overleftarrow{\nabla}^k_\alpha \bigr\rangle_{lj} - \bigl\langle x^k_\alpha \overrightarrow{\nabla}^i_\alpha \bigr\rangle_{lj} + \bigl\langle x^k_\alpha \overleftarrow{\nabla}^i_\alpha \bigr\rangle_{lj} \Bigr) \, ,
\end{equation}
which after integration by parts can be simplified to 
\begin{equation}
    \bigl\langle x^i_\alpha v^k_{\psi,\alpha} \bigr\rangle_{lj} = -\frac{i}{2m_\psi} \bigl\langle x^i \overrightarrow{\nabla}^k_\alpha - x^k \overrightarrow{\nabla}^i_\alpha \bigr\rangle_{lj} = \frac{1}{2m_\psi} \epsilon_{ikk'} \langle L^{k'}_\alpha\rangle_{lj} \, ,
\end{equation}
where $\vect{L}_\alpha$ is the angular momentum operator. We therefore have
\begin{eqnarray}
\sum_\alpha\bigl\langle e^{i\vect{q}\cdot\vect{x}_\alpha} \vect{v}_{\psi,\alpha} \bigr\rangle_{lj} &\simeq& -\frac{i\vect{q}}{2m_\psi} \times \sum_\alpha \langle \vect{L}_{\psi,\alpha} \rangle_{lj} = -\frac{i\vect{q}}{2m_\psi} \times\hl{\langle\vect{L}_\psi\rangle}_{lj} \,,\\
\sum_\alpha\bigl\langle e^{i\vect{q}\cdot\vect{x}_\alpha}\, \vect{v}_{\psi,\alpha} \times \vect{S}_{\psi,\alpha} \bigr\rangle_{lj} &\simeq& -\frac{i}{2m_\psi} \sum_\alpha \langle (\vect{q}\times\vect{L}_{\psi,\alpha})\times \vect{S}_{\psi,\alpha} \rangle_{lj} 
\nonumber\\
&=& -\frac{i}{2m_\psi} \Bigl( \sum_\alpha \langle \vect{L}_{\psi,\alpha}\otimes \vect{S}_{\psi,\alpha}\rangle_{lj}\cdot\vect{q} - \sum_\alpha \langle \vect{L}_{\psi,\alpha}\cdot \vect{S}_{\psi,\alpha}\rangle_{lj} \,\vect{q}\Bigr)
\nonumber\\
&=& -\frac{i}{2m_\psi} \bigl(\langle\vect{L}_\psi \otimes \vect{S}_\psi \rangle_{lj}\cdot\vect{q} -\langle\vect{L}_\psi \cdot \vect{S}_\psi \rangle_{lj}\,\vect{q} \bigr) \nonumber\\
&=& -\frac{i}{2m_\psi} \Bigl[\hl{\langle\vect{L}_\psi \otimes \vect{S}_\psi \rangle}_{lj}\cdot\vect{q} -\text{tr}\bigl(\hl{\langle\vect{L}_\psi \otimes \vect{S}_\psi \rangle}_{lj}\bigr)\,\vect{q} \Bigr] \,.
\end{eqnarray}
where
\begin{equation}
\bigl(\hl{\langle\vect{L}_\psi \otimes \vect{S}_\psi \rangle}_{lj}\bigr)^{ik} = \langle L_\psi^i S_\psi^k\rangle_{lj} \equiv \sum_\alpha \langle L_{\psi,\alpha}^i S_{\psi,\alpha}^k\rangle_{lj} 
\label{eq:LS_def}
\end{equation}
are Cartesian components of the spin-orbit coupling tensor. So we finally obtain
\begin{eqnarray}
&& c_{8b}^{(\psi)}\, \vect{S}_\chi \cdot \sum_\alpha \bigl\langle e^{i\vect{q}\cdot\vect{x}_\alpha} \vect{v}_{\psi,\alpha} \bigr\rangle_{lj} =
-c_{8b}^{(\psi)}\, \vect{S}_\chi \cdot\biggl(\frac{i\vect{q}}{2m_\psi} \times\hl{\langle\vect{L}_\psi\rangle}_{lj}\biggr)\,,\\
&& c_{3b}^{(\psi)}\,\frac{i\vect{q}}{m_\psi} \cdot\sum_\alpha \bigl\langle e^{i\vect{q}\cdot\vect{x}_\alpha}\, \vect{v}_{\psi,\alpha}\times\vect{S}_{\psi,\alpha} \bigr\rangle_{lj} 
 = -c_{3b}^{(\psi)}\, \frac{1}{2m_\psi^2} (\vect{q}^2\delta^{ik} -q^iq^k) \bigl(\hl{\langle\vect{L}_\psi \otimes \vect{S}_\psi \rangle}_{lj}\bigr)^{ik}\,.
\end{eqnarray}
As alluded to previously, the $\vect{v}_\psi$-dependent operators $\OO_{8b}^{(\psi)}$ and $\OO_{3b}^{(\psi)}$ induce DM couplings to the $\psi$ particles' total orbital angular momentum and spin-orbit coupling.

We can carry out the same calculation for the other operators in Table~\ref{tab:operators}. The result is
\begin{eqnarray}
\tilde{\cal V}_{lj}(-\vect{q},\vect{v}) &=& \sum_{\psi=p,n,e} 
c_1^{(\psi)} \hl{\langle N_\psi\rangle}_{lj}
\Tstrut\Bstrut\nonumber\\
&&\quad
-\,c_{3a}^{(\psi)}\, \frac{iq}{m_\psi} \,\vect{v}_\chi\cdot\bigl(\vect{\hat q}\times\hl{\langle\vect{S}_\psi\rangle}_{lj}\bigr)
-\,c_{3b}^{(\psi)}\, \frac{q^2}{2m_\psi^2} \,(\delta^{ik} -\hat q^i \hat q^k) \bigl(\hl{\langle \vect{L}_\psi \otimes \vect{S}_\psi \rangle}_{lj} \bigr)^{ik}
\Tstrut\Bstrut\nonumber\\
&&\quad
+\,c_4^{(\psi)}\, \vect{S}_\chi\cdot\hl{\langle\vect{S}_\psi\rangle}_{lj} 
\Tstrut\Bstrut\nonumber\\
&&\quad
+\,c_{5a}^{(\psi)} \,\frac{i\vect{q}}{m_\psi} \cdot\bigl(\vect{v}_\chi\times\vect{S}_\chi\bigr)\hl{\langle N_\psi\rangle}_{lj}
-\,c_{5b}^{(\psi)} \,\frac{q^2}{2m_\psi^2}\, \vect{S}_\chi\cdot \bigl(\mathbb{1}-\vect{\hat{q}}\vect{\hat{q}}\bigr)\cdot\hl{\langle\vect{L}_\psi\rangle}_{lj} 
\Tstrut\Bstrut\nonumber\\
&&\quad
+\, c_6^{(\psi)} \frac{q^2}{m_\psi^2}\, \bigl(\vect{\hat{q}}\cdot\vect{S}_\chi\bigr) \bigl(\vect{\hat{q}}\cdot\hl{\langle\vect{S}_\psi\rangle}_{lj}\bigr)
\Tstrut\Bstrut\nonumber\\
&&\quad
+\, c_{7a}^{(\psi)} \,\vect{v}_\chi\cdot\hl{\langle\vect{S}_\psi\rangle}_{lj}
-\, c_{7b}^{(\psi)} \,\epsilon^{ikk'}\frac{iq^{k'}}{2m_\chi} \bigl(\hl{\langle \vect{L}_\psi \otimes \vect{S}_\psi \rangle}_{lj} \bigr)^{ik}
\Tstrut\Bstrut\nonumber\\
&&\quad
+\,c_{8a}^{(\psi)} \,\bigl(\vect{v}_\chi\cdot\vect{S}_\chi\bigr) \hl{\langle N_\psi\rangle}_{lj}
-\,c_{8b}^{(\psi)} \,\frac{iq}{2m_\psi}\vect{S}_\chi\cdot\bigl(\vect{\hat q}\times \hl{\langle\vect{L}_\psi\rangle}_{lj}\bigr)
\Tstrut\Bstrut\nonumber\\
&&\quad 
+\,c_9^{(\psi)} \frac{iq}{m_\psi}\vect{S}_\chi\cdot\bigl(\hl{\langle\vect{S}_\psi\rangle}_{lj} \times \vect{\hat q} \bigr)
\Tstrut\Bstrut\nonumber\\
&&\quad
+\,c_{10}^{(\psi)} \frac{i\vect{q}}{m_\psi}\cdot\hl{\langle\vect{S}_\psi\rangle}_{lj} 
\Tstrut\Bstrut\nonumber\\
&&\quad
+\,c_{11}^{(\psi)} \frac{i\vect{q}}{m_\psi}\cdot\vect{S}_\chi\,\hl{\langle N_\psi\rangle}_{lj} 
\Tstrut\Bstrut\nonumber\\
&&\quad
+\,c_{12a}^{(\psi)}\, \bigl(\vect{v}_\chi\times \vect{S}_\chi\bigr)\cdot \hl{\langle\vect{S}_\psi\rangle}_{lj}
-\,c_{12b}^{(\psi)}\,\frac{iq}{2m_\psi} \bigl((\vect{\hat q}\cdot\vect{S}_\chi)\delta^{ik}-\hat q^k S_\chi^i\bigr) \bigl(\hl{\langle \vect{L}_\psi \otimes \vect{S}_\psi \rangle}_{lj} \bigr)^{ik}
\Tstrut\Bstrut\nonumber\\
&&\quad
+\,c_{13a}^{(\psi)} \,\frac{iq}{m_\psi} \bigl(\vect{v}_\chi\cdot \vect{S}_\chi\bigr) \bigl(\vect{\hat q}\cdot\hl{\langle\vect{S}_\psi\rangle}_{lj} \bigr)
-\,c_{13b}^{(\psi)} \,\frac{q^2}{2m_\psi^2}\, \bigl(\vect{\hat q}\times\vect{S}_\chi\bigr)\cdot\hl{\langle \vect{L}_\psi \otimes \vect{S}_\psi \rangle}_{lj} \cdot\vect{\hat q}
\Tstrut\Bstrut\nonumber\\
&&\quad
+\,c_{14a}^{(\psi)} \,\frac{iq}{m_\psi}\bigl(\vect{\hat q}\cdot\vect{S}_\chi\bigr) \bigl(\vect{v}_\chi\cdot\hl{\langle\vect{S}_\psi\rangle}_{lj}\bigr)
+\,c_{14b}^{(\psi)} \,\epsilon^{ikk'}\frac{q^2}{2m_\psi^2}\, {\hat q}^{k'}\bigl(\vect{\hat q}\cdot\vect{S}_\chi\bigr) \bigl(\hl{\langle \vect{L}_\psi \otimes \vect{S}_\psi \rangle}_{lj}\bigr)^{ik}
\Tstrut\Bstrut\nonumber\\
&&\quad
-\,c_{15a}^{(\psi)} \,\frac{q^2}{m_\psi^2}\bigl(\vect{\hat q}\cdot(\vect{v}_\chi\times \vect{S}_\chi)\bigr) \bigl(\vect{\hat q}\cdot\hl{\langle\vect{S}_\psi\rangle}_{lj} \bigr)
-\,c_{15b}^{(\psi)} \,\frac{iq^3}{2m_\psi^3}\,
\vect{S}_\chi\cdot \bigl(\mathbb{1}-\vect{\hat{q}}\vect{\hat{q}}\bigr)
\cdot\hl{\langle \vect{L}_\psi \otimes \vect{S}_\psi \rangle}_{lj} \cdot\vect{\hat q}
\,,\nonumber\\
\label{eq:Vlj}
\end{eqnarray}
where $\vect{v}_\chi = \vect{v} -\frac{\vect{q}}{2m_\chi}$ (with the incoming DM's velocity $\vect{v}$ and momentum transfer $\vect{q}$ to be integrated over when calculating detection rates), and summation over repeated Cartesian indices is implicit. 
Here and in what follows, we denote $q\equiv|\vect{q}|$ (so that $q^2\equiv\vect{q}^2\ne q^\mu q_\mu$), and $\vect{\hat q}\equiv \vect{q}/q$.

To summarize, in the long wavelength limit, the DM-ion scattering potential $\tilde{\cal V}_{lj}$ involves a set of quantities that characterize properties of the ion: the total numbers $\hl{\langle N_\psi\rangle}$, spins $\hl{\langle \vect{S}_\psi\rangle}$ and orbital angular momenta $\hl{\langle \vect{L}_\psi\rangle}$ of the constituent particles $\psi=p,n,e$, as well as the spin-orbit coupling tensors $\hl{\bigl\langle \vect{L}_\psi \otimes \vect{S}_\psi \bigr\rangle}$. We will refer to these as different types of  {\it crystal responses}, as DM couplings to these quantities drive collective excitations in the crystal; they play a similar role to the nuclear responses in nuclear recoil calculations (which similarly reduce to the total nucleon numbers, spins, etc.\ in the long wavelength limit~\cite{Fitzpatrick:2012ix,Anand:2013yka,Gresham:2014vja,Anand:2014kea}). 
We emphasize, however, that in contrast to standard nuclear recoil where nuclei are treated as free -- a valid approximation at energy depositions $\gtrsim 500$\,meV~\cite{Trickle:2019nya} -- collective excitations arise in a lower energy regime where inter-ionic interactions become important; the EFT therefore involves different degrees of freedom and the calculation proceeds differently.\footnote{Technically, Refs.~\cite{Fitzpatrick:2012ix,Anand:2013yka} defined a few ``nuclear response functions,'' $W_M^{\tau\tau'}$, $W_{\Sigma'}^{\tau\tau'}$ etc., which the nuclear recoil rate is proportional to, from the unpolarized average of nuclear matrix element squared. No such averaging is involved in the calculation of collective excitations, and the rate formulae derived below do not depend on the same functions $W_M^{\tau\tau'}$, $W_{\Sigma'}^{\tau\tau'}$ etc.\ even in the absence of coupling to electrons. Here we are simply borrowing the terminology ``response'' in the sense that it refers to a type of coupling, just as $M$, $\Sigma'$, etc., usually called ``nuclear responses,'' are different types of couplings to the nucleus.} We will sometimes abbreviate the crystal responses introduced above as $\hl{N}$, $\hl{\vect{S}}$, $\hl{\vect{L}}$, $\hl{\vect{L}\otimes\vect{S}}$, or simply $\hl{N}$, $\hl{S}$, $\hl{L}$, $\hl{L\otimes S}$, when there is no confusion. The crystal responses generated by each NR operator and in each benchmark DM model have been summarized in Tables~\ref{tab:operators} and \ref{tab:benchmarks}, respectively.

We reiterate that, among the four types of crystal responses, $\hl{\langle N_\psi\rangle}$ and $\hl{\langle \vect{S}_\psi\rangle}$ are induced by DM couplings to point-like ionic degrees of freedom (which do not involve internal motions of nucleons or electrons inside an ion), while $\hl{\langle \vect{L}_\psi\rangle}$ and $\hl{\bigl\langle \vect{L}_\psi \otimes \vect{S}_\psi \bigr\rangle}$ are induced by DM couplings to composite degrees of freedom. We therefore refer to them as point-like and composite responses respectively. $\vect{v}_\psi$-independent operators (the first two categories in Table~\ref{tab:operators}) generate point-like responses, while $\vect{v}_\psi$-dependent operators (the last two categories in Table~\ref{tab:operators}) generate composite responses. 
It is worth noting that operators related by $\vect{v}_\chi\leftrightarrow\vect{v}_\psi$ ({\it e.g.}\ $\OO_{3a}$ and $\OO_{3b}$, $\OO_{5a}$ and $\OO_{5b}$) are usually generated with similar coefficients. 
%
For each such pair of operators, the ratio of composite versus\ point-like responses ({\it i.e.}\ coefficients of $\hl{\langle \vect{L}_\psi\rangle}$ versus\ $\hl{\langle N_\psi\rangle}$, or $\hl{\bigl\langle \vect{L}_\psi \otimes \vect{S}_\psi \bigr\rangle}$ versus\ $\hl{\langle \vect{S}_\psi\rangle}$ in Eq.~\eqref{eq:Vlj}) is, parametrically, $\frac{q}{m_\psi v}$. This is  generic, as point-like and composite responses result from the leading two terms in the expansion $e^{i\vect{q}\cdot\vect{x}_\alpha} = 1+i\vect{q}\cdot\vect{x}_\alpha+\dots$, and $qx\sim \frac{q}{m_\psi v} L$, with $L\sim\OO(1)$. For nuclear recoils (where the two operators have exactly equal and opposite coefficients), $\frac{q}{m_\psi v} \sim \frac{\mu_{\chi N}}{m_{p,n}}$ with $\mu_{\chi N}$ the reduced mass of the DM and the target nucleus, so composite responses can be significant, as emphasized in Refs.~\cite{Anand:2013yka,Gresham:2014vja}. In contrast, in the present case of collective excitations induced by light DM, we have $\frac{q}{m_\psi v} \lesssim \frac{m_\chi}{m_\psi}$. For couplings to nucleons, $\psi = p,n$, this ratio is always smaller than one for sub-GeV DM, so for a given type of excitation, point-like responses tend to dominate; for couplings to electrons, $\psi=e$, both point-like and composite responses, if present, can be important. From the bottom-up point of view, it is useful to keep in mind this interplay between point-like and composite responses for the purpose of organizing the effects of various operators, although from the top-down point of view, it seems difficult to construct well-motivated simple models that dominantly generate a composite response ($\hl{L}$ or $\hl{L\otimes S}$) without being accompanied by a point-like response ($\hl{N}$ or $\hl{S}$) of at least comparable size, similar to the case of nuclear recoil as highlighted in Ref.~\cite{Gresham:2014vja}. We will elaborate on this in Sec.~\ref{sec:benchmarks}.

\subsection{Quantization of Lattice Potential for Phonons and Magnons}
\label{subsec:quantization}

Now that we have obtained $\tilde{\cal V}_{lj}$ in terms of the lattice degrees of freedom, Eq.~\eqref{eq:Vlj}, it remains to compute the matrix elements
\begin{equation}
\langle \nu,\vect{k} |\,\tilde{\cal V} (-\vect{q},\vect{v}) |0\rangle = \sum_{l,j} \langle \nu,\vect{k} |\,e^{i\vect{q}\cdot\vect{x}_{lj}} \,\tilde{\cal V}_{lj} (-\vect{q},\vect{v}) |0\rangle
\label{eq:Vtilde_ME}
\end{equation}
by quantizing the lattice potential in terms of phonon or magnon modes. The simplest cases, where phonon excitations in a crystal proceed through $\hl{\langle N_\psi\rangle}$ (via the SI operator $\OO_1=\mathbb{1}$) and magnon excitations proceed through $\hl{\langle \vect{S}_e\rangle}$ were considered previously in Refs.~\cite{Knapen:2017ekk,Griffin:2018bjn,Trickle:2019nya, Griffin:2019mvc} and Ref.~\cite{Trickle:2019ovy}, respectively. Here we extend those calculations to include all four crystal responses ($\hl{\langle N_\psi \rangle}$, $\hl{\langle \vect{S}_\psi \rangle}$, $\hl{\langle \vect{L}_\psi \rangle}$, $\hl{\langle \vect{L}_\psi \otimes \vect{S}_\psi \rangle}$) identified in the previous subsection, which can be generated by the full set of effective operators.

Phonons arise from the ions' displacements with respect to their equilibrium positions $\vect{x}_{lj}^0$:
\begin{equation}
\vect{u}_{lj} = \vect{x}_{lj}-\vect{x}_{lj}^0 = \sum_{\nu=1}^{3n} \sum_{\vect{k}\in\text{1BZ}} \frac{1}{\sqrt{2Nm_j\omega_{\nu,\vect{k}}}} \Bigl( \,\hat a_{\nu,\vect{k}}\,\vect{\epsilon}_{\nu,\vect{k},j} \,e^{i\vect{k}\cdot\vect{x}_{lj}^0} +\hat a^\dagger_{\nu,\vect{k}}\, \vect{\epsilon}_{\nu,\vect{k},j}^* \,e^{-i\vect{k}\cdot\vect{x}_{lj}^0} \Bigr)\,.
\label{eq:ulj}
\end{equation}
Recall that $N$ (without subscript, not to be confused with $\hl{\langle N_\psi\rangle}$) is the total number of primitive cells in the crystal lattice, to be sent to infinity at the end of the calculation. The phonon creation and annihilation operators satisfy the canonical commutation relations, $[\hat a_{\nu,\vect{k}}, \hat a_{\nu',\vect{k}'}^\dagger]=\delta_{\nu\nu'}\delta_{\vect{k},\vect{k}'}$ with all others vanishing. The eigenenergies $\omega_{\nu,\vect{k}}$ and eigenvectors $\vect{\epsilon}_{\nu,\vect{k},j}$ (normalized such that $\sum_j|\vect{\epsilon}_{\nu,\vect{k},j}|^2=1$) are solved for by diagonalizing the quadratic crystal potential. The quadratic crystal potential, and equilibrium positions, are computed with DFT~\cite{Martin2004a} (see Refs.~\cite{Griffin:2018bjn,Griffin:2019mvc} for details) and the diagonalization is performed with \textsf{phonopy} \cite{phonopy}.  At leading order, dependence of the matrix element in Eq.~\eqref{eq:Vtilde_ME} on $\vect{u}_{lj}$ comes only from the phase factor $e^{i\vect{q}\cdot\vect{x}_{lj}}$; we assume the DM-ion scattering potentials $\tilde{\cal V}_{lj} (-\vect{q},\vect{v})$ are not significantly affected by ionic displacements and can thus be pulled out of the matrix element.\footnote{If $\tilde{\cal V}_{lj}$ receives contributions from DM-electron couplings, the scattering potential can depend on $\vect{u}_{lj}$ directly, as ionic displacements distort the electron wavefunctions. This correction can be taken into account via the Born effective charges in the case of SI interactions in the long wavelength limit, as discussed in Ref.~\cite{Trickle:2019nya}.} Then, evaluating the matrix element of the phase factor, $\langle \nu,\vect{k} |\,e^{i\vect{q}\cdot\vect{x}_{lj}} |0\rangle$, follows the standard procedure of expanding $\vect{x}_{lj}$ as in Eq.~\eqref{eq:ulj} and applying the Baker-Campbell-Hausdorff formula to normal-order the phonon creation and annihilation operators~\cite{Trickle:2019nya}. As a result,
\begin{equation}
\langle \nu,\vect{k} |\,\tilde{\cal V} (-\vect{q},\vect{v}) |0\rangle = \frac{1}{\sqrt{N}}\sum_{\nu,\vect{k},j} \Biggl[\sum_l \tilde{\cal V}_{lj} (-\vect{q},\vect{v}) \,e^{i(\vect{q}-\vect{k})\cdot\vect{x}_{lj}^0}\Biggr] e^{-W_j(\vect{q})} \,\frac{i(\vect{q}\cdot\vect{\epsilon}^*_{\nu,\vect{k},j})}{\sqrt{2m_j\omega_{\nu,\vect{k}}}} \,,
\label{eq:phonon_ME}
\end{equation}
where $W_j(\vect{q}) = \frac{1}{4Nm_j}\sum_{\nu,\vect{k}}\frac{|\vect{q}\cdot\vect{\epsilon}_{\nu,\vect{k},j}|^2}{\omega_{\nu,\vect{k}}}$ is the Debye-Waller factor. Crucially, the $\frac{1}{\sqrt{N}}$ factor (which originates from Eq.~\eqref{eq:ulj} and is to be squared when computing the rate), together with the prefactor $\frac{1}{V}$ in the rate formula Eq.~\eqref{eq:Gamma_phononmagnon}, indicates that the rate $\Gamma$ would scale as $\frac{1}{N^2}\to0$ {\it unless} the $l$ sum in Eq.~\eqref{eq:phonon_ME} scales with $N$. This in turn requires the $N$ terms in the $l$ sum to add up coherently, which is possible only when {\it i)} the phonon momentum $\vect{k}$ matches the momentum transfer $\vect{q}$ up to reciprocal lattice vectors, which is the statement of lattice momentum conservation, and {\it ii)} $\sum_l \tilde{\cal V}_{lj}\sim N$, {\em i.e.}\ the DM couples coherently across the crystal lattice. The second requirement is trivially satisfied for DM couplings to the scalar quantities $\hl{\langle N_\psi\rangle}$, $\text{tr}(\hl{\langle \vect{L}_\psi\otimes\vect{S}_\psi\rangle})$. For couplings to the vector and tensor quantities $\hl{\langle \vect{S}_\psi\rangle}$, $\hl{\langle \vect{L}_\psi\rangle}$, $\hl{\langle\vect{L}_\psi\otimes\vect{S}_\psi\rangle}$ (modulo the trace part), on the other hand, coherence is possible only when they are ordered (or polarized), so that they point in the same directions in all primitive cells; in the case of $\hl{\langle \vect{S}_\psi \rangle}$, this can be achieved by spontaneous magnetic ordering for $\psi=e$, or by applying an external magnetic field for $\psi=p,n$.

Up to possible small corrections due to the presence of different isotopes, we can set $\tilde{\cal V}_{lj}=\tilde{\cal V}_j$, which is independent of $l$.  We then obtain the single phonon excitation rate:
\begin{equation}
\Gamma(\vect{v}) = \frac{1}{\Omega} \int \frac{d^3q}{(2\pi)^3} \sum_{\nu=1}^{3n}\, 2\pi\,\delta\bigl(\omega_{\nu,\vect{k}}-\omega_{\vect{q}}\bigr) \frac{1}{2\omega_{\nu,\vect{k}}} \biggl|\sum_j e^{-W_j(\vect{q})} e^{i\vect{G}\cdot\vect{x}_j^0} \,\frac{\vect{q}\cdot\vect{\epsilon}_{\nu,\vect{k},j}^*}{\sqrt{m_j}}\,\tilde{\cal V}_j (-\vect{q},\vect{v}) \biggr|^2 \,,
\label{eq:phonon_rate}
\end{equation}
where $\Omega$ is the volume of the primitive cell, $\vect{x}_j^0$ is the equilibrium position of the $j$th ion with respect to the cell center, and it is implicit that $\vect{q} = \vect{k}+\vect{G}$ where $\vect{G}$ is a reciprocal lattice vector. To map $\vect{q}$ to a vector $\vect{k}$ within the 1BZ, we first write $\vect{q} = \sum_{i=1}^3 a_i \vect{b}_i$, with $\vect{b}_i$ the basis vectors of the reciprocal lattice, then construct a set of eight candidate $\vect{G}$ vectors whose components in reduced coordinates take the floor and ceiling integer values of $a_i$, and finally choose the correct $\vect{G}$ vector to be the one that minimizes $|\vect{q}-\vect{G}|$.

The DM-ion scattering potential $\tilde{\cal V}_j$ that enters Eq.~\eqref{eq:phonon_rate} is simply given by Eq.~\eqref{eq:Vlj} above, with the $l$ subscripts dropped, assuming $\hl{\langle\vect{S}_\psi\rangle}$, $\hl{\langle\vect{L}_\psi\rangle}$, $\hl{\langle\vect{L}_\psi\otimes\vect{S}_\psi\rangle}$ are ordered, as explained above; in the absence of ordering, the corresponding terms should be dropped (with $\hl{\langle\vect{L}_\psi\otimes\vect{S}_\psi\rangle}$ set to its scalar component $\frac{1}{3}\,\text{tr}(\hl{\langle\vect{L}_\psi\otimes\vect{S}_\psi\rangle})\,\mathbb{1} = \frac{1}{3}\,\langle\vect{L}_\psi\cdot\vect{S}_\psi\rangle\,\mathbb{1}$). In the special case of SI interactions, one has only $c_1^{(\psi)}$, so $\tilde V_j =\sum_\psi c_1^{(\psi)} \hl{\langle N_\psi\rangle}_j$, reproducing the results in Ref.~\cite{Trickle:2019nya}, whereas in the full EFT, all four crystal responses can contribute to phonon excitations.

Next we move on to magnons. They are collective spin excitations in a magnetically ordered phase, and can thus respond to DM scattering only if the potentials $\tilde{\cal V}_{lj}$ depend on the magnetic ions' effective spins $\vect{S}_{lj}$. Generally, $\vect{S}_{lj}$ can come from the electrons' spin and orbital angular momenta, $\hl{\langle \vect{S}_e\rangle}_{lj}$ and $\hl{\langle \vect{L}_e\rangle}_{lj}$, respectively. When projected onto the Hilbert space spanned by $\vect{S}_{lj}$, they become
\begin{equation}
   \hl{ \langle \vect{S}_e\rangle}_{lj} \to \lambda_{S,j} \vect{S}_{lj}\,,
\qquad
\hl{\langle \vect{L}_e\rangle}_{lj} \to \lambda_{L,j} \vect{S}_{lj}\,,
\label{eq:proj}
\end{equation}
where $\lambda_{S,j}$, $\lambda_{L,j}$ are numbers (which we will say more about shortly). Therefore, from Eq.~\eqref{eq:Vlj} we obtain the  matrix element for exciting a magnon mode $|\nu,\vect{k}\rangle$:
\begin{equation}
\langle \nu,\vect{k} |\,\tilde{\cal V} (-\vect{q},\vect{v}) |0\rangle = \sum_{l,j} e^{i\vect{q}\cdot\vect{x}_{lj}} \vect{f}_j(-\vect{q},\vect{v}) \cdot \langle \nu,\vect{k} | \vect{S}_{lj}|0\rangle\,,
\end{equation}
where
 \begin{eqnarray}
 \vect{f}_j(-\vect{q},\vect{v}) &=& \lambda_{S,j} \biggl[ 
 c_{3a}^{(e)} \frac{iq}{m_e}\bigl(\vect{\hat q}\times\vect{v}_\chi\bigr) 
 +c_4^{(e)}\vect{S}_\chi 
 + c_6^{(e)} \frac{q^2}{m_e^2} \bigl(\vect{\hat{q}}\cdot\vect{S}_\chi\bigr)\, \vect{\hat{q}}
 + c_{7a}^{(e)}\vect{v}_\chi
 + c_9^{(e)} \frac{iq}{m_e}\bigl(\vect{\hat q}\times\vect{S}_\chi\bigr)
 +c_{10}^{(e)} \frac{i\vect{q}}{m_e} 
  \nonumber\\
 &&\;\;
 + c_{12a}^{(e)} \bigl(\vect{v}_\chi \times \vect{S}_\chi\bigr)
 +  c_{13a}^{(e)} \frac{i\vect{q}}{m_e} \bigl( \vect{v}_\chi \cdot \vect{S}_\chi \bigr)
 + c_{14a}^{(e)} \frac{i q}{m_e} \left(\vect{\hat{q}} \cdot \vect{S}_\chi \right) \vect{v}_\chi
 - c_{15a}^{(e)} \frac{q^2}{m_e^2} \bigl( \vect{\hat{q}} \cdot ( \vect{v}_\chi \times \vect{S}_\chi ) \bigr) \vect{\hat{q}}
 \biggr] 
  \nonumber\\
 && +\frac{\lambda_{L,j}}{2} \biggl[ -c_{5b}^{(e)}\frac{q^2}{m_e^2} \bigl(\mathbb{1}-\vect{\hat{q}}\vect{\hat{q}}\bigr)\cdot\vect{S}_\chi
 +c_{8b}^{(e)}\frac{iq}{m_e}\bigl(\vect{\hat q}\times\vect{S}_\chi\bigr) \biggr]\,.
 \label{eq:fj}
 \end{eqnarray}
As in Eq.~\eqref{eq:Vlj}, we have defined $q\equiv|\vect{q}|$, $\vect{\hat q}\equiv \vect{q}/q$, and $\vect{v}_\chi =\vect{v} - \frac{\vect{q}}{2m_\chi}$.

Now we need to compute $\langle \nu,\vect{k} | \vect{S}_{lj}|0\rangle$. The calculation follows Ref.~\cite{Trickle:2019ovy}, which we encourage the reader to consult for more details. The magnetic order is captured by a set of rotation matrices $\tens{R}_j$ that take each $\vect{S}_{lj}$ to a local coordinate system where it points in the $+z$ direction:
\begin{equation}
\vect{S}_{lj} = \tens{R}_j\cdot \vect{S}'_{lj}\,,\qquad\quad
\langle\vect{S}'_{lj} \rangle = \bigl( \langle S'^x_{lj}\rangle,\, \langle S'^y_{lj}\rangle,\, \langle S'^z_{lj}\rangle  \bigr)
= \bigl( 0,\, 0,\, S_j\bigr)\,.
\end{equation}
We restrict ourselves to commensurate orders, in which case the rotation matrices $\tens{R}_j$ do not depend on the primitive cell label $l$. We then apply the Holstein-Primakoff transformation and expand $\vect{S}_{lj}$ around the ground state in terms of bosonic creation and annihilation operators:
\begin{equation}
S'^x_{lj} = \bigl(2S_j-\hat a^\dagger_{lj} \hat a_{lj}\bigr)^{1/2}\,  \hat a_{lj}\,,\qquad
S'^y_{lj} = \hat a^\dagger_{lj}\, \bigl(2S_j-\hat a^\dagger_{lj} \hat a_{lj}\bigr)^{1/2} \,,\qquad
S'^z_{lj} = S_j -\hat a^\dagger_{lj} \hat a_{lj} \,.
\end{equation}
Magnon eigenstates are obtained by diagonalizing the spin Hamiltonian, which is specific to the target material; in the simplest cases, the target can be modeled by Heisenberg exchange interactions $\vect{S}_{lj}\cdot\vect{S}_{l'j'}$ between neighboring sites, while more complicated model descriptions are needed in other cases. For a general spin Hamiltonian, the diagonalization can be achieved by a Bogoliubov transformation in momentum space: 
\begin{equation}
\hat a_{lj} = \frac{1}{\sqrt{N}} \sum_{\vect{k}\in\text{1BZ}} \hat a_{j,\vect{k}} \,e^{i\vect{k}\cdot\vect{x}_{lj}}\,,\qquad
\left(\begin{matrix}
	\hat a_{j,\vect{k}} \\
	\hat a_{j,-\vect{k}}^\dagger
\end{matrix}\right) =
\left(\begin{matrix}
	\mathbb{U}_{j\nu,\vect{k}} & \mathbb{V}_{j\nu,\vect{k}} \\
	\mathbb{V}_{j\nu,-\vect{k}}^* & \mathbb{U}_{j\nu,-\vect{k}}^*
\end{matrix}\right)
\left(\begin{matrix}
	\hat b_{j,\vect{k}} \\
	\hat b_{j,-\vect{k}}^\dagger
\end{matrix}\right) \,,
\end{equation}
where $\mathbb{U}$, $\mathbb{V}$ are $n\times n$ matrices (with $n$ the number of magnetic ions per cell), and $\hat b_{j,\vect{k}}^\dagger$, $\hat b_{j,\vect{k}}$ are the creation and annihilation operators for the magnon eigenstates satisfying canonical commutation relations, $[\hat b_{\nu,\vect{k}}, \hat b_{\nu',\vect{k}'}^\dagger]=\delta_{\nu\nu'}\delta_{\vect{k},\vect{k}'}$ with all others vanishing. An efficient algorithm for the diagonalization can be found in Ref.~\cite{Toth-Lake} (see also Refs.~\cite{Trickle:2019ovy,Mitridate:2020kly}). Now computing the magnon excitation matrix element $\langle\nu,\vect{k}| \vect{S}_{lj} |0\rangle$, and hence the DM scattering rate, is reduced to standard algebra. We obtain~\cite{Trickle:2019ovy,Mitridate:2020kly}
\begin{equation}
\Gamma(\vect{v}) = \frac{1}{\Omega} \int \frac{d^3q}{(2\pi)^3} \sum_{\nu=1}^{n}\, 2\pi\,\delta\bigl(\omega_{\nu,\vect{k}}-\omega_{\vect{q}}\bigr) 
\,\frac{1}{2} \,\biggl|\sum_j e^{i\vect{G}\cdot\vect{x}_j^0} \sqrt{S_j}\, \bigl(\mathbb{U}^*_{j\nu,\vect{k}} \vect{r}_j +\mathbb{V}_{j\nu,-\vect{k}}\vect{r}_j^*\bigr)   \cdot \vect{f}_j(-\vect{q},\vect{v})  \biggr|^2 \,,
\label{eq:magnon_rate}
\end{equation}
where $\vect{r}_j = (R_j^{xx},\, R_j^{yx},\, R_j^{zx}) +i \, (R_j^{xy},\, R_j^{yy},\, R_j^{zy})$. As in the phonon case, it is implicit that $\vect{k}$ matches $\vect{q}$ up to a reciprocal lattice vector, $\vect{q} = \vect{k}+\vect{G}$, due to lattice momentum conservation.

A comment is in order about the target choice. In the case where the total $\vect{S}_{lj}$ involve only spin degrees of freedom (as is the case for yttrium iron garnet (YIG) discussed in Ref.~\cite{Trickle:2019ovy}),  $\lambda_{S,j}=1$, $\lambda_{L,j}=0$, and only the first two lines of Eq.~\eqref{eq:fj} are relevant. Targets for which $\lambda_{L,j} \ne 0$ are more exotic. One class of materials with $\lambda_{L,j} \ne 0$ is spin-orbit-entangled Mott insulators \cite{Jackeli_2009,Witczak_Krempa_2014,trebst2017kitaev}, where the combined effect of crystal fields and spin-orbit coupling results in effective spins $S_j=\frac{1}{2}$, and we can show that $\lambda_{S,j}=-\frac{1}{3}$, $\lambda_{L,j}=-\frac{4}{3}$ (see Appendix~\ref{app:angular_momentum} for details, and Refs.~\cite{Kim_2008,Witczak_Krempa_2014,trebst2017kitaev,Winter_2017} for related discussion), so the magnetic ions' effective spins are in fact dominated by their orbital components. Perovskite irridates such as Sr$_2$IrO$_4$~\cite{Kim_2008,Jackeli_2009} and Kitaev materials Na$_2$IrO$_3$, $\alpha$-RuCl$_3$ \cite{Plumb_2014,trebst2017kitaev,Janssen_2017,Winter_2017} are among the materials with this feature that have been actively studied recently by the condensed matter physics community. While perhaps futuristic as DM detectors, such materials have the novel feature of being sensitive to DM couplings with electrons' orbital angular momenta.

As a final remark, we note from the derivation above that when the same crystal response, $\hl{\langle\vect{S}_e\rangle}$ or $\hl{\langle\vect{L}_e\rangle}$, excites both phonons and magnons, the phonon excitation rate is parametrically suppressed by $\frac{q^2}{m_\text{ion}\omega}\sim 10^{-2} \bigl(\frac{q}{\text{keV}}\bigr)^2\bigl(\frac{10\,\text{GeV}}{m_\text{ion}}\bigr)\bigl(\frac{10\,\text{meV}}{\omega}\bigr)$. Thus, for example, for the second group of operators in Table~\ref{tab:operators} with $\psi=e$, which generates $\hl{\langle\vect{S}_e\rangle}$ response, single magnon excitation is expected to achieve better sensitivity than single phonon excitation for the same exposure and detector efficiency. On the other hand, since phonons can be excited also by other crystal responses, they have a broader coverage of the DM theory space. We will investigate the interplay between single phonon and magnon excitations in the context of our benchmark models in the next section.

\section{Application to Benchmark Models}
\label{sec:benchmarks}

We now apply the general results of the previous section to the set of benchmark models in Table~\ref{tab:benchmarks}. The first step of the calculation -- matching the relativistic model onto the NR EFT -- was already done in Sec.~\ref{subsec:NR_matching}. The results are the operator coefficients $c_i^{(\psi)}$ listed in the second to last column of Table~\ref{tab:benchmarks}. We then need to substitute these operator coefficients into the formulae derived in Secs.~\ref{subsec:lattice_matching} and \ref{subsec:quantization} to compute direct detection rates $\Gamma(\vect{v})$ --- Eq.~\eqref{eq:phonon_rate} together with Eq.~\eqref{eq:Vlj} for single phonon excitations, and Eq.~\eqref{eq:magnon_rate} together with Eq.~\eqref{eq:fj} for single magnon excitations. 

In order to present the results in a concise way, let us introduce the following definitions. For single phonon excitation, we define (cf.\ Eq.~\eqref{eq:phonon_rate})
\begin{equation}
F_{X,\nu}^{(\psi)} (\vect{q}) \equiv \sum_j e^{-W_j(\vect{q})} e^{i\vect{G}\cdot\vect{x}_j^0} \,\frac{\vect{q}\cdot\vect{\epsilon}_{\nu,\vect{k},j}^*}{\sqrt{2m_j\omega_{\nu,\vect{k}}}}\,\hl{\langle X_\psi\rangle}_j \,,
\label{eq:FX_def}
\end{equation}
where $X$ represents one of the crystal responses, $X = \hl{N}, \hl{S}, \hl{L}, \hl{L\otimes S}$; note that $F_{X,\nu}^{(\psi)}$ are vector (tensor) quantities when $X=\hl{S},\hl{L}$ ($X=\hl{L\otimes S}$), and will be written as $\vect{F}_{X,\nu}^{(\psi)}$ ($\tens{F}_{X,\nu}^{(\psi)}$). These $F_{X,\nu}^{(\psi)}$ play the role of form factors for exciting a single phonon via a certain type of response. For single magnon excitation, we define (cf.\ Eq.~\eqref{eq:magnon_rate})
\begin{equation}
\vect{E}_{X,\nu}(\vect{q}) \equiv \sum_j e^{i\vect{G}\cdot\vect{x}_j^0} \,\sqrt{\frac{S_j}{2}}\, \bigl(\mathbb{U}^*_{j\nu,\vect{k}} \vect{r}_j +\mathbb{V}_{j\nu,-\vect{k}}\vect{r}_j^*\bigr) \,\lambda_{X,j}\,,
\label{eq:EX_def}
\end{equation}
where $X=\hl{S},\hl{L}$. These are formally analogous to polarization vectors of a vector field. In both Eqs.~\eqref{eq:FX_def} and \eqref{eq:EX_def}, $\vect{k}$ is the phonon momentum inside the 1BZ that satisfies $\vect{q} = \vect{k}+\vect{G}$ for some reciprocal lattice vector $\vect{G}$; as emphasized below Eq.~\eqref{eq:phonon_rate}, $\vect{k}$ is uniquely determined by mapping $\vect{q}$ into the 1BZ through reciprocal lattice vectors. We further define a set of quantities $\Sigma_\nu(\vect{q})$, for both single phonon and single magnon excitations, by (cf.\ Eq.~\eqref{eq:Gamma_phononmagnon})
\begin{equation}
\Gamma(\vect{v}) \equiv \frac{1}{\Omega} \int \frac{d^3q}{(2\pi)^3} \sum_{\nu}\, 2\pi\,\delta\bigl(\omega_{\nu,\vect{k}}-\omega_{\vect{q}}\bigr) \,\Sigma_\nu(\vect{q})\,.
\label{eq:Gamma}
\end{equation}
We will refer to $\Sigma_\nu(\vect{q})$, which have mass dimension $-4$, as ``differential rates.'' Practically, $\Sigma_\nu(\vect{q})$ are obtained simply by taking $\tilde{\cal V}_{lj}$ in Eq.~\eqref{eq:Vlj}, substituting $\hl{\langle X_\psi\rangle}_{lj}$ by $F_{X,\nu}^{(\psi)}$ (for $\psi=p,n,e$ and $X=\hl{N},\hl{S},\hl{L},\hl{L\otimes S}$) or $\vect{E}_{X,\nu}$ (for $\psi=e$ only, and $X=\hl{S},\hl{L}$), squaring it and averaging over the DM's spin (which amounts to replacing $S_\chi^i S_\chi^k \to \frac{1}{4} \delta^{ik}$). As we will see, written in terms of the dimensionless quantities $F_{X,\nu}^{(\psi)}$ and $\vect{E}_{X,\nu}$ defined above, $\Sigma_\nu(\vect{q})$ can be expressed in a concise form for each benchmark model. This will be convenient when we compare the rates between different models, and between phonon and magnon excitations.

Our final results will be presented in terms of the rate per unit target mass,
\begin{equation}
R = \frac{1}{\rho_T}\frac{\rho_\chi}{m_\chi} \int d^3v\, f_\chi(\vect{v}) \,\Gamma(\vect{v})\,,
\label{eq:total_rate}
\end{equation}
where $\rho_T$ is the target's mass density that we take from Ref.~\cite{MaterialsProject}, $\rho_\chi=0.4\,\text{GeV}/\text{cm}^3$ is the local DM mass density, and $f_\chi(\vect{v})$ is the DM's velocity distribution, taken to be a boosted and truncated Maxwell-Boltzmann distribution --- see Appendix~\ref{app:v_int} for technical details of evaluating the velocity integrals. For the projected reach, we assume 3 events per kilogram-year exposure, corresponding to 95\% C.L.\ exclusion in a background-free counting experiment, and assume a detector energy threshold of 1\,meV. While we will present full numerical results, the main features can usually be understood by simple parametric estimates. Generally, noting that the velocity integral over the energy conserving delta function $\delta(\omega_{\nu,\vect{k}}-\omega_{\vect{q}})$ yields a function that scales as $q^{-1}$ (see Appendix~\ref{app:v_int}), we have from Eqs.~\eqref{eq:Gamma} and \eqref{eq:total_rate}, parametrically, 
\begin{equation}
R\sim \frac{\rho_\chi}{m_\chi}\frac{1}{m_\text{cell}} \int dq \, q\, \Sigma\,,
\label{eq:R_est}
\end{equation}
where $m_\text{cell}=\rho_T\Omega$ is the mass of the primitive cell, and as before, $q=|\vect{q}|$. Then, from the formulas for $\Sigma_\nu(\vect{q})$ presented below for each model in terms of $F_{X,\nu}^{(\psi)}$ and $\vect{E}_{X,\nu}$ defined in Eqs.~\eqref{eq:FX_def} and \eqref{eq:EX_def}, we can estimate the rate $R$ by
\begin{equation}
F_{X,\nu}^{(\psi)} \sim \frac{q}{\sqrt{m_\text{ion}\omega}}\langle X_\psi\rangle \,,\qquad
\vect{E}_{X,\nu} \sim \sqrt{S_\text{ion}} \,.
\label{eq:FX_EX_est}
\end{equation}
In the case of single phonon excitations, we should further note that $\omega$, which appears in $F_{X,\nu}^{(\psi)}$ above, can scale differently with $q$ for different models and DM masses. Typically, either acoustic phonons (associated with in-phase oscillations) or optical phonons (associated with out-of-phase oscillations) dominate the total rate, depending on whether the DM model couples to different ions in a correlated or anti-correlated way. For acoustic phonons, and for $q$ within the 1BZ, $\omega\sim c_s q$ (with $c_s$ the sound speed that is typically $\OO(10^{-5})$), whereas for optical phonons or for $q$ beyond the 1BZ, $\omega\sim q^0$. The size of the 1BZ is set by the inverse lattice spacing $a^{-1}$, and is typically $\OO(\text{keV})$. Since $v\sim\OO(10^{-3})$, contributions from outside the 1BZ are possible for DM masses above around an MeV. We will see below that in several cases, the curves scale differently for $m_\chi\lesssim$ MeV and $m_\chi\gtrsim$ MeV for this reason. 

On the target side, we will consider the following representative set of materials:
\begin{itemize}
    \item {\bf GaAs} [phonons, subject of R\&D]. As the first-studied target for DM detection via phonons, GaAs is already in R\&D as a target for both electron excitations and phonon excitations~\cite{tesseract}.  Phonons in GaAs form 3 acoustic and 3 optical branches, and have energies up to $\sim$35\,meV. 
	\item {\bf SiO$_2$} (quartz) [phonons, optimal sensitivity]. Based on our previous theoretical study comparing the phonon reach of a variety of target materials~\cite{Griffin:2019mvc}, we have advocated quartz as having good sensitivity to DM couplings to both acoustic and optical phonons. Also, quartz has complementary features compared to GaAs: while GaAs has a simple crystal structure and relatively low phonon energies, quartz has a large number of phonon branches (3 acoustic, 24 optical), with energies up to $\sim150$\,meV.
    \item Y$_3$Fe$_5$O$_{12}$ ({\bf YIG}) [mostly magnons, also phonons for comparison]. YIG is a well studied material with ferrimagnetic order, and is already used in an axion DM detection experiment QUAX~\cite{Ruoso:2015ytk,Barbieri:2016vwg,Crescini:2018qrz,Alesini:2019ajt,Crescini:2020cvl}. The effective spin Hamiltonian is a Heisenberg model, with $S_j=\frac{5}{2}$ for the magnetic Fe$^{3+}$ ions coming entirely from electron spins \hl{$\langle \vect{S}_e\rangle$} ({\it i.e.}\ $\lambda_{S,j}=1$, $\lambda_{L,j}=0$ in Eq.~\eqref{eq:EX_def}). We take the antiferromagnetic exchange coupling parameters from Ref.~\cite{Cherepanov:1993}, together with the crystal parameters from Ref.~\cite{MaterialsProject}, to compute the magnon spectrum and rotation matrices. YIG has 20 magnon branches, one of which is gapless and has a quadratic dispersion at small $k$. The gapped magnons have energies up to $\sim90$\,meV. We will mostly consider YIG as a candidate material for DM detection via magnon excitations, but will also consider phonon excitations in YIG via DM couplings to the ordered electron spins in Sec.~\ref{subsec:SD} for comparison; in this case the scattering potential is determined by $\hl{\langle \vect{S}_e\rangle}_{lj}$ of the Fe$^{3+}$ ions, which have magnitude $\frac{5}{2}$ and directions set by the ferrimagnetic order. YIG has 80 ions in total in the primitive cell and therefore 240 phonon branches (3 acoustic, 237 optical), with energies up to $\sim 120$ meV.
	\item {\bf $\alpha$-RuCl$_3$} [small-gap magnons with orbital component]. As discussed below Eq.~\eqref{eq:fj}, $\alpha$-RuCl$_3$ is one of the materials where the effective ionic spins involve orbital degrees of freedom, and is therefore sensitive to DM couplings to the electrons' orbital angular momenta. The magnetic ions Ru$^{3+}$ have $S_{lj}=\frac{1}{2}$, coming from both \hl{$\langle \vect{S}_e\rangle$} and \hl{$\langle \vect{L}_e\rangle$} with $\lambda_{S,j}=-\frac{1}{3}$, $\lambda_{L,j}=-\frac{4}{3}$, as discussed in Appendix~\ref{app:angular_momentum}. The effective spin Hamiltonian features Kitaev-type bond-directional exchange couplings. We use the Hamiltonian parameters derived from neutron scattering data in Ref.~\cite{Banerjee2016}, which also includes an antiferromagnetic Heisenberg exchange; see Ref.~\cite{Janssen_2017} for a summary of some alternative model parameterizations derived from a variety of experimental and numerical techniques. The resulting magnetic order is zig-zag antiferromagnetic. Magnons in $\alpha$-RuCl$_3$, of which there are 4 branches, are at very low energy, below 7\,meV, and can thus probe lighter DM than YIG. Also, since all magnon branches are gapped at zero momentum, the sensitivity is not significantly affected by the finite detector threshold. This is in contrast with YIG, where the assumed 1\,meV energy threshold limits the momentum transfer to be greater than $\sim80$\,eV in order to excite magnons on the gapless branch. Therefore, even though the experimental prospects of $\alpha$-RuCl$_3$ itself are unclear, it can be regarded as a useful benchmark which highlights the generic advantage of small-gap targets.
\end{itemize}

Our main results are Figs.~\ref{fig:SD}-\ref{fig:LS}. We give a brief summary here and discuss them in more detail in the following subsections. A major issue of interest is the comparison of sensitivity to various types of DM interactions, via single phonon and magnon excitations induced by various crystal responses. First, we consider the standard SD interaction in Fig.~\ref{fig:SD}, where we see that magnons outperform phonons, typically, by more than an order of magnitude in terms of the coupling reach. Next, in Fig.~\ref{fig:scalar}, we compare the phonon and magnon rates for the four combinations of scalar mediator couplings; the phonon production rate is larger, if the scalar and pseudoscalar couplings are of the same order, while magnons allow access to the models where the mediator dominantly couples to the pseudoscalar currents of SM fermions.  Next, we compare the reach of phonons and magnons to multipole models in Fig.~\ref{fig:multipole}; for the magnetic dipole and anapole models we expect the magnon reach to be better, and indeed it is. However, the phonon reach from quartz is sufficiently strong that, given the greater experimental challenges currently associated with magnon read-out, quartz should be considered a competitor for these models. Lastly, in Fig.~\ref{fig:LS}, we compare theoretical reach in the $(\vect{L}\cdot\vect{S})$-interacting model, where magnons outperform phonons for sub-MeV DM with the same exposure; however, the $(\vect{L}\cdot\vect{S})$-interacting model is difficult to UV complete, and our calculation is perhaps somewhat an academic exercise that demonstrate aspects of the EFT.  

We now discuss each benchmark model in turn.

\subsection{Standard Spin-Dependent Interaction}
\label{subsec:SD}

For the standard SD interaction there is only one operator, $\OO_4$, which generates the $\hl{S}$ response, and can excite both phonons and magnons in a magnetically ordered target. Here, only couplings to electrons (whose spins are ordered) are relevant, and we obtain, for the differential rates, 
\begin{eqnarray}
\Sigma_\nu(\vect{q})_\text{phonon} &=& \frac{4g_\chi^2g_e^2}{m_V^4} \,\bigl| \vect{F}_{S,\nu}^{(e)}\bigr|^2\,,\label{eq:Sigma_SD_phonon}\\
\Sigma_\nu(\vect{q})_\text{magnon} &=& \frac{4g_\chi^2g_e^2}{m_V^4} \,\bigl| \vect{E}_{S,\nu}\bigr|^2\,.
\end{eqnarray}
%

\begin{figure}
	\includegraphics[width=0.75\textwidth]{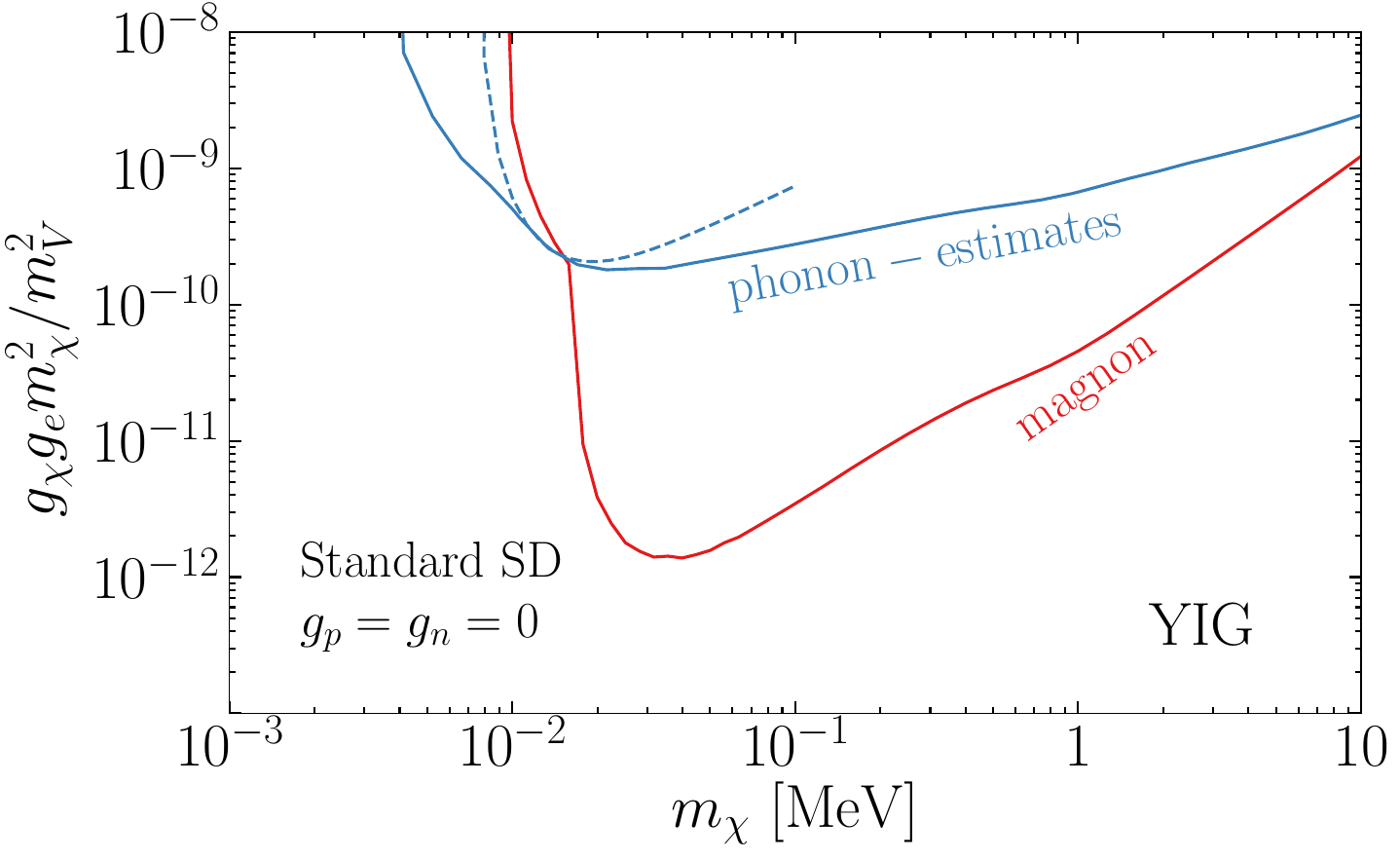}
	\caption{\label{fig:SD}
		Projected reach on the standard SD model listed in Table~\ref{tab:benchmarks} from single magnon (red) and phonon (blue) excitations in YIG. The phonon rate is estimated in two ways, as discussed in the text, which lead to the solid and dashed curves, respectively. Since this model generates only the $S$ response, magnons are seen to have better sensitivity than phonons.
	}
\end{figure}

In Fig.~\ref{fig:SD}, we compare the phonon and magnon reach with YIG. As a technical note, in the absence of a DFT calculation for the crystal potential in YIG which is necessary for computing the phonon eigenmodes, we estimate the rate in two ways. First, we carry out an approximate analytic calculation taking into account long-wavelength acoustic phonons, as explained in Appendix~\ref{app:YIG_phonon}. This results in the dashed reach curve in Fig.~\ref{fig:SD}, which is truncated at the DM mass for which the maximum momentum transfer reaches the edge of the 1BZ, so that the approximations we make cease to hold. Second, we borrow the crystal potential of Y$_3$Ga$_5$O$_{12}$ (YGG) which is publicly available~\cite{phonondb}. YGG has the same crystal structure as YIG, with Fe replaced by Ga, and the phonon dispersions we obtain for YGG are very similar to those of YIG~\cite{Wang_2020}. The resulting reach is shown by the solid blue curve in Fig.~\ref{fig:SD}. We see from the figure that both estimates are in good agreement near $m_\chi\sim 10^{-2}$\,MeV, where acoustic phonons dominate, while including optical phonon contributions in the second approach improves the reach at lower and higher $m_\chi$.

We can understand these curves by estimating the rates using Eqs.~\eqref{eq:R_est} and \eqref{eq:FX_EX_est}. The $q$ integrals are dominated by $q_\text{max}\sim m_\chi v$. As a result,
\begin{eqnarray}
R_\text{phonon} &\sim& \frac{\rho_\chi}{m_\chi} \frac{1}{m_\text{cell}} \frac{g_\chi^2 g_e^2}{m_V^4} \frac{S_\text{ion}^2}{m_\text{ion}} \int dq\, \frac{q^3}{\omega} \nonumber\\
&\sim&
\begin{cases}
\frac{g_\chi^2 g_e^2}{m_V^4}\frac{\rho_\chi}{m_\chi} \frac{S_\text{ion}^2}{m_\text{cell}m_\text{ion}c_s}(m_\chi v)^3 & (\text{acoustic, $m_\chi v\lesssim a^{-1}$}) \,,\\
\frac{g_\chi^2 g_e^2}{m_V^4}\frac{\rho_\chi}{m_\chi} \frac{S_\text{ion}^2}{m_\text{cell}m_\text{ion}\langle\omega\rangle} (m_\chi v)^4 & (\text{otherwise}) \,,
\end{cases} \label{eq:Rphonon_SD}\\
R_\text{magnon} &\sim& \frac{\rho_\chi}{m_\chi} \frac{1}{m_\text{cell}} \frac{g_\chi^2 g_e^2}{m_V^4} \,S_\text{ion}\int dq\,q
\sim
\frac{g_\chi^2 g_e^2}{m_V^4}\frac{\rho_\chi}{m_\chi} \frac{S_\text{ion}}{m_\text{cell}}(m_\chi v)^2 \,.\label{eq:Rmagnon_SD}
\end{eqnarray}
Fixing $R$, the coupling plotted in Fig.~\ref{fig:SD}, $g_\chi g_e\frac{m_\chi^2}{m_V^2}$ scales as $m_\chi$, $m_\chi^{1/2}$ and $m_\chi^{3/2}$, respectively, in the three cases, in agreement with the high-$m_\chi$ behaviors of the dashed blue, solid blue and red curves in Fig.~\ref{fig:SD}, respectively. Also, magnons have better sensitivity than phonons to the SD coupling by a factor of $\sqrt{\frac{R_\text{magnon}}{R_\text{phonon}}} \sim \frac{\sqrt{m_\text{ion}\omega/S_\text{ion}}}{m_\chi v}$, and the advantage becomes more significant at smaller $m_\chi$ (though the magnon curve hits the kinematic threshold at higher $m_\chi$ due to the dispersion being quadratic).

\subsection{Scalar Mediator Models}

We next consider scalar mediator models with both scalar and pseudoscalar couplings. We take the mediator couplings to SM fermions to be proportional to their masses, $g_\psi\propto m_\psi$ (motivated by Higgs-portal hidden sector theories, see Ref.~\cite{Arcadi:2019lka} for a recent review), and consider each of the four combinations of currents, which we denote by $S \times S$, $P \times S$, $S \times P$ and $P \times P$. Among them, $S \times S$ ({\it i.e.}\ standard SI considered previously in Refs.~\cite{Knapen:2017ekk,Griffin:2018bjn,Trickle:2019nya,Griffin:2019mvc}) and $P \times S$ can excite phonons via the $\hl{N}$ response,\footnote{These models generate additional operators when matched onto the NR EFT beyond leading order, which could excite magnons. We do not consider magnon excitation here due to the severely suppressed rate. The same applies to the SI and electric dipole DM models in Sec.~\ref{subsec:multipole}.} while $S \times P$ and $P \times P$ can excite both phonons and magnons in a magnetically ordered target via the $\hl{S}$ response. However, similar to the standard SD interaction in Sec.~\ref{subsec:SD}, the phonon excitation rate will be suppressed relative to the magnon excitation rate, so we focus on the latter here. We obtain the following expressions for the differential rates defined in Eq.~\eqref{eq:Gamma}:
\begin{eqnarray}
\Sigma_\nu(\vect{q})_\text{phonon}^{S\times S} &=& \frac{g_\chi^2}{(q^2+m_\phi^2)^2} \,\Bigl|\sum_\psi g_\psi^\text{eff} F_{N,\nu}^{(\psi)}\Bigr|^2 \,,\\
\Sigma_\nu(\vect{q})_\text{phonon}^{P\times S} &=& \frac{g_\chi^2}{(q^2+m_\phi^2)^2} \frac{q^2}{4m_\chi^2} \,\Bigl| \sum_\psi g_\psi^\text{eff} F_{N,\nu}^{(\psi)}\Bigr|^2 \,,\\
\Sigma_\nu(\vect{q})_\text{magnon}^{S\times P} &=& \frac{g_\chi^2g_e^2}{(q^2+m_\phi^2)^2} \frac{q^2}{m_e^2} \,\Bigl| \vect{\hat q}\cdot \vect{E}_{S,\nu} \Bigr|^2 \,,\\
\Sigma_\nu(\vect{q})_\text{magnon}^{P\times P} &=& \frac{g_\chi^2g_e^2}{(q^2+m_\phi^2)^2} \frac{q^4}{4m_\chi^2m_e^2}\,\Bigl| \vect{\hat q}\cdot \vect{E}_{S,\nu} \Bigr|^2\,.
\end{eqnarray}
Note that for the $S\times S$ and $P\times S$ models, screening effects have been taken into account by using $g_\psi^\text{eff}$ in place of $g_\psi$, as discussed around Eq.~\eqref{eq:in-medium}; the dielectric tensors $\epstens_\infty$ of the phonon targets GaAs and SiO$_2$ are obtained from DFT calculations~\cite{Griffin:2020lgd}.

\begin{figure}
	\includegraphics[width=\textwidth]{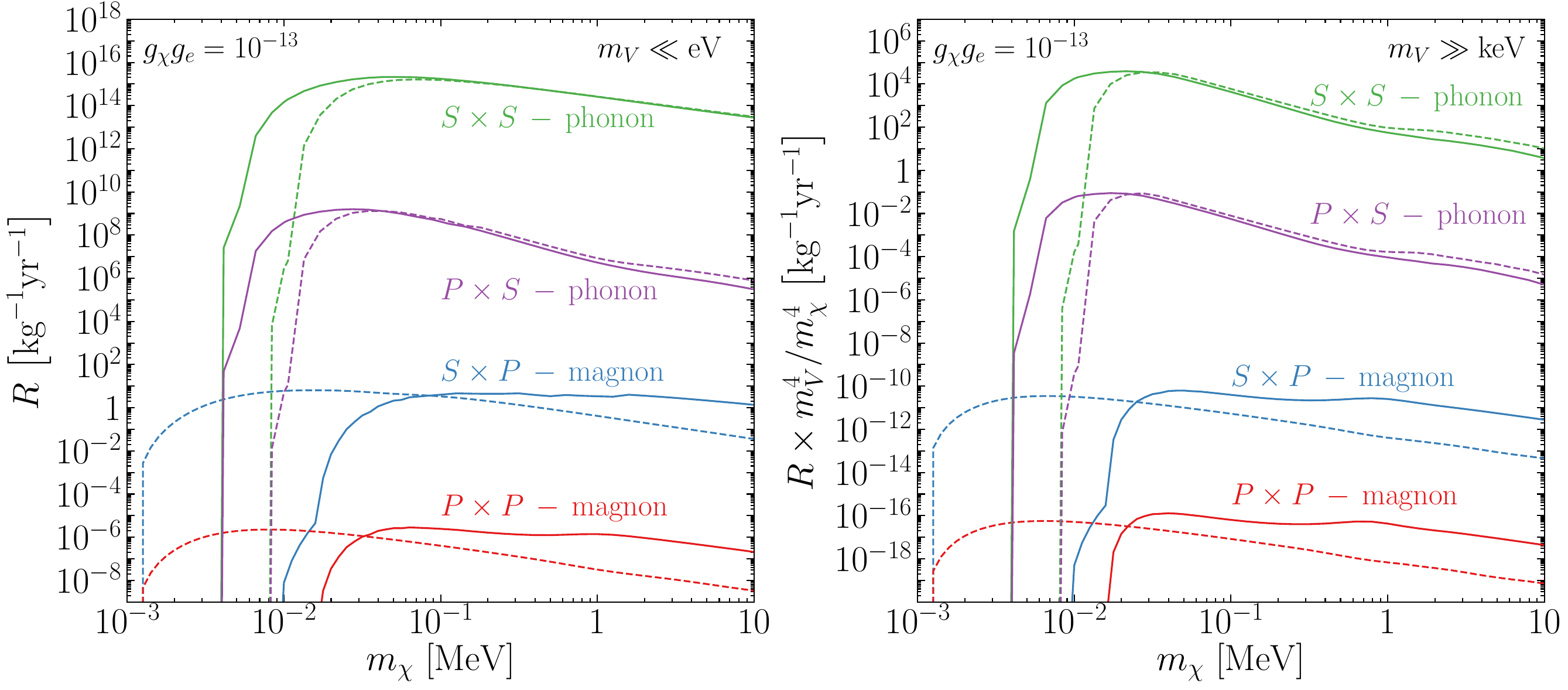}
	\caption{\label{fig:scalar}
		Comparison of the total detection rate in models with a light (left panel) or heavy (right panel) scalar mediator. The couplings to SM fermions are taken proportional to their masses, $g_p = g_n = \frac{m_p}{m_e} g_e$, and we fix $g_\chi g_e = 10^{-13}$. Each curve is labeled with the model type as in Table \ref{tab:benchmarks} and the excitation type (phonon or magnon) that can probe each model. The phonon curves assume \ce{SiO2} (solid) and \ce{GaAs} (dashed) targets, and the magnon curves assume YIG (solid) and $\alpha$-RuCl$_3$ (dashed) targets.
	}
\end{figure}

In Fig.~\ref{fig:scalar}, we plot the expected rate for each of the four coupling combinations, for a common value for the product of couplings, to illustrate the hierarchy between the rate from the different interactions. We have chosen to show the rate instead of projected reach here so that the general case where more than one types of interactions are present, it would be straightforward to rescale the curves to see which one is dominant. For example, if $g_\chi^{(S)}\sim g_\chi^{(P)}$, $g_\psi^{(S)}\sim g_\psi^{(P)}$, we have the highest rate from phonon excitations via the $S \times S$ coupling, {\em i.e.}\ the standard SI interaction, as expected. On the other hand, if the couplings to SM fermions are dominantly pseudoscalar, $g_\psi^{(P)}/g_\psi^{(S)}\gtrsim 10^{7}$, magnon excitations have better sensitivity than phonon excitations for the same exposure; this is one of the benchmark models considered previously in Ref.~\cite{Trickle:2019ovy}. 
The hierarchy seen in Fig.~\ref{fig:scalar}, and also some main features of the curves, can be understood following Eqs.~\eqref{eq:R_est} and \eqref{eq:FX_EX_est}, as we now explain.

First consider the light mediator case, $m_V\ll q$ (left panel of Fig.~\ref{fig:scalar}). For phonon excitations in the $S \times S$ and $P \times S$ models, since the couplings to all ions have the same sign, the rate is dominated by acoustic phonons. For $q$ within the 1BZ, setting $\omega\sim c_s q$, we obtain
\begin{eqnarray}
R_\text{phonon}^{S\times S} &\sim& \frac{\rho_\chi}{m_\chi} \frac{1}{m_\text{cell}} \,g_\chi^2 g_p^2\, \frac{\langle N_{p,n}\rangle^2}{m_\text{ion}}\int dq\, \frac{1}{q\omega}
\sim g_\chi^2 g_p^2\,\frac{\rho_\chi}{m_\chi} \frac{\langle N_{p,n}\rangle^2}{m_\text{cell}m_\text{ion}} \frac{1}{\omega_\text{min}} \,,\\
R_\text{phonon}^{P\times S} &\sim& \frac{\rho_\chi}{m_\chi} \frac{1}{m_\text{cell}} \,g_\chi^2 g_p^2\, \frac{\langle N_{p,n}\rangle^2}{m_\text{ion}m_\chi^2}\int dq\, \frac{q}{\omega}
\sim g_\chi^2 g_p^2\,\frac{\rho_\chi}{m_\chi} \frac{\langle N_{p,n}\rangle^2}{m_\text{cell}m_\text{ion}}\frac{v}{m_\chi c_s} \,,\label{eq:Rphonon_PxS}
\end{eqnarray}
where $\omega_\text{min}=c_s q_\text{min}$. These are consistent with the $m_\chi^{-1}$ and $m_\chi^{-2}$ scaling of the green and purple curves for $m_\chi$ up to $\sim$ MeV. Also, consistent with the figure, the ratio between them is $\frac{R_\text{phonon}^{P\times S}}{R_\text{phonon}^{S\times S}}\sim \frac{\omega_\text{min}}{m_\chi}\frac{v}{c_s} \sim 10^{-6} \,\frac{\omega_\text{min}}{1\,\text{meV}} \frac{10^{-1}\,\text{MeV}}{m_\chi} \frac{v}{10^{-3}}\frac{10^{-5}}{c_s}$ for couplings of the same size. For heavier DM, on the other hand, momentum transfers beyond the 1BZ are allowed. For the $S\times S$ model, this is irrelevant since the integral is dominated by small $q$, so the $m_\chi^{-1}$ trend continues past MeV. For the $P\times S$ model, since the integral is dominated by high $q$ where $\omega$ no longer scales with $q$, we have $\frac{v^2}{\langle\omega\rangle}$ in place of $\frac{v}{m_\chi c_s}$ in Eq.~\eqref{eq:Rphonon_PxS}. This explains the $m_\chi^{-1}$ scaling of the purple curves past $m_\chi\sim$ MeV in the left panel of Fig.~\ref{fig:scalar}.

For magnon excitations in the $S \times P$ and $P \times P$ models, we have
\begin{eqnarray}
R_\text{magnon}^{S\times P} &\sim& \frac{\rho_\chi}{m_\chi} \frac{1}{m_\text{cell}} g_\chi^2 g_e^2 \frac{S_\text{ion}}{m_e^2}\int dq\,\frac{1}{q}
\sim g_\chi^2 g_e^2\,\frac{\rho_\chi}{m_\chi} \frac{S_\text{ion}}{m_\text{cell}m_e^2} \,,\\
R_\text{magnon}^{P\times P} &\sim& \frac{\rho_\chi}{m_\chi} \frac{1}{m_\text{cell}} g_\chi^2 g_e^2 \frac{S_\text{ion}}{m_e^2 m_\chi^2}\int dq\,q
\sim g_\chi^2 g_e^2\,\frac{\rho_\chi}{m_\chi} \frac{S_\text{ion}}{m_\text{cell}m_e^2} \,v^2 \,,
\end{eqnarray}
again consistent with the $m_\chi^{-1}$ scaling of the corresponding curves in Fig.~\ref{fig:scalar} (though the YIG curves have a bump near MeV due to the gapped magnons starting to contribute, as discussed in Ref.~\cite{Trickle:2019ovy}, which slightly obscures the overall scaling with $m_\chi$). Comparing the two models, we see that $\frac{R_\text{magnon}^{P\times P}}{R_\text{magnon}^{S\times P}}\sim v^2$. Also, comparing with phonon excitation in the $S\times S$ model, we have $\frac{R_\text{magnon}^{S\times P}}{R_\text{phonon}^{S\times S}}\sim \frac{g_e^2}{g_p^2}\frac{S_\text{ion}m_\text{ion}\omega_\text{min}}{\langle N_{p,n}\rangle^2 m_e^2} \sim \frac{\omega_\text{min}}{m_\text{ion}} \sim 10^{-14} \,\frac{\omega_\text{min}}{1\,\text{meV}}\frac{100\,\text{GeV}}{m_\text{ion}}$, assuming similar values of $m_\text{cell}$, $m_\text{ion}$ for the targets and $S_\text{ion}\sim\OO(1)$, and noting that $g_p^\text{eff}\simeq g_p$ and $g_e/g_p = m_e/m_p$. This is consistent with what we see in Fig.~\ref{fig:scalar}.

The heavy mediator case, $m_V\gg q$ (right panel of Fig.~\ref{fig:scalar}), follows a similar analysis. All the $q$ integrals are now peaked at $q_\text{max}\sim m_\chi v$, and we find
\begin{eqnarray}
R_\text{phonon}^{S\times S} &\sim& \frac{\rho_\chi}{m_\chi} \frac{1}{m_\text{cell}} \frac{g_\chi^2 g_p^2}{m_V^4} \frac{\langle N_{p,n}\rangle^2}{m_\text{ion}}\int dq\, \frac{q^3}{\omega} \sim
\begin{cases}
\frac{g_\chi^2 g_p^2}{m_V^4} \frac{\rho_\chi}{m_\chi}\frac{\langle N_{p,n}\rangle^2}{m_\text{cell}m_\text{ion} c_s} (m_\chi v)^3 & (m_\chi v \lesssim a^{-1}) \,,\\
\frac{g_\chi^2 g_p^2}{m_V^4} \frac{\rho_\chi}{m_\chi}\frac{\langle N_{p,n}\rangle^2}{m_\text{cell}m_\text{ion} \langle\omega\rangle} (m_\chi v)^4 & (m_\chi v \gtrsim a^{-1})\,,
\end{cases} \\
R_\text{phonon}^{P\times S} &\sim& v^2 \,R_\text{phonon}^{S\times S} \,,\\
R_\text{magnon}^{S\times P} &\sim& \frac{\rho_\chi}{m_\chi} \frac{1}{m_\text{cell}} \frac{g_\chi^2 g_p^2}{m_V^4} \frac{S_\text{ion}}{m_e^2} \int dq\, q^3
\sim \frac{g_\chi^2 g_e^2}{m_V^4} \frac{\rho_\chi}{m_\chi}\frac{S_\text{ion}}{m_\text{cell}m_e^2} (m_\chi v)^4 \,,\\
R_\text{magnon}^{P\times P} &\sim& v^2 \,R_\text{magnon}^{S\times P} \,.
\end{eqnarray}
These equations explain both the hierarchy of the rates for the four models, and the $m_\chi$ scaling: in all cases, $R\,\frac{m_V^4}{m_\chi^4}\sim m_\chi^{-1}$ at large $m_\chi$, while the phonon curves switch to $m_\chi^{-2}$ scaling below $\sim$ MeV.

\subsection{Multipole Dark Matter Models}
\label{subsec:multipole}

\begin{figure}
    \includegraphics[width=\textwidth]{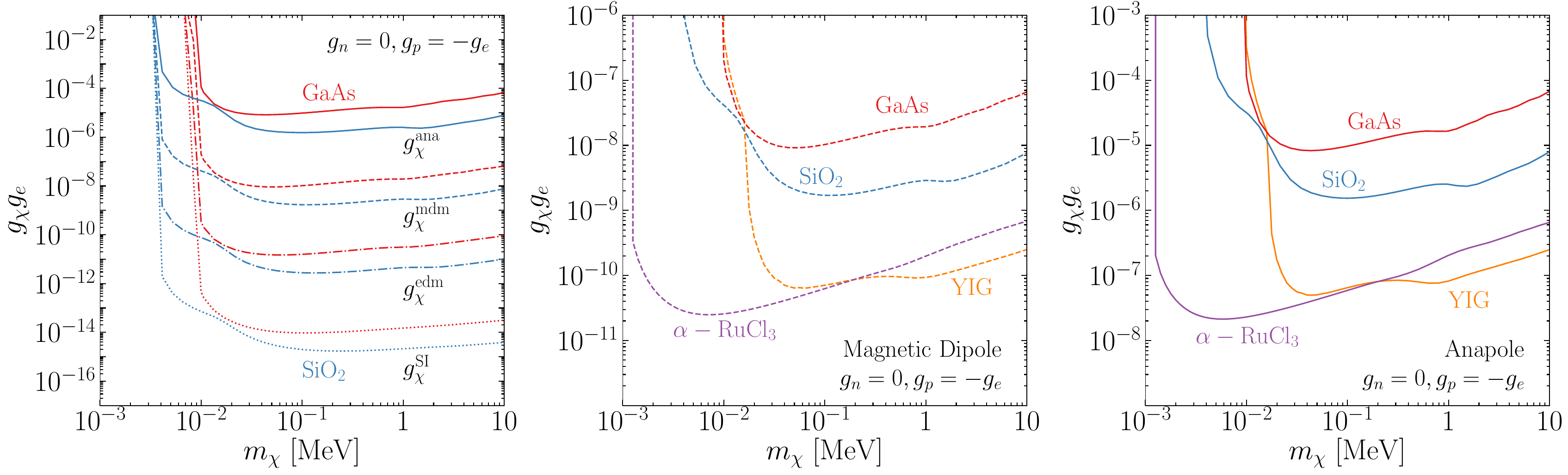}
    \caption{\label{fig:multipole}
    	Projected reach on the multipole DM models listed in Table~\ref{tab:benchmarks}, assuming dark photon-like couplings to SM particles: $g_p = -g_e, g_n = 0$. The left panel shows the hierarchy of sensitivities of single phonon excitations, in GaAs and in SiO$_2$, to the three multipole DM models, together with the SI interaction model for comparison. The center and right panels focus on the magnetic dipole and anapole DM models, respectively, and compare the phonon reach of GaAs and SiO$_2$ (via the $N$ response), and the magnon reach of YIG (via the $S$ response) and $\alpha$-RuCl$_3$ (via both $S$ and $L$ responses); these models are best probed by magnons, though the phonon sensitivity with an optimal target like SiO$_2$ may be competitive.
    }
\end{figure}

We now turn to the electric dipole, magnetic dipole, and anapole DM models in Table~\ref{tab:benchmarks}. For comparison, we also include the SI interaction model with a vector mediator. Motivated by the kinetic mixing benchmark, we assume the mediator couples to electric charge, $g_p=-g_e$, $g_n=0$, and is much lighter than the smallest momentum transfer, $m_V \ll$ eV. The SI and electric dipole DM models generate $\OO_1$ and $\OO_{11}$ at leading order, respectively, both of which induce only the $\hl{N}$ response, which can be probed by single phonon excitation. The differential rates are
\begin{eqnarray}
\Sigma_\nu(\vect{q})_\text{phonon}^\text{SI} &=& \frac{g_\chi^2 g_e^2}{(\vect{q}\cdot\epstens_\infty\cdot\vect{q})^2} \,\Bigl| F_{N,\nu}^{(p)} - F_{N,\nu}^{(e)}\Bigr|^2 \,,\label{eq:SigmaSI}\\
\Sigma_\nu(\vect{q})_\text{phonon}^\edm &=& \frac{g_\chi^2 g_e^2}{(\vect{q}\cdot\epstens_\infty\cdot\vect{q})^2} \frac{q^2}{4m_\chi^2} \,\Bigl| F_{N,\nu}^{(p)} - F_{N,\nu}^{(e)}\Bigr|^2 \,.
\end{eqnarray}
Eq.~\eqref{eq:SigmaSI} is in agreement with previous results in Refs.~\cite{Griffin:2018bjn,Trickle:2019nya,Griffin:2019mvc}. The magnetic dipole and the anapole DM models generate, in addition to $\hl{N}$, also $\hl{S}$ and $\hl{L}$ responses, and can therefore be probed by both phonons and magnons. For single phonon excitation, we have 
\begin{eqnarray}
\Sigma_\nu(\vect{q})_\text{phonon}^\mdm &=& \frac{g_\chi^2 g_e^2}{4m_\chi^2q^2} \biggl\{ \frac{q^4}{(\vect{q}\cdot\epstens_\infty\cdot\vect{q})^2}\frac{q^2}{4m_\chi^2}\Bigl| F_{N,\nu}^{(p)} - F_{N,\nu}^{(e)}\Bigr|^2 
\nonumber\\
&&\qquad
+ \biggl| \frac{q^4}{(\vect{q}\cdot\epstens_\infty\cdot\vect{q})^2} (\vect{\hat q}\times \vect{v})\Bigl(F_{N,\nu}^{(p)} - F_{N,\nu}^{(e)}\Bigr) 
\nonumber\\
&&\qquad\quad
-(\mathbb{1}-\vect{\hat q}\vect{\hat q})\cdot
\biggl[\frac{iq}{2m_p}\Bigl(2\tilde\mu_p \vect{F}_{S,\nu}^{(p)} +\vect{F}_{L,\nu}^{(p)} \Bigr) 
-\frac{iq}{2m_e}\Bigl(2\tilde\mu_e \vect{F}_{S,\nu}^{(e)} +\vect{F}_{L,\nu}^{(e)} \Bigr)\biggr] 
\biggr|^2\biggr\}\,,\quad\\
\Sigma_\nu(\vect{q})_\text{phonon}^\ana &=& \frac{g_\chi^2 g_e^2}{16m_\chi^4}  \,\biggl| \frac{q^4}{(\vect{q}\cdot\epstens_\infty\cdot\vect{q})^2}\biggl(\vect{v}-\frac{\vect{q}}{2m_\chi}\biggr) \Bigl(F_{N,\nu}^{(p)} - F_{N,\nu}^{(e)}\Bigr)
\nonumber\\
&&\qquad\qquad\qquad
+\frac{i\vect{q}}{2m_p}\times\Bigl(2\tilde\mu_p \vect{F}_{S,\nu}^{(p)} +\vect{F}_{L,\nu}^{(p)} \Bigr) 
-\frac{i\vect{q}}{2m_e}\times\Bigl(2\tilde\mu_e \vect{F}_{S,\nu}^{(e)} +\vect{F}_{L,\nu}^{(e)} \Bigr)
\biggr|^2 \,.
\end{eqnarray}
Note that for an unordered/unpolarized target, $\vect{F}_{S,\nu}^{(\psi)}=\vect{F}_{L,\nu}^{(\psi)}=\vect{0}$. For single magnon excitation, we have 
\begin{eqnarray}
\Sigma_\nu(\vect{q})_\text{magnon}^\mdm &=& \frac{g_\chi^2 g_e^2}{16m_\chi^2 m_e^2} \,\Bigl| (\mathbb{1}-\vect{\hat q}\vect{\hat q}) \cdot (2\tilde\mu_e\vect{E}_{S,\nu}+\vect{E}_{L,\nu}) \Bigr|^2 \,,\\
\Sigma_\nu(\vect{q})_\text{magnon}^\ana &=& \frac{g_\chi^2 g_e^2}{64m_\chi^4m_e^2} \,\Bigl|\vect{q}\times (2\tilde\mu_e\vect{E}_{S,\nu}+\vect{E}_{L,\nu}) \Bigr|^2 \,,
\end{eqnarray}
which extend the results in Ref.~\cite{Trickle:2019ovy}. 

A comparison of the phonon reach in these models is shown in the left panel of Fig.~\ref{fig:multipole}. The center and right panels of Fig.~\ref{fig:multipole} zoom in on the magnetic dipole and anapole DM models, respectively, and compare the reach of phonon and magnon excitations.

We can carry out a similar analysis as in the previous subsections to understand the main features in Fig.~\ref{fig:multipole}. For single phonon excitation in GaAs and SiO$_2$, we keep only the $F_{N,\nu}^{(\psi)}$ terms in the $\Sigma_\nu(\vect{q})$ formulae above, and note that, as in the SI case discussed previously in Refs.~\cite{Knapen:2017ekk,Griffin:2018bjn,Trickle:2019nya, Griffin:2019mvc}, the DM-ion couplings, being proportional to $\langle N_p\rangle-\langle N_e\rangle=Q_\text{ion}$, have opposite signs for oppositely charged ions, so the optical phonon modes with $\omega\sim q^0$ give the dominant contributions. Using Eqs.~\eqref{eq:R_est} and \eqref{eq:FX_EX_est}, we obtain the following parametric estimates:
\begin{eqnarray}
R_\text{phonon}^\text{SI} &\sim& \frac{\rho_\chi}{m_\chi} \frac{1}{m_\text{cell}} \frac{g_\chi^2 g_e^2}{\varepsilon_\infty^2} \frac{Q_\text{ion}^2}{m_\text{ion}\omega} \int dq \,\frac{1}{q}
\sim g_\chi^2 g_e^2\, \frac{\rho_\chi}{m_\chi} \biggl(\frac{Q_\text{ion}^2}{\varepsilon_\infty^2 m_\text{cell}m_\text{ion}\omega}\biggr) \,,\label{eq:Rphonon_mcp}\\
\frac{R_\text{phonon}^\edm}{R_\text{phonon}^\text{SI}} &\sim& \frac{R_\text{phonon}^\mdm}{R_\text{phonon}^\edm}
\sim \frac{R_\text{phonon}^\ana}{R_\text{phonon}^\mdm} \sim v^2 \,,\label{eq:Rphonon_multipole}\\
R_\text{magnon}^\mdm &\sim& \frac{\rho_\chi}{m_\chi} \frac{S_\text{ion}}{m_\text{cell}}\frac{g_\chi^2 g_e^2}{m_\chi^2 m_e^2} \int dq\, q
\sim g_\chi^2 g_e^2\,\frac{\rho_\chi}{m_\chi}\frac{S_\text{ion}v^2}{m_\text{cell} m_e^2} \,,\label{eq:Rmagnon_mdm}\\
\frac{R_\text{magnon}^\ana}{R_\text{magnon}^\mdm} &\sim& v^2\,.\label{eq:Rmagnon_ana}
\end{eqnarray}

Several comments are in order. First, Eq.~\eqref{eq:Rphonon_multipole} explains the hierarchy of sensitivity of phonon excitations to the four models in the left panel of Fig.~\ref{fig:multipole}, while Eq.~\eqref{eq:Rmagnon_ana} shows a similar hierarchy of sensitivity of magnon excitations to the magnetic dipole and anapole DM models. Also, note that in all cases, $R\sim m_\chi^{-1}$, so the reach on $g_\chi g_e$ scales as $m_\chi^{1/2}$, as seen in Fig.~\ref{fig:multipole}.

Next, let us compare the reach of different target materials, and via phonons versus magnons. For phonon excitations, the factor in parentheses in Eq.~\eqref{eq:Rphonon_mcp} reproduces the ``quality factor'' identified in Ref.~\cite{Griffin:2019mvc}, up to $\OO(1)$ factors we have dropped here. It captures the material properties that determine the sensitivity to the SI model with a dark photon mediator, and is the quantity to maximize in order to optimize target choice. For example, SiO$_2$ has a quality factor that is about 80 times that of GaAs, which explains its significantly better reach, by almost an order of magnitude on the coupling $g_\chi g_e$, as seen in Fig.~\ref{fig:multipole} (and also previously in Ref.~\cite{Griffin:2019mvc}).

For magnon excitations for the magnetic dipole and anapole DM models, we have considered YIG, which probes only the $\hl{S}$ response, and $\alpha$-RuCl$_3$, which probes both $\hl{S}$ and $\hl{L}$. Since for these models, DM couples to the linear combination $2\vect{S}_e+\vect{L}_e$  -- the spin of an elementary particle has a Land\'e $g$-factor of 2 -- the additional $\hl{L}$ response that $\alpha$-RuCl$_3$ has does not qualitatively improve the sensitivity. Indeed, we see from Fig.~\ref{fig:multipole} that YIG and $\alpha$-RuCl$_3$ have very similar reach around $m_\chi \sim 0.1\,$MeV.  At higher $m_\chi$, YIG performs better due to additional contributions from the large number of gapped magnon modes. On the other hand, $\alpha$-RuCl$_3$ extends the reach down to much lower $m_\chi\sim$ keV. As discussed previously, this is because the magnon modes at zero momentum are gapped at a few meV (in contrast to YIG which has a gapless magnon branch that dominates the coupling to DM in the low momentum transfer limit).

Finally, we can compare the magnon and phonon excitation rates for the two models (magnetic dipole and anapole DM) where both are available. Let us denote ${\cal Q} \equiv \frac{Q_\text{ion}^2}{\varepsilon_\infty^2} \frac{m_p^2}{m_\text{cell}m_\text{ion}}\frac{1\,\text{meV}}{\omega}$, which is the phonon quality factor with the dimensionful parameters normalized in a way close to Ref.~\cite{Griffin:2019mvc}. Its values are typically $\OO(10^{-7}$-$10^{-5})$, with GaAs and SiO$_2$ residing on the lower and higher ends of the interval, respectively.  We find
\begin{equation}
\frac{R_\text{phonon}^\mdm}{R_\text{magnon}^\mdm} \sim \frac{R_\text{phonon}^\ana}{R_\text{magnon}^\ana} \sim \frac{{\cal Q}\, m_\text{cell} m_e^2v^2}{S_\text{ion}m_p^2 \cdot 1\,\text{meV}} \sim 10^{-4}\,\biggl(\frac{{\cal Q}}{1.4\times 10^{-7}}\biggr)\,,
\end{equation}
where $m_\text{cell}$ is for the target for magnon excitations, and we have substituted the numbers for YIG in the last equation. We see that, for the magnetic dipole and anapole DM models, magnons are indeed more sensitive than phonons, though choosing high phonon quality factor targets, such as SiO$_2$ with ${\cal Q}\sim 10^{-5}$ can approach the magnon sensitivity. Up to $\OO(1)$ factors, this is consistent with the center and right panels of Fig.~\ref{fig:multipole}.

\subsection{$(\vect{L}\cdot\vect{S})$-Interacting Dark Matter}

\begin{figure}
    \includegraphics[width=\textwidth]{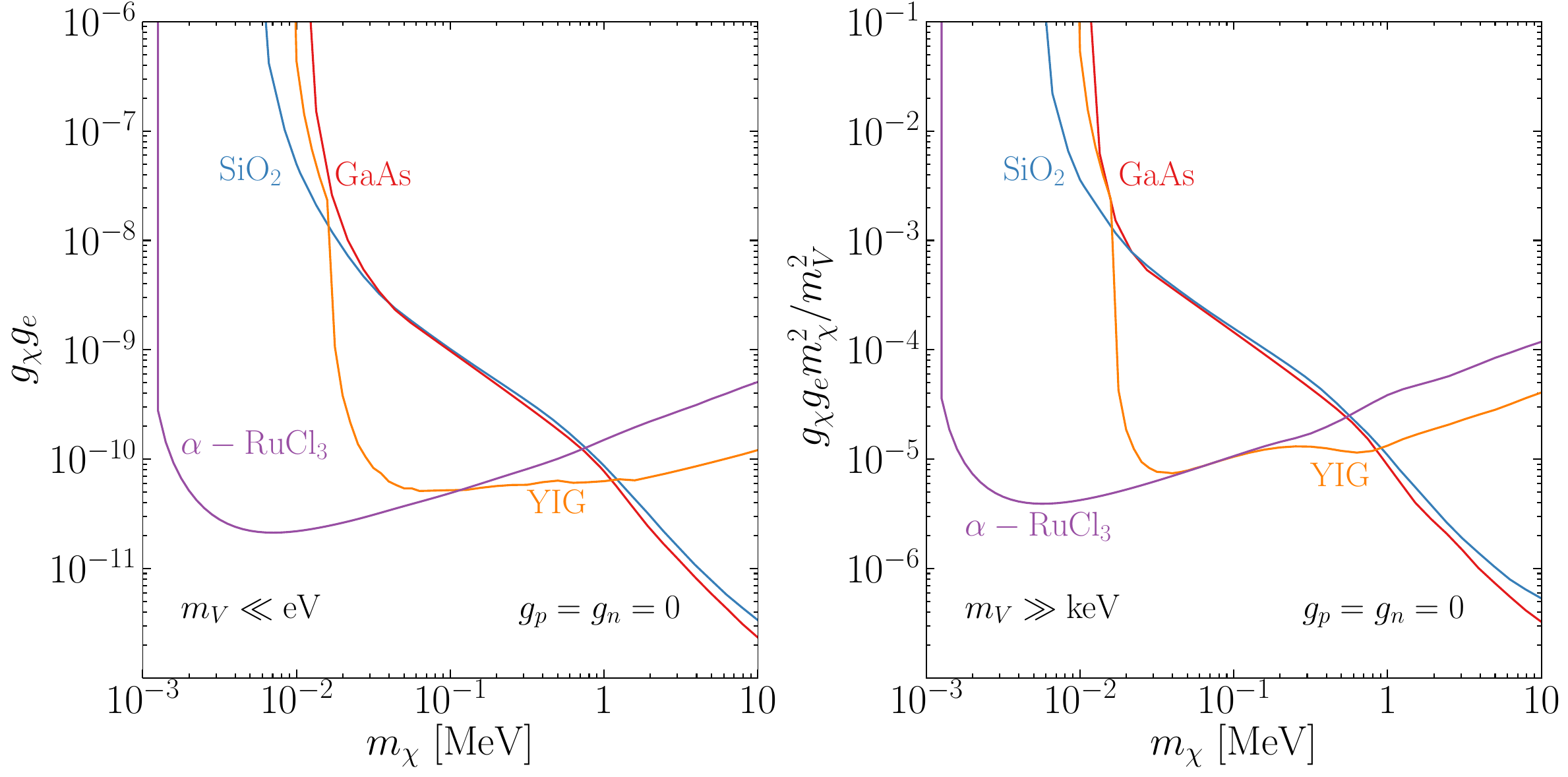}
    \caption{\label{fig:LS}
    	Projected reach on the $(\vect{L}\cdot\vect{S})$-interacting DM model in Table~\ref{tab:benchmarks}, assuming coupling only to electrons, and $\kappa=0$. Single phonon excitations in GaAs and SiO$_2$ targets (via the $N$ response) and single magnon excitations in YIG and $\alpha$-RuCl$_3$ targets (via the $S$ response) are seen to cover complementary regions of parameter space.
    }
\end{figure}

We finally consider the $(\vect{L}\cdot\vect{S})$-interacting DM model, which induces $\hl{N}$, $\hl{S}$ and $\hl{L\otimes S}$ responses. Taking the mediator to couple only to electrons for simplicity, we obtain the differential rates: 
\begin{eqnarray}
\Sigma_\nu(\vect{q})_\text{phonon} &=&  \frac{g_\chi^2g_e^2}{(q^2+m_V^2)^2} \biggl\{ \,\biggl|\frac{q^2}{4m_e^2} \Bigl[(1+\kappa) F_{N,\nu}^{(e)} +2\,\text{tr} \bigl((\mathbb{1}-\vect{\hat q}\vect{\hat q}) \cdot \tens{F}_{L\otimes S,\nu}^{(e)}\bigr)\Bigr]
+\biggl(\frac{i\vect{q}}{m_e}\times\vect{v}\biggr)\cdot\vect{F}_{S,\nu}^{(e)}\biggr|^2 \nonumber\\
&&\qquad\qquad\qquad\;
+\frac{q^4}{4m_\chi^2 m_e^2}\Bigl|(\mathbb{1}-\vect{\hat q}\vect{\hat q}) \cdot\vect{F}_{S,\nu}^{(e)}\Bigr|^2 \biggr\}\,,\label{eq:Sigma_phonon_LS}\\
\Sigma_\nu(\vect{q})_\text{magnon} &=&  \frac{g_\chi^2g_e^2}{(q^2+m_V^2)^2} \frac{q^2}{m_e^2} \biggl[\frac{q^2}{4m_\chi^2} \Bigl|(\mathbb{1}-\vect{\hat q}\vect{\hat q}) \cdot\vect{E}_{S,\nu}\Bigr|^2 + \Bigl| \bigl(\vect{\hat q}\times\vect{v}\bigr)\cdot\vect{E}_{S,\nu}\Bigr|^2 \biggr]\,.
\end{eqnarray}
In the absence of magnetic order, $\vect{F}_{S,\nu}^{(e)}=0$. Also, unless $\kappa$ is tuned to be very close to $-1$, we do not expect the $\tens{F}_{L\otimes S,\nu}^{(e)}$ term in Eq.~\eqref{eq:Sigma_phonon_LS} to dominate --- the total spin-orbit coupling vanishes for full shells, and is otherwise often suppressed by crystal fields, especially for lighter elements. Thus, while an interesting feature of this model, the coupling to $\vect{L}\cdot\vect{S}$ does not suggest a better probe than searching for phonon excitations via the $\hl{N}$ response with already proposed target materials. In Fig.~\ref{fig:LS}, we include only the $F_{N,\nu}^{(e)}$ term when computing phonon reach for GaAs and SiO$_2$, and for concreteness set $\kappa=0$. Since the total electron numbers of ions are all positive, the rate is dominated by acoustic phonons, with $\omega\sim c_s q$. Again using Eqs.~\eqref{eq:R_est} and \eqref{eq:FX_EX_est}, we can estimate
\begin{equation}
R_\text{phonon} \sim \frac{\rho_\chi}{m_\chi}\frac{1}{m_\text{cell}} \,g_\chi^2 g_e^2\,\frac{\langle N_e\rangle^2}{m_\text{ion} m_e^4} \int dq \,\frac{q^3}{\omega} \sim
\begin{cases}
g_\chi^2 g_e^2\,\frac{\rho_\chi}{m_\chi}\frac{\langle N_e\rangle^2 (m_\chi v)^3}{m_\text{cell}m_\text{ion} m_e^4 c_s} & (m_\chi v \lesssim a^{-1}) \,, \\
g_\chi^2 g_e^2\,\frac{\rho_\chi}{m_\chi}\frac{\langle N_e\rangle^2 (m_\chi v)^4}{m_\text{cell}m_\text{ion} m_e^4\langle\omega\rangle} & (m_\chi v \gtrsim a^{-1}) \,,
\end{cases}
\end{equation}
for a light mediator ($m_V\ll q$), and
\begin{equation}
R_\text{phonon} \sim \frac{\rho_\chi}{m_\chi}\frac{1}{m_\text{cell}} \frac{g_\chi^2 g_e^2}{m_V^4} \frac{\langle N_e\rangle^2}{m_\text{ion} m_e^4} \int dq \,\frac{q^7}{\omega} \sim
\begin{cases}
\frac{g_\chi^2 g_e^2}{m_V^4} \frac{\rho_\chi}{m_\chi}\frac{\langle N_e\rangle^2 (m_\chi v)^7}{m_\text{cell}m_\text{ion} m_e^4 c_s} & (m_\chi v \lesssim a^{-1}) \,, \\
\frac{g_\chi^2 g_e^2}{m_V^4} \frac{\rho_\chi}{m_\chi}\frac{\langle N_e\rangle^2 (m_\chi v)^8}{m_\text{cell}m_\text{ion} m_e^4\langle\omega\rangle} & (m_\chi v \gtrsim a^{-1}) \,,
\end{cases}
\end{equation}
for a heavy mediator ($m_V\gg q$). These equations explain the scaling of the phonon curves in Fig.~\ref{fig:LS}: fixing $R$, we obtain $g_\chi g_e \sim m_\chi^{-1}$ ($m_\chi^{-3/2}$) for $m_\chi$ below (above) about an MeV in the light mediator case, and the same for $g_\chi g_e \,\frac{m_\chi^2}{m_V^2}$ in the heavy mediator case.

The magnon reach curves for YIG and $\alpha$-RuCl$_3$ can be understood in a similar way. We have
\begin{equation}
R_\text{magnon} \sim \frac{\rho_\chi}{m_\chi}\frac{1}{m_\text{cell}} \,g_\chi^2 g_e^2\, \frac{S_\text{ion}}{m_e^2 m_\chi^2}\int dq \, q \sim g_\chi^2 g_e^2\,\frac{\rho_\chi}{m_\chi}\frac{S_\text{ion}v^2}{m_\text{cell} m_e^2}
\end{equation}
for a light mediator ($m_V\ll q$), and
\begin{equation}
R_\text{magnon} \sim \frac{\rho_\chi}{m_\chi}\frac{1}{m_\text{cell}} \frac{g_\chi^2 g_e^2}{m_V^4} \frac{S_\text{ion}}{m_e^2 m_\chi^2}\int dq \, q^5 \sim \frac{g_\chi^2 g_e^2}{m_V^4} \frac{\rho_\chi}{m_\chi}\frac{S_\text{ion}m_\chi^4v^6}{m_\text{cell} m_e^2}
\end{equation}
for a heavy mediator ($m_V\gg q$). In contrast to the phonon case, the reach on $g_\chi g_e$ ($g_\chi g_e \,\frac{m_\chi^2}{m_V^2}$) in the light (heavy) mediator case scales as $m_\chi^{1/2}$. So the magnon reach becomes better at lower $m_\chi$, as we can see in Fig.~\ref{fig:LS}. In particular, magnons outperform phonons for $m_\chi$ below about an MeV, which we can understand from the estimate: $\frac{R_\text{phonon}}{R_\text{magnon}}\sim \frac{\langle N_e\rangle^2 m_\chi^3 v}{S_\text{ion}m_\text{ion} m_e^2 c_s} \sim \bigl(\frac{m_\chi}{\text{MeV}}\bigr)^3 \bigl(\frac{\langle N_e\rangle}{10}\bigr)^2 \frac{10\,\text{GeV}}{m_\text{ion}} \frac{v}{10^{-3}}\frac{10^{-5}}{c_s}$, assuming similar $m_\text{cell}$ and $m_\text{ion}$ for the targets and $S_\text{ion}\sim\OO(1)$.

\section{Conclusions}

We have formulated an EFT framework for computing direct detection rates via single phonon and magnon excitations for general DM interactions, and illustrated its application with a set of benchmark models, listed in Table~\ref{tab:benchmarks}, that cover a wide range of possibilities for a spin-$\frac{1}{2}$ DM particle interacting with SM fermions $\psi = p,n,e$ (proton, neutron and electron). The procedure consists of first matching a relativistic DM model onto a set of NR effective operators, listed in Table~\ref{tab:operators}, and then matching these operators onto lattice degrees of freedom, including particle numbers $\hl{\langle N_\psi\rangle}$, spins $\hl{\langle \vect{S}_\psi\rangle}$, orbital angular momenta $\hl{\langle \vect{L}_\psi\rangle}$ and spin-orbit couplings $\hl{\bigl\langle \vect{L}_\psi \otimes \vect{S}_\psi \bigr\rangle}$ for the $\psi=p,n,e$ particles in an ion. These define the four types of {\it crystal responses} and enter the rate formula for single phonon excitation, while a subset of them -- $\hl{\langle\vect{S}_e\rangle}$ and $\hl{\langle\vect{L}_e\rangle}$ -- also enter the rate formula for single magnon excitation.

A practical prescription for computing direct detection rates, as explained around Eq.~\eqref{eq:Gamma}, utilizes the central formula, Eq.~\eqref{eq:Vlj}, which gives the lattice scattering potential in terms of the effective operator coefficients $c_i^{(\psi)}$. Upon plugging in the $c_i^{(\psi)}$'s generated by a relativistic theory of DM (listed in Table~\ref{tab:benchmarks} for our benchmark models), one simply replaces the ionic expectation values $\hl{\langle X_\psi\rangle}_{lj}$ by $F_{X,\nu}^{(\psi)}$ defined in Eq.~\eqref{eq:FX_def} (for $\psi=p,n,e$ and $X=\hl{N},\hl{S},\hl{L},\hl{L\otimes S}$) or $\vect{E}_{X,\nu}$ defined in Eq.~\eqref{eq:EX_def} (for $\psi=e$ and $X=\hl{S},\hl{L}$), squares the expression and takes the DM spin average. This gives the differential rates $\Sigma_\nu(\vect{q})$, which are then substituted into Eqs.~\eqref{eq:Gamma} and \eqref{eq:total_rate} for the total rate of single phonon or magnon excitation. 

The set of crystal responses that we have identified point to various possibilities of optimizing detector target choice. However, a general observation from our calculations in Sec.~\ref{sec:benchmarks} is that, among the four types of crystal responses, $\hl{\langle N_\psi \rangle}$ and $\hl{\langle \vect{S}_\psi \rangle}$, which are associated with point-like degrees of freedom, tend to dominate the rate, compared to the composite responses $\hl{\langle \vect{L}_\psi\rangle}$ and $\hl{\bigl\langle \vect{L}_\psi \otimes \vect{S}_\psi \bigr\rangle}$. This implies that, purely from the point of view of maximizing the rate, exotic materials with orbital orders or strong spin-orbit couplings are not necessary for improving the reach to a broad class of DM models.

Meanwhile, as phonon DM experiments focused on crystal targets, such as SPICE (Sub-eV Polar Interactions Cryogenic Experiment), which is part of the TESSERACT (Transition Edge Sensors with Sub-EV Resolution And Cryogenic Targets) project~\cite{tesseract}, move into R\&D, it is important to note that their discovery potential extends well beyond the simplest models with spin-independent interactions studied previously. As we showed in Sec.~\ref{sec:benchmarks}, 
phonon excitations have broad sensitivity to light DM models. Perhaps surprisingly, with judicious choice of target material, phonon excitations may even be competitive with magnon excitations for some DM models where the latter is expected to have a parametrically higher rate, such as the magnetic dipole and anapole DM models. Given the greater challenges associated with single magnon detection relative to phonons, this is encouraging for phonon-based experiments in the near term.

\vspace{10pt}
\paragraph*{Acknowledgments.}
We thank Jason Alicea, Sin\'ead Griffin, Thomas Harrelson, David Hsieh, Katherine Inzani, Chunxiao Liu, Andrea Mitridate and Mengxing Ye for useful discussion. 
Special thanks to Andrea Mitridate for discussions and collaboration on related work that helped clarify the treatment of in-medium effects. 
This work is supported by the Quantum Information Science Enabled Discovery (QuantISED) for High Energy Physics (KA2401032).


\vspace{20pt}

\appendix

\section{Nonrelativistic Matching for a Fermion Field}
\label{app:fermion}

In this appendix, we review the procedure of decomposing a Dirac fermion field $\psi$ in the NR limit. Consider the following unperturbed relativistic Lagrangian:
\begin{equation}
\LL_0 = \bar\psi\, i(\partial_\mu-iA_\mu) \gamma^\mu \psi -m_\psi\bar\psi\psi\,.
\end{equation}
In free space, we would expand the $\psi$ field in plane waves multiplied by the usual $u$, $v$ spinors satisfying the free particle Dirac equation. Here, we allow the presence of an external gauge potential $A^\mu = (\Phi, \vect{A})$, which may not be a small perturbation. For example, if $\psi$ is an electron in a crystal, it is bound by the electromagnetic potential from the ions, and the bound state wavefunctions are very different from plane waves. Generally, we can expand the $\psi$ field in the basis of energy eigenstate wavefunctions. Dropping the antiparticle part, we have
\begin{equation}
\psi(\vect{x}, t) = \sum_I u_I(\vect{x},t)\,\hat b_I 
= \sum_I  e^{-iE_It}\,u_I(\vect{x})\,\hat b_I \,,
\end{equation}
where the $c$-number $u_I$ spinors satisfy 
\begin{equation}
\bigl( E_I \gamma^0 - i\vect{\gamma}\cdot\nabla -m_\psi -\gamma^0\Phi(\vect{x}) +\vect{\gamma}\cdot\vect{A}(\vect{x}) \bigr) \,u_I(\vect{x}) = 0\,.
\label{eq:dirac_u}
\end{equation}
Writing
\begin{equation}
u_I(\vect{x}) = \frac{1}{\sqrt{2}}\left(
\begin{matrix}
\Psi_I(\vect{x})+\Theta_I(\vect{x})\\
\Psi_I(\vect{x})-\Theta_I(\vect{x})
\end{matrix}
\right)\,,
\end{equation}
with $\Psi_I$, $\Theta_I$ two-component wavefunctions, we see that Eq.~\eqref{eq:dirac_u} is solved by
\begin{equation}
\Theta_I(\vect{x}) = \frac{i\vect{\sigma}\cdot\bigl(\nabla -i\vect{A}(\vect{x})\bigr)}{E_I+m_\psi-\Phi(\vect{x})} \,\Psi_I(\vect{x})\,.
\label{eq:eta_sol}
\end{equation}
This immediately leads to Eq.~\eqref{eq:NR_matching_fermion}, repeated here for easy reference:
\begin{equation}
\psi(\vect{x},t) = e^{-im_\psi t} \,\frac{1}{\sqrt{2}}
\left(\begin{matrix}
\Bigl(1-\frac{\vect{\sigma}\cdot\vect{k}}{2m_\psi+\varepsilon}\Bigr)\, \psi^+(\vect{x},t) \\
\Bigl(1+\frac{\vect{\sigma}\cdot\vect{k}}{2m_\psi+\varepsilon}\Bigr)\, \psi^+(\vect{x},t)
\end{matrix}\right),
\end{equation}
where $\vect{k} = -i\nabla-\vect{A}$, $\varepsilon= i\partial_t-\Phi$, and $\psi^+(\vect{x}, t) = \sum_I e^{-i\varepsilon_I t}\, \Psi_I(\vect{x})\,\hat b_I$ with $\varepsilon_I=E_I-m_\psi$. The prefactor has been chosen such that the NR field $\psi^\pm$'s kinetic term is normalized at leading order as in Eq.~\eqref{eq:EFT_Lag}.

In the NR limit, $|\Theta_I|\ll|\Phi_I|$. The large component $\Psi_I$ satisfies
\begin{equation}
\biggl[-\vect{\sigma}\cdot\bigl(\nabla -i\vect{A}(\vect{x})\bigr) \,\frac{1}{2m_\psi+\varepsilon_I-\Phi(\vect{x})}\, \vect{\sigma}\cdot\bigl(\nabla -i\vect{A}(\vect{x})\bigr) + \Phi(\vect{x}) \biggr] \Psi_I(\vect{x}) = \varepsilon_I \Psi_I(\vect{x})\,.
\end{equation}
At leading order, we replace $\frac{1}{2m_\psi+\varepsilon_I-\Phi(\vect{x})}\to\frac{1}{2m_\psi}$, and recover the NR Schr\"odinger equation:
\begin{equation}
\biggl[-\frac{\bigl(\nabla -i\vect{A}(\vect{x})\bigr)^2}{2m_\psi} +\frac{1}{2m_\psi} \vect{\sigma}\cdot\bigl(\nabla\times\vect{A}(\vect{x})\bigr) +\Phi(\vect{x})\biggr] \Psi_I(\vect{x}) = \varepsilon_I \Psi_I(\vect{x})\,.
\end{equation}
Corrections to this equation can be incorporated order by order if needed.

\section{Projection of Angular Momentum Operators}
\label{app:angular_momentum}

In this appendix, we detail the steps that lead to the numbers $\lambda_{S,j}=-\frac{1}{3}$, $\lambda_{L,j}=-\frac{4}{3}$ in the case of $\alpha$-RuCl$_3$, following the projection of angular momentum operators $\vect{S}_e$, $\vect{L}_e$ in Eq.~\eqref{eq:proj}. The formation of effective ionic spins $S_j=\frac{1}{2}$ is due to the combined effect of crystal fields and spin-orbit coupling~\cite{trebst2017kitaev}. First, octahedral crystal fields split the five degenerate 3$d$ orbitals ($\ell=2$) of Ru$^{3+}$ into two higher-energy $e_g$ orbitals and three lower-energy $t_{2g}$ orbitals with an effective orbital moment $\ell_\text{eff}=1$. The energy difference between the $e_g$ and $t_{2g}$ orbitals is $\OO(\text{eV})$, rendering the (unoccupied) $e_g$ orbitals irrelevant for the discussion. For the $t_{2g}$ orbitals, spin-orbit coupling further splits these $\ell_\text{eff}=1$ states into $j_\text{eff}=\frac{3}{2}$ and $\frac{1}{2}$. With five 3$d$ electrons, the lower-energy $j_\text{eff}=\frac{3}{2}$ states are fully occupied, while the higher-energy $j_\text{eff}=\frac{1}{2}$ Kramers doublet is occupied by a single electron --- it is this electron that contributes to the magnetic order. Therefore, the goal is to project the angular momentum operators $\vect{S}$, $\vect{L}$ (dropping subscript $e$ from here on for simplicity) onto the $j_\text{eff}=\frac{1}{2}$ subspace.

The first step is to project $\vect{L}$ onto the $t_{2g}$ subspace. The $t_{2g}$ states are denoted by $d_{yz}$, $d_{zx}$, $d_{xy}$. The angular part of their wavefunctions are linear combinations of spherical harmonics $Y_{\ell=2}^m(\theta,\phi)$ (see e.g.\ Ref.~\cite{Coey_2001}); equivalently, these $t_{2g}$ states are linear combinations of $|\ell, m_\ell\rangle$ states with $\ell=2$:
\begin{equation}
|d_{yz}\rangle = \frac{i}{\sqrt{2}}\bigl(|2,1\rangle +|2,-1\rangle\bigr)\,,\quad
|d_{zx}\rangle = -\frac{1}{\sqrt{2}} \bigl(|2,1\rangle -|2,-1\rangle\bigr)\,,\quad
|d_{xy}\rangle = -\frac{i}{\sqrt{2}} \bigl(|2,2\rangle -|2,-2\rangle\bigr) \,.
\end{equation}
To compute $\mathcal{P}_{t_{2g}}\vect{L}\,\mathcal{P}_{t_{2g}}$, with the projection operator
\begin{equation}
\mathcal{P}_{t_{2g}} = |d_{yz}\rangle\langle d_{yz}| + |d_{zx}\rangle\langle d_{zx}| + |d_{xy}\rangle\langle d_{xy}|\,,
\end{equation}
we make use of the familiar formulae
\begin{equation}
\langle \ell',m'_\ell| L_z |\ell, m_\ell\rangle = m_\ell\,\delta_{\ell',\ell}\delta_{m'_\ell,m_\ell} \,,\qquad
\langle \ell',m'_\ell| L_\pm |\ell, m_\ell\rangle = \sqrt{(\ell\mp m_\ell)(\ell\pm m_\ell+1)} \,\delta_{\ell',\ell}\delta_{m'_\ell,m_\ell\pm 1}\,,
\end{equation}
where $L_\pm = L_x \pm iL_y$, and obtain, for the matrix representation in the $|d_{yz}\rangle$, $|d_{zx}\rangle$, $|d_{xy}\rangle$ basis:
\begin{equation}
\mathcal{P}_{t_{2g}}\, L_z \,\mathcal{P}_{t_{2g}} \;\dot{=}\; 
\begin{pmatrix}
0 & \;i\; & \;\;0\;\; \\
-i & \;0\; & \;\;0\;\; \\
0 & \;0\; & \;\;0\;\;
\end{pmatrix}\,,\qquad\quad
\mathcal{P}_{t_{2g}}\, L_\pm \,\mathcal{P}_{t_{2g}} \;\dot{=}\; 
\begin{pmatrix}
0 & \;0\; & \pm 1 \\
0 & \;0\; & \;\;i\;\; \\
\mp 1 & -i & \;\;0\;\;
\end{pmatrix}\,.
\label{eq:proj_t2g}
\end{equation}
These might not look familiar, but they are nothing but $\ell=1$ angular momentum operators in the $|p_x\rangle$, $|p_y\rangle$, $|p_z\rangle$ basis, which is related to the $|\ell, m_\ell\rangle$ basis with $\ell=1$ by~\cite{Coey_2001}
\begin{equation}
|p_x\rangle = -\frac{1}{\sqrt{2}}\bigl( |1,1\rangle -|1,-1\rangle\bigr)\,,\qquad
|p_y\rangle = \frac{i}{\sqrt{2}} \bigl( |1,1\rangle +|1,-1\rangle\bigr)\,,\qquad
|p_z\rangle = |1,0\rangle\,.
\end{equation}
The angular momentum operators in this basis read
\begin{equation}
L_z \;\dot{=}\; 
\begin{pmatrix}
\;\;0\;\; & -i & \;0\;\; \\
\;\;i\;\; & 0 & \;0\;\; \\
\;\;0\;\; & 0 & \;0\;\;
\end{pmatrix}\,,\qquad\quad
L_\pm \;\dot{=}\; 
\begin{pmatrix}
0 & \;0\; & \mp 1 \\
0 & \;0\; & \;\;-i\;\; \\
\pm 1 & \;i\; & \;\;0\;\;
\end{pmatrix}\,.
\label{eq:l1}
\end{equation}
Comparing Eq.~\eqref{eq:proj_t2g} and \eqref{eq:l1}, we see that $\vect{L}$ acts as an effective angular momentum with $\ell=1$ on the $t_{2g}$ subspace:
\begin{equation}
\mathcal{P}_{t_{2g}}\,\vect{L}\,\mathcal{P}_{t_{2g}}=-\vect{L}_\text{eff}^{(\ell=1)}\,.
\end{equation}

The second step is to combine this effective orbital angular momentum $\ell_\text{eff}=1$ with the electron's spin $s=\frac{1}{2}$. This follows the standard angular momentum addition, and we obtain, for the $j_\text{eff}=\frac{1}{2}$ states:
\begin{eqnarray}
\bigl|j_\text{eff} = \tfrac{1}{2}, m_{j_\text{eff}} = \tfrac{1}{2}\bigr\rangle &=& \sqrt{\tfrac{2}{3}} \,\bigl|m_{\ell_\text{eff}} = 1, m_s = -\tfrac{1}{2}\bigr\rangle -\sqrt{\tfrac{1}{3}} \,\bigl|m_{\ell_\text{eff}} = 0, m_s = \tfrac{1}{2}\bigr\rangle \,,\\
\bigl|j_\text{eff} = \tfrac{1}{2}, m_{j_\text{eff}} = -\tfrac{1}{2}\bigr\rangle &=& \sqrt{\tfrac{1}{3}} \,\bigl|m_{\ell_\text{eff}} = 0, m_s = -\tfrac{1}{2}\bigr\rangle -\sqrt{\tfrac{2}{3}} \,\bigl|m_{\ell_\text{eff}} = -1, m_s = \tfrac{1}{2}\bigr\rangle \,,
\end{eqnarray}
where the coefficients are Clebsch-Gordan coefficients. It is now straightforward to project $\vect{L}_\text{eff}$ and $\vect{S}$ onto the $j_\text{eff}=\frac{1}{2}$ subspace:
\begin{eqnarray}
&&
\mathcal{P}_{j_\text{eff}=\frac{1}{2}}\, L_z^\text{eff}\,\mathcal{P}_{j_\text{eff}=\frac{1}{2}} \dot{=} 
\begin{pmatrix}
\;\frac{2}{3}\;\; & 0\; \\
\;0\;\; & -\frac{2}{3}\;
\end{pmatrix}
\,,\quad
\mathcal{P}_{j_\text{eff}=\frac{1}{2}}\, L_+^\text{eff}\,\mathcal{P}_{j_\text{eff}=\frac{1}{2}} \dot{=} 
\begin{pmatrix}
\;0\; & \;\;\frac{4}{3}\; \\
\;0\; & \;\;0\;
\end{pmatrix} \,,\\
&&
\mathcal{P}_{j_\text{eff}=\frac{1}{2}}\, S_z\,\mathcal{P}_{j_\text{eff}=\frac{1}{2}} \dot{=} 
\begin{pmatrix}
-\frac{1}{6} & \;0\; \\
0 & \;\frac{1}{6}\;
\end{pmatrix}
\,,\quad
\mathcal{P}_{j_\text{eff}=\frac{1}{2}}\, S_+\,\mathcal{P}_{j_\text{eff}=\frac{1}{2}} \dot{=} 
\begin{pmatrix}
\;0\;\; & -\frac{1}{3}\; \\
\;0\;\; & 0\;
\end{pmatrix} \,.
\end{eqnarray}
We see that both $\vect{L}_\text{eff}$ and $\vect{S}$ are proportional to $\vect{J}_\text{eff} = \frac{\vect{\sigma}}{2}$ (identified as the total ionic spin as discussed above) when acting on the $j_\text{eff}=\frac{1}{2}$ subspace. So finally, we obtain
\begin{equation}
\mathcal{P}_{j_\text{eff}=\frac{1}{2}}\,\vect{L}\,\mathcal{P}_{j_\text{eff}=\frac{1}{2}} 
= -\mathcal{P}_{j_\text{eff}=\frac{1}{2}}\,\vect{L}_\text{eff}\,\mathcal{P}_{j_\text{eff}=\frac{1}{2}}
= -\frac{4}{3} \vect{J}_\text{eff} \,,\qquad
\mathcal{P}_{j_\text{eff}=\frac{1}{2}}\,\vect{S}\,\mathcal{P}_{j_\text{eff}=\frac{1}{2}} 
= -\frac{1}{3} \vect{J}_\text{eff} \,.
\end{equation}
%

\section{Velocity Integrals}
\label{app:v_int}

When the velocity dependent rate $\Gamma(\vect{v})$, given by Eq.~\eqref{eq:Gamma}, is convoluted with the incoming DM's velocity distribution $f_\chi(\vect{v})$ to yield the total rate, Eq.~\eqref{eq:total_rate}, we encounter the following scalar, vector and tensor velocity integrals:
\begin{eqnarray}
g_0(\vect{q},\omega) &\equiv& \int d^3v\, f_\chi(\vect{v}) \,2\pi\delta(\omega-\omega_{\vect{q}}) \,,\\
\vect{g}_1(\vect{q},\omega) &\equiv& \int d^3v\, f_\chi(\vect{v}) \,2\pi\delta(\omega-\omega_{\vect{q}}) \,\vect{v}_\chi \,,\\
\tens{g}_2(\vect{q},\omega) &\equiv& \int d^3v\, f_\chi(\vect{v}) \,2\pi\delta(\omega-\omega_{\vect{q}})\, \vect{v}_\chi\vect{v}_\chi \,,
\end{eqnarray}
where $\vect{v}_\chi= \vect{v}-\frac{\vect{q}}{2m_\chi}$, and $\omega_{\vect{q}} = \vect{q}\cdot\vect{v}-\frac{q^2}{2m_\chi}$. From the expressions of differential rates $\Sigma_\nu(\vect{q})$ throughout Sec.~\ref{sec:benchmarks}, it should be easy to see how these integrals emerge. Note that for velocity-independent interactions, only the scalar integral $g_0$ appears~\cite{Griffin:2018bjn,Coskuner:2019odd,Trickle:2019nya}.

As we now show, all three velocity integrals above can be evaluated analytically for a boosted and truncated Maxwell-Boltzmann distribution, which we assume in this work:
\begin{equation}
f_\chi(\vect{v}) = \frac{1}{N_0} e^{-(\vect{v}+\vve)^2/v_0^2} \,\Theta \bigl(\vesc-|\vect{v} +\vve|\bigr) \,,
\label{eq:fchi_MB}
\end{equation}
where
\begin{equation}
N_0 = \pi^{3/2} v_0^2 \Biggl[ v_0\,\text{erf} \bigl(\vesc/v_0\bigr) -\frac{2\,\vesc}{\sqrt{\pi}} \exp\bigl(-\vesc^2/v_0^2\bigr)\Biggr],
\end{equation}
and we take $v_0=230\,\text{km}/\text{s}$, $\vesc=600\,\text{km}/\text{s}$, $\ve=240\,\text{km}/\text{s}$. For all the target materials considered in Sec.~\ref{sec:benchmarks}, the rates are insensitive to the direction of $\vve$. The analytic results obtained here are key to efficient rate calculations, as they reduce the six-dimensional integral $\int d^3v \int d^3q$ to just a three-dimensional integral $\int d^3q$, which we then compute numerically.

First, the scalar integral $g_0$ follows from Refs.~\cite{Griffin:2018bjn,Coskuner:2019odd,Trickle:2019nya}. Shifting $\vect{v}\to\vect{v}-\vve$, we obtain
\begin{eqnarray}
g_0(\vect{q},\omega) &=& \frac{2\pi}{N_0} \int d^3v\, e^{-v^2/v_0^2} \,\Theta(\vesc-v) \,\delta\biggl(\vect{q}\cdot\vect{v} -\vect{q}\cdot\vve -\frac{q^2}{2m_\chi} -\omega\biggr) \nonumber\\
&=& \frac{4\pi^2}{N_0} \int_0^{\vesc} dv\, v^2\, e^{-v^2/v_0^2} \int_{-1}^1 d\cos\theta \,\delta\biggl(qv\cos\theta -\vect{q}\cdot\vve-\frac{q^2}{2m_\chi} -\omega \biggr) \,.
\end{eqnarray}
Let us define
\begin{equation}
v_* \equiv \frac{1}{q} \biggl(\vect{q}\cdot\vve+\frac{q^2}{2m_\chi} +\omega\biggr)\,,
\qquad
v_- \equiv \min \bigl( |v_*| \,,\, \vesc \bigr) \,.
\label{eq:vstar_def}
\end{equation}
We then obtain\\[2pt]
\begin{eqnarray}
g_0(\vect{q},\omega) &=&  \frac{4\pi^2}{N_0 q} \int_0^{\vesc} dv\, v\, e^{-v^2/v_0^2} \int_{-1}^1 d\cos\theta \,\delta\biggl(\cos\theta-\frac{v_*}{v}\biggr) \nonumber\\
&=& \frac{4\pi^2}{N_0 q} \int_{v_-}^{\vesc} dv\, v\, e^{-v^2/v_0^2} \nonumber\\
&=& \frac{2\pi^2 v_0^2}{N_0 q} \Bigl(e^{-v_-^2/v_0^2} -e^{-\vesc^2/v_0^2}\Bigr)\,.
\end{eqnarray}

Next, the vector integral $\vect{g}_1$ can be decomposed as
\begin{eqnarray}
\vect{g}_1(\vect{q},\omega) &=& \int d^3v\, f_\chi(\vect{v}) (\vect{v}+\vve) -\biggl(\vve+\frac{\vect{q}}{2m_\chi}\biggr) \,g_0(\vect{q},\omega) \,.
\end{eqnarray}
The first term can be computed by shifting $\vect{v}\to\vect{v}-\vve$ as before, but this time the integrand also depends on the azimuthal angle $\phi$:
\begin{eqnarray}
\int d^3v\, f_\chi(\vect{v}) (\vect{v}+\vve) &=&
\frac{2\pi}{N_0} \int d^3v\, e^{-v^2/v_0^2} \,\Theta(\vesc-v) \,\delta\biggl(\vect{q}\cdot\vect{v} -\vect{q}\cdot\vve -\frac{q^2}{2m_\chi} -\omega\biggr)\,\vect{v} \nonumber\\
&=& \frac{4\pi^2}{N_0} \int_0^{\vesc} dv\, v^3\, e^{-v^2/v_0^2} \int_{-1}^1 d\cos\theta \,\delta\biggl(qv\cos\theta -\vect{q}\cdot\vve-\frac{q^2}{2m_\chi} -\omega \biggr) \nonumber\\
&&\qquad \int_0^{2\pi}\frac{d\phi}{2\pi} \bigl[ \cos\theta \,\vect{\hat q} +\sin\theta \,(\cos\phi \,\vect{\hat n}_1 +\sin\phi \,\vect{\hat n}_2)\bigr] \nonumber\\
&=& \frac{4\pi^2}{N_0} \,\vect{\hat q}\int_0^{\vesc} dv\, v^3\, e^{-v^2/v_0^2} \int_{-1}^1 d\cos\theta \,\delta\biggl(qv\cos\theta -\vect{q}\cdot\vve-\frac{q^2}{2m_\chi} -\omega \biggr) \cos\theta \nonumber\\
&=& \frac{4\pi^2}{N_0q} \,\vect{\hat q}\int_0^{\vesc} dv\, v^2\, e^{-v^2/v_0^2} \int_{-1}^1 d\cos\theta \,\delta\biggl(\cos\theta-\frac{v_*}{v}\biggr) \cos\theta \nonumber\\
&=& \frac{4\pi^2v_*}{N_0q} \,\vect{\hat q}\int_{v_-}^{\vesc} dv\, v\, e^{-v^2/v_0^2} = v_*\,\vect{\hat q}\,g_0(\vect{q},\omega)\,,
\label{eq:v_int_vect}
\end{eqnarray}
where $\vect{\hat n}_1$, $\vect{\hat n}_2$ are orthogonal unit vectors in the plane perpendicular to $\vect{q}$. Plugging in the definition of $v_*$ in Eq.~\eqref{eq:vstar_def}, we obtain
\begin{equation}
\vect{g}_1(\vect{q},\omega) = \biggl[\frac{\omega}{q}\,\vect{\hat q} -(\mathbb{1}-\vect{\hat q}\vect{\hat q})\cdot\vve\biggr] \,g_0(\vect{q},\omega)\,.
\end{equation}

Finally, we compute the tensor integral $\tens{g}_2$, which can be similarly decomposed as
\begin{eqnarray}
\tens{g}_2(\vect{q},\omega) &=& \int d^3v\, f_\chi(\vect{v}) (\vect{v}+\vve)(\vect{v}+\vve) -\biggl[\biggl(\vve+\frac{\vect{q}}{2m_\chi}\biggr)\,\vect{\hat q} +\vect{\hat q}\,\biggl(\vve+\frac{\vect{q}}{2m_\chi}\biggr)\biggr]\, v_* \,g_0(\vect{q},\omega) \nonumber\\
&& +\biggl(\vve+\frac{\vect{q}}{2m_\chi}\biggr)\biggl(\vve+\frac{\vect{q}}{2m_\chi}\biggr)\,g_0(\vect{q},\omega) \nonumber\\
&=& \int d^3v\, f_\chi(\vect{v}) (\vect{v}+\vve)(\vect{v}+\vve) \nonumber\\
&&
+\biggl\{ \biggl[\frac{\omega}{q}\,\vect{\hat q} -(\mathbb{1}-\vect{\hat q}\vect{\hat q})\cdot\vve\biggr] \biggl[\frac{\omega}{q}\,\vect{\hat q} -(\mathbb{1}-\vect{\hat q}\vect{\hat q})\cdot\vve\biggr] -v_*^2\, \vect{\hat q}\vect{\hat q}\biggr\}\, g_0(\vect{q},\omega)\,,
\end{eqnarray}
where we have used Eq.~\eqref{eq:v_int_vect}. The remaining integral can be evaluated similarly to Eq.~\eqref{eq:v_int_vect}:
\begin{eqnarray}
&&\int d^3v\, f_\chi(\vect{v}) (\vect{v}+\vve)(\vect{v}+\vve) \nonumber\\[4pt]
&=&
\frac{4\pi^2}{N_0} \int_0^{\vesc} dv\, v^4\, e^{-v^2/v_0^2} \int_{-1}^1 d\cos\theta \,\delta\biggl(qv\cos\theta -\vect{q}\cdot\vve-\frac{q^2}{2m_\chi} -\omega \biggr) \nonumber\\
&&\qquad \int_0^{2\pi}\frac{d\phi}{2\pi} \bigl[ \cos\theta \,\vect{\hat q} +\sin\theta (\cos\phi \,\vect{\hat n}_1 +\sin\phi \,\vect{\hat n}_2)\bigr]\otimes \bigl[ \cos\theta \,\vect{\hat q} +\sin\theta (\cos\phi \,\vect{\hat n}_1 +\sin\phi \,\vect{\hat n}_2)\bigr] \nonumber\\[4pt]
&=& \frac{4\pi^2}{N_0q} \int_0^{\vesc} dv\, v^3\, e^{-v^2/v_0^2}  \int_{-1}^1 d\cos\theta \,\delta\biggl(\cos\theta-\frac{v_*}{v}\biggr)\biggl[\cos^2\theta \,\vect{\hat q}\vect{\hat q} +\frac{1}{2}\sin^2\theta(\mathbb{1}-\vect{\hat q}\vect{\hat q}) \biggr] \nonumber\\[4pt]
&=&  \frac{4\pi^2}{N_0q} \int_{v_-}^{\vesc} dv\, v\, e^{-v^2/v_0^2} \,\biggl[ v_*^2\,\vect{\hat q}\vect{\hat q} +\frac{1}{2}(v^2-v_*^2)(\mathbb{1}-\vect{\hat q}\vect{\hat q}) \biggr] \nonumber\\[4pt]
&=& v_*^2\,\vect{\hat q}\vect{\hat q} \,g_0(\vect{q},\omega) +(\mathbb{1}-\vect{\hat q}\vect{\hat q})\, \frac{\pi^2 v_0^2}{N_0q} \Bigl[ (v_0^2-v_*^2+v_-^2)\,e^{-v_-^2/v_0^2} -(v_0^2-v_*^2+\vesc^2)\,e^{-\vesc^2/v_0^2} \Bigr] \nonumber\\[4pt]
&=& v_*^2\,\vect{\hat q}\vect{\hat q} \,g_0(\vect{q},\omega) +(\mathbb{1}-\vect{\hat q}\vect{\hat q})\, \frac{\pi^2 v_0^2}{N_0q} \Bigl[ v_0^2\,e^{-v_-^2/v_0^2} -(v_0^2-v_-^2+\vesc^2)\,e^{-\vesc^2/v_0^2} \Bigr]\,,
\end{eqnarray}
where we have used $\vect{\hat n}_1\vect{\hat n}_1+\vect{\hat n}_2\vect{\hat n}_2 = \mathbb{1}-\vect{\hat q}\vect{\hat q}$. Therefore,
\begin{eqnarray}
\tens{g}_2(\vect{q},\omega) &=& \biggl[\frac{\omega}{q}\,\vect{\hat q} -(\mathbb{1}-\vect{\hat q}\vect{\hat q})\cdot\vve\biggr] \biggl[\frac{\omega}{q}\,\vect{\hat q} -(\mathbb{1}-\vect{\hat q}\vect{\hat q})\cdot\vve\biggr]\, g_0(\vect{q},\omega) \nonumber\\
&& +(\mathbb{1}-\vect{\hat q}\vect{\hat q})\, \frac{\pi^2 v_0^2}{N_0q} \Bigl[ v_0^2\,e^{-v_-^2/v_0^2} -(v_0^2-v_-^2+\vesc^2)\,e^{-\vesc^2/v_0^2} \Bigr] \,.
\end{eqnarray}

The following relations between the velocity integrals often help simplify the calculation:
\begin{equation}
\vect{q}\cdot\vect{g}_1(\vect{q},\omega)=\omega\,g_0(\vect{q},\omega) \,,\qquad
\vect{q}\cdot\tens{g}_2(\vect{q},\omega)=\tens{g}_2(\vect{q},\omega)\cdot\vect{q}=\omega\,\vect{g}_1(\vect{q},\omega) \,.
\end{equation}
They follow from $\vect{q}\cdot\vect{v}_\chi=\omega_{\vect{q}}$, and can be easily checked using the explicit expressions above.

\section{Estimation of Single Phonon Excitation Rate in YIG}
\label{app:YIG_phonon}

In this appendix, we explain the analytic estimation that results in the dashed curve in Fig.~\ref{fig:SD}. For the standard SD interaction considered in Sec.~\ref{subsec:SD}, the single phonon excitation rate is 
\begin{equation}
\Gamma(\vect{v}) = \frac{4g_\chi^2g_e^2}{m_V^4}\frac{1}{\Omega} \int \frac{d^3q}{(2\pi)^3} \sum_{\nu}\, 2\pi\,\delta\bigl(\omega_{\nu,\vect{k}}-\omega_{\vect{q}}\bigr)
 \,\bigl| \vect{F}_{S,\nu}^{(e)}\bigr|^2\,,
\end{equation}
where
\begin{equation}
\vect{F}_{S,\nu}^{(e)} (\vect{q}) =\sum_{j=\text{Fe}^{3+}} e^{-W_j(\vect{q})} e^{i\vect{G}\cdot\vect{x}_j^0} \,\frac{\vect{q}\cdot\vect{\epsilon}_{\nu,\vect{k},j}^*}{\sqrt{2m_j\omega_{\nu,\vect{k}}}}\,\langle \vect{S}_e\rangle_j \,.
\end{equation}
See Eqs.~\eqref{eq:Gamma}, \eqref{eq:Sigma_SD_phonon} and \eqref{eq:FX_def}. For YIG, $\nu$ runs from 1 to 240. However, since DM has same-sign couplings to all the Fe$^{3+}$ ions (and zero couplings to the other ions), we expect acoustic phonons to give an $\OO(1)$ contribution to the total rate at low momentum transfer. Further, the dot product $\vect{q}\cdot\vect{\epsilon}_{\nu,\vect{k},j}^*$ in $\vect{F}_{S,\nu}^{(e)}$ singles out the longitudinal acoustic branch, $\nu=3$, which has the following general properties at low momentum~\cite{Cox:2019cod}:
\begin{equation}
\omega_{\nu=3,\vect{k}}\simeq c_s k \,,\qquad
\vect{\epsilon}_{\nu=3,\vect{k},j}\simeq\sqrt{\frac{m_j}{m_\text{cell}}}\,\vect{\hat{k}}\,,
\end{equation}
where $c_s$ is the longitudinal acoustic sound speed. Also, we can set $\vect{G}=\vect{0}$, $\vect{k}=\vect{q}$, and $W_j\simeq 0$ at low $q$. Therefore,
\begin{equation}
\vect{F}_{S,\nu=3}^{(e)} (\vect{q}) \simeq \sqrt{\frac{q}{2m_\text{cell} c_s}} \sum_{j=\text{Fe}^{3+}} \langle \vect{S}_e\rangle_j
= \sqrt{\frac{q}{2m_\text{cell} c_s}}\,\vect{S}_\text{cell}\,,
\end{equation}
and the velocity-dependent rate becomes
\begin{eqnarray}
\Gamma(\vect{v}) &\simeq& \frac{g_\chi^2g_e^2}{m_V^4}\frac{2S_\text{cell}^2}{\Omega m_\text{cell} c_s} \int \frac{d^3q}{(2\pi)^2}\, \delta\bigl(c_s q-\omega_{\vect{q}}\bigr)\,q \nonumber\\
&=& \frac{g_\chi^2g_e^2}{m_V^4}\frac{S_\text{cell}^2}{\pi\Omega m_\text{cell} c_s} \int dq\, q^3 \int d\cos\theta \,\delta\biggl(c_sq -qv\cos\theta +\frac{q^2}{2m_\chi}\biggr) \nonumber\\
&=& \frac{g_\chi^2g_e^2}{m_V^4}\frac{S_\text{cell}^2}{\pi\Omega m_\text{cell} c_s}\frac{1}{v} \int dq\, q^2 \,\Theta\bigl(v-v_\text{min}(q)\bigr)\,,
\end{eqnarray}
where
\begin{equation}
v_\text{min}(q) \equiv \frac{q}{2m_\chi}+c_s\,.
\end{equation}
Now we can write the total rate per unit target mass in terms of the commonly used $\eta$ function, defined by
\begin{equation}
\eta(v_\text{min}) \equiv \int d^3v \frac{f(\vect{v})}{v} \,\Theta(v-v_\text{min})\,.
\end{equation}
The result is
\begin{eqnarray}
R &\simeq& \frac{1}{\rho_T} \frac{\rho_\chi}{m_\chi} \frac{g_\chi^2g_e^2}{m_V^4}\frac{S_\text{cell}^2}{\pi\Omega m_\text{cell} c_s}\int dq\, q^2 \,\eta\bigl(v_\text{min}(q)\bigr) \nonumber\\
&=& \frac{1}{\pi c_s} \biggl(\frac{S_\text{cell}}{m_\text{cell}}\biggr)^2 \frac{g_\chi^2g_e^2}{m_V^4}\frac{\rho_\chi}{m_\chi} \int dq\, q^2 \,\eta\bigl(v_\text{min}(q)\bigr)\,.
\end{eqnarray}
This is the formula we use to estimate the single phonon excitation rate in YIG in Sec.~\ref{subsec:SD}. The material parameters are $c_s=7.2$\,km/s~\cite{Clark:1961}, $S_\text{cell}=10$, $m_\text{cell}=\rho_T\Omega$, with $\rho_T = 4.95$\,g/cm$^3$, $\Omega= 990.683$\,\AA$^3$. The analytic expression for the $\eta(v_\text{min})$ function for the Maxwell-Boltzmann distribution of Eq.~\eqref{eq:fchi_MB} can be found in {\it e.g.}\ Ref.~\cite{Trickle:2019nya}. Since the $\eta$ function has support up to $q_\text{max}\simeq 2m_\chi (\ve+\vesc)$, we cut off the dashed curve in Fig.~\ref{fig:SD} at the $m_\chi$ value for which $q_\text{max}$ reaches $\frac{\pi}{\Omega^{1/3}}$, roughly the edge of the 1BZ.
\vspace{10pt}

\bibliography{refs_EFT}

\end{document}